\documentclass[10pt, english,onecolumn,prd,superscriptaddress,nofootinbib,preprintnumbers,showpacs,floatfix]{revtex4-1}
\usepackage{amsmath}
\usepackage{amsfonts}
\usepackage{amssymb}
\usepackage{graphicx}
\usepackage{ulem}
\usepackage[usenames,dvipsnames]{xcolor}
\usepackage{hyperref}
\hypersetup{colorlinks=true,
citecolor=MidnightBlue, linkcolor=MidnightBlue, urlcolor=MidnightBlue}
\usepackage{tikz}
\definecolor{purple}{rgb}{1,0,1}
\definecolor{lime}{HTML}{A6CE39} % needs xcolor

\newcommand{\R}{\mathcal{R}}

\newcommand{\h}{\mathcal{H}}
\newcommand{\OA}{\Omega_A}

\newcommand{\lp}{\left(}
\newcommand{\rp}{\right)}
\newcommand{\lb}{\left[}
\newcommand{\rb}{\right]}

\newcommand{\lsim}   {\mathrel{\mathop{\kern 0pt \rlap
  {\raise.2ex\hbox{$<$}}}
  \lower.9ex\hbox{\kern-.190em $\sim$}}}
\newcommand{\gsim}   {\mathrel{\mathop{\kern 0pt \rlap
  {\raise.2ex\hbox{$>$}}}
  \lower.9ex\hbox{\kern-.190em $\sim$}}}
\newcommand{\bw}{\begin{widetext}\begin{equation}}
\newcommand{\ew}{\end{equation}\end{widetext}}
\newcommand{\be}{\begin{equation}}
\newcommand{\ee}{\end{equation}}
\newcommand{\bea}{\begin{eqnarray}}
\newcommand{\eea}{\end{eqnarray}}

\newcommand{\e}{\mathrm{e}}

\newcommand{\nn}{\nonumber}

\newcommand{\F}{\OA+\phi}
\newcommand{\Fp}{\lp\OA+\phi\rp}

\newcommand{\f}[2]{\frac{#1}{#2}}

\usepackage{ulem}
\usepackage[usenames,dvipsnames]{xcolor}
\usepackage{hyperref}
\hypersetup{colorlinks=true,
citecolor=MidnightBlue, linkcolor=MidnightBlue, urlcolor=MidnightBlue}
\usepackage{tikz}
\newcommand{\orcidicon}{%
	\begin{tikzpicture}
	\draw[lime, fill=lime] (0,0)
		circle [radius=0.16]
		node[white] {{\fontfamily{qag}\selectfont \tiny ID}};
	\draw[white, fill=white] (-0.0625,0.095)
		circle [radius=0.007];
	\end{tikzpicture}	\hspace{-2mm}
}
\newcommand\orcidTiberiu{{\href{https://orcid.org/0000-0002-1990-9172}{\orcidicon}}}
\newcommand\orcidFrancisco{{\href{https://orcid.org/0000-0002-9388-8373}{\orcidicon}}}
\begin{document}

\title{Beyond Einstein's General Relativity: \\
Hybrid metric-Palatini gravity and curvature-matter couplings}

%=================================================================
\author{Tiberiu Harko\orcidTiberiu\!\!}
\email{tiberiu.harko@aira.astro.ro}
\affiliation{Astronomical Observatory, 19 Ciresilor Street, 400487 Cluj-Napoca, Romania,}
\affiliation{Faculty of Physics, Babes-Bolyai University, 1 Kogalniceanu Street,
400084 Cluj-Napoca, Romania}
\affiliation{School of Physics, Sun Yat-Sen University,  Xingang  Road, 510275 Guangzhou, People's
Republic of China}
%-----------------------------------------------------------------
\author{Francisco S. N. Lobo\orcidFrancisco\!\!}
\email{fslobo@fc.ul.pt}
\affiliation{Instituto de Astrofísica e Ci\^encias do Espa\c{c}o, Faculdade de Ci\^encias da Universidade de Lisboa, Edif\'icio C8, Campo Grande, P-1749-016, Lisbon, Portugal}
%-----------------------------------------------------------------

\date{\today}

%%%%%%%%%%%%%%%%%%%%%%%%%%%%%%%%%%%%%%%%%%%%%%%%%%%%%%%%%%%%%%%%%%%%%%%%%%
\begin{abstract}
Einstein's General Relativity (GR) is possibly one of the greatest intellectual achievements ever conceived by the human mind. In fact, over the last century, GR has proven to be an extremely successful theory, with a well established experimental footing, at least for weak gravitational fields. Its predictions range from the existence of black holes and gravitational radiation (now confirmed) to the cosmological models. Indeed, a central theme in modern Cosmology is the perplexing fact that the Universe is undergoing an accelerating expansion, which represents a new imbalance in the governing gravitational equations. The cause of the late-time cosmic acceleration remains an open and tantalizing question, and has forced theorists and experimentalists to question whether GR is the correct relativistic theory of gravitation.
This has spurred much research in modified theories of gravity, where extensions of the Hilbert-Einstein action describe the gravitational field, in particular, $f(R)$ gravity, where $R$ is the curvature scalar. In this review, we perform a detailed theoretical and phenomenological analysis of specific modified theories of gravity and investigate their astrophysical and cosmological applications.

We present essentially two largely explored extensions of $f(R)$ gravity, namely: (i) the hybrid metric-Palatini theory; (ii) and modified gravity with curvature-matter couplings.
Relative to the former, it has been established that both metric and Palatini versions of $f(R)$ gravity possess interesting features but also manifest severe drawbacks. A hybrid combination, containing elements from both of these formalisms, turns out to be very successful in accounting for the observed phenomenology and avoids some drawbacks of the original approaches.
Relative to the curvature-matter coupling theories, these offer interesting extensions of $f(R)$ gravity, where the explicit nonminimal couplings between an arbitrary function of the scalar curvature $R$ and the Lagrangian density of matter, induces a non-vanishing covariant derivative of the energy-momentum tensor, which implies nongeodesic motion and consequently leads to the appearance of an extra force.
We extensively explore both theories in a plethora of applications, namely, the weak-field limit, galactic and extragalactic dynamics, cosmology, stellar-type compact objects, irreversible matter creation processes and the quantum cosmology of a specific curvature-matter coupling theory.
\end{abstract}
%%%%%%%%%%%%%%%%%%%%%%%%%%%%%%%%%%%%%%%%%%%%%%%%%%%%%%%%%%%%%%%%%%%%%%%%%%

\maketitle

%%%%%%%%%%%%%%%%%%%%%%%%%%%%%%%%%%%%%%%%%%%%%%%%%%%%%%%%%%%%%%%%%%%%%%%%%%
\tableofcontents
%%%%%%%%%%%%%%%%%%%%%%%%%%%%%%%%%%%%%%%%%%%%%%%%%%%%%%%%%%%%%%%%%%%%%%%%%%

%\mainmatter

\newpage

%%%%%%%%%%%%%%%%%%%%%%%%%%%%%%%%%%%%%%%%%%%%%%%%%%%%%%%%%%%%%%%%%%%%%%%%%%
\part{Introduction}
%%%%%%%%%%%%%%%%%%%%%%%%%%%%%%%%%%%%%%%%%%%%%%%%%%%%%%%%%%%%%%%%%%%%%%%%%%

%%%%%%%%%%%%%%%%%%%%%%%%%%%%%%%%%%%%%%%%%%%%%%%%%%%%%%%%%%%%%%%%%%%%%%%%%%
\subsection{Going beyond Einstein's General Relativity}
%%%%%%%%%%%%%%%%%%%%%%%%%%%%%%%%%%%%%%%%%%%%%%%%%%%%%%%%%%%%%%%%%%%%%%%%%%

Einstein's General Relativity (GR) is one of the greatest intellectual achievements ever conceived by the human mind. In GR, the basic assumption is that the gravitational field can be interpreted geometrically, and it is fully determined by the quantities that describe the intrinsic geometrical properties and structure of spacetime. This important idea has the fundamental implication that the spacetime geometry itself is locally induced by physical phenomena and processes, such as the distribution of matter and energy, or the motion of gravitating objects, and that space and time are not a priori determined absolute concepts.
In this context, one of the most remarkable evolutions in modern physics was the natural emergence of the ideas of the Riemannian geometry (already developed by the mid nineteenth century) as introduced in the framework of gravitational fields.  This evolution took place in the years 1909--1916, and is due mainly to Albert Einstein and his collaborator Marcel Grossmann, with David Hilbert also making important contributions to the theory.
Thus, in order to describe the properties of the gravitational field in the framework of GR, one must resort to the Einstein gravitational field equations, defined in a Riemannian geometry. These equations establish a deep connection between the geometric properties of the spacetime, and its matter content, providing a full description of both geometric characteristics of the spacetime, and of the dynamics of gravitating objects. Moreover, in addition to its theoretical beauty, GR provides an excellent description of the gravitational effects and phenomena. In particular, the weak field limit within the boundaries of the Solar System, and applications to Cosmology, have provided an excellent testing ground on both astrophysical and large, cosmological scales.

Indeed, during the last few decades Cosmology has evolved from being mainly a theoretical area of physics to a scientific domain supported by high precision observational data. Recent experiments and observations call upon state of the art technology in Astronomy and Astrophysics to provide detailed information on the contents and history of the Universe. The observational advances have led to the measurement of parameters that describe our Universe with increasing precision. The standard model of Cosmology, having its theoretical basis in GR, is remarkably successful in accounting for the observed features of the Universe.  However, a number of fundamental open questions still remain at the foundation of the standard model. In particular, we lack a fundamental understanding of the physical processes behind the  recent acceleration of the Universe \cite{Perlmutter:1998np,Riess:1998cb}.
Hence, in this context, modern astrophysical and cosmological models are plagued with two severe theoretical problems, namely, the dark energy and the dark matter enigmas. Relative to the latter, the dynamics of test particles around galaxies, as well as the corresponding mass discrepancy in galactic clusters, is commonly explained by postulating the existence of a hypothetical form of substance, called  dark matter, with a corresponding particle. Relative to the dark energy problem, high precision observational data has confirmed with startling evidence that the universe is undergoing a phase of accelerated expansion. This phase is one of the most important and challenging current problems in cosmology, and represents a new imbalance in the governing gravitational equations. The simplest scenario to explain the late-time cosmic speedup is to invoke the $\Lambda$CDM paradigm. However, if we assume that the cosmological constant constitutes the vacuum energy of the gravitational field, we are faced with an extremely embarrassing discrepancy of 120 orders of magnitude between the observed value and that predicted by quantum field theory. This is the celebrated cosmological constant problem.
Several other candidates, responsible for this expansion, have been proposed in the literature, in particular, dark energy models and modified theories of gravity, among others \cite{Avelino:2016lpj}.

In the framework of Riemannian geometry the simplest modifications of GR  are the $f(R)$ modified theories of gravity, in which the Ricci scalar $R$ is replaced in the Einstein-Hilbert action with an arbitrary function of the curvature scalar \cite{Nojiri:2010wj,Nojiri:2006ri,Nojiri:2003ft,Sotiriou:2008rp,DeFelice:2010aj,Capozziello:2011et,Capozziello:2002rd,Capozziello:2003tk,Capozziello:2007ec}. The main goal and physical justification of this extension is to give a reason to the recent cosmological observations that cannot find a satisfactorily explanation within the framework of GR, and especially for providing a credible alternative to the cosmological constant. In the standard cosmological model the accelerated expansion of the Universe is interpreted by the presence of the cosmological constant $\Lambda$, introduced by Einstein in 1917, and which represents around 68.3\% of the total matter-energy density of the Universe \cite{Aghanim:2018eyx}. However, the true nature of the dark energy, or of the cosmological constant, remains elusive.
As mentioned above, the cosmic acceleration may also be explained by considering some other modifications of GR,  which are different from the simple approach consisting in the formal addition of the cosmological constant to the field equations, and to the action principle. The cosmological observational evidences can also be explained by modifying the Einstein-Hilbert action, in which $f(R)$ theories provide such a framework that allows the straightforward generalization of the Einstein gravity.
An important property of $f(R)$ gravity is its relation to some classes of scalar-tensor theories. After performing a Legendre transformation, a scalar field non-minimally coupled to the curvature can be introduced. This procedure significantly simplifies the analysis of the theory, since it allows the extensive use of the mathematical and physical methods developed in scalar-tensor gravity.

%%%%%%%%%%%%%%%%%%%%%%%%%%%%%%%%%%%%%%%%%%%%%%%%%%%%%%%%%%%%%%%%%%%%%%%%%%
\subsection{Motivations for hybrid metric-Palatini gravity}
%%%%%%%%%%%%%%%%%%%%%%%%%%%%%%%%%%%%%%%%%%%%%%%%%%%%%%%%%%%%%%%%%%%%%%%%%%

$f(R)$ theories of gravity have been intensively investigated in both their metric \cite{Sotiriou:2006hs,Cognola:2007zu,Capozziello:2008rq,Motohashi:2017vdc,Motohashi:2019tyj,Lobo:2008sg} and Palatini formulations \cite{Li:2006ag,Borowiec:2011wd,Stachowski:2016zio}. For a review of the Palatini formalism in $f(R)$ gravity theories see \cite{Olmo:2011uz}.  In the metric approach to gravitational field theories one considers the metric tensor as the only dynamical variable in the action. On the other hand in the Palatini approach the connection is considered as an independent variable, independent of the metric tensor.  Hence in the Palatini approach the field equations are obtained by varying the action
with respect to both the metric and the connection.  It turns out that in case of Palatini $f(R)$ theory the scalar field is not dynamical, which implies that the Palatini formalism does not introduce additional degrees of freedom \cite{Olmo:2011uz}. This represents a basic difference between the metric and the Palatini formulations for this type of theories.

Despite representing a promising avenue for the extension of general relativity, $f(R)$ gravity possesses some very serious drawbacks. As we have already mentioned, $f(R)$ theories introduce in their formalism an additional degree of freedom that behaves like a scalar field. In order to be relevant for the large scale dynamics of the Universe, the scalar field must have a low mass. On the other hand the presence of such a scalar field would influence the dynamics on smaller scales, like, for example, at the level of the Solar System. Since no such small scale effects were detected, one must find different physical explanations for this situation, such as the introduction of  a screening mechanism \cite{Capozziello:2007eu,Khoury:2003rn}. Moreover, in the Palatini formalism, the scalar field is an algebraic function of the trace of the energy-momentum tensor, and hence no additional degrees of freedom are introduced. This fact has very serious consequences for the theory, leading to the presence of infinite tidal forces on the surface of massive astrophysical type objects \cite{Olmo:2011uz}.
In order to circumvent some of the shortcomings of the Palatini and metric $f(R)$ theories of gravity the so-called hybrid metric-Palatini gravity (HMPG) theories were proposed initially in \cite{Harko:2011nh}, and further investigated and developed recently.  One of the main advantages of the HMPG theory is that in its scalar-tensor representation it introduces long-range forces that automatically pass the Solar System tests, and thus no contradiction between the theory and the local measurements arise.

In Part \ref{Part:HMPLG} we will review the {\it hybrid} variation of these theories, in which the (purely metric) Einstein-Hilbert action is supplemented with (metric-affine) correction terms constructed \`a la Palatini \cite{Capozziello:2013uya,Harko:2011nh,Capozziello:2012ny,Capozziello:2015lza,Harko:2018ayt,Capozziello:2013wq}.
Given that both metric and Palatini $f(R)$ theories allow the construction of simple extensions of GR with interesting properties, and, at the same time, suffer from different types of drawbacks, we initiated a program to establish bridges between these two seemingly disparate approaches, hoping to find ways to cure or improve their individual deficiencies. For that purpose, in a number of works we have considered a hybrid combination of metric and Palatini elements to construct the gravity Lagrangian, and we have found that viable models sharing properties of both formalisms are possible.

An interesting aspect of these theories is the possibility to generate long-range forces without conflicts with local tests of gravity, and without invoking any kind of screening mechanism (which would however require that at the present time the cosmological evolution reduces to GR). The possibility of expressing these hybrid $f(R)$ metric-Palatini theories using a scalar-tensor representation significantly simplifies the analysis of the field equations, and the construction of their solutions.
In some sense, considering a theory like $R+f({\cal R})$ means that we retain all the positive results of GR, represented by the Einstein-Hilbert part of the action $R$, while the further ``gravitational budget'' is endowed in the metric-affine $f({\cal R})$ component, where ${\cal R}$ is the Palatini curvature scalar constructed in terms of an independent connection. In fact, it is well known that metric-affine and purely metric formalisms coincide in GR, i.e., considering the action $R$.  On the  contrary, the two formalisms lead to different dynamical theories considering more generic functions $f({\cal R})$ \cite{Olmo:2011uz}.

%%%%%%%%%%%%%%%%%%%%%%%%%%%%%%%%%%%%%%%%%%%%%%%%%%%%%%%%%%%%%%%%%%%%%%%%%%
\subsection{Couplings between geometry and matter}
%%%%%%%%%%%%%%%%%%%%%%%%%%%%%%%%%%%%%%%%%%%%%%%%%%%%%%%%%%%%%%%%%%%%%%%%%%

In Part \ref{PartIII:CMcoup}, we consider another interesting possibility for an extended theory of gravity, which includes nonminimal couplings between the scalar curvature and the matter Lagrangian density, introduced in \cite{Bertolami:2007gv}, and its extensions and generalizations, which we follow closely \cite{Harko:2014gwa}. This nonminimal coupling between matter and the gravitational field was first studied by Goenner, in 1984 \cite{Gonner:1984zx}, who considered a departure from GR, by breaking the principle of minimal coupling. Recall that the latter is incorporated into Einstein's theory and states that in the Lagrangian density of the variational principle describing the gravitational field, the dynamical variables describing matter do not couple to the curvature tensor, which represents the gravitational field. An explicit curvature-dark matter coupling was also considered in \cite{Nojiri:2004fw}.
The specific cases of a curvature coupling to dark energy \cite{Nojiri:2004fw}, the Maxwell field \cite{Bamba:2008ja} and an explicit curvature-Yang-Mills coupling \cite{Bamba:2008xa} was also explored in the context of inflation and of the late-time cosmic acceleration.
Indeed, in these theories, it was shown that an explicit coupling between an arbitrary function of the scalar curvature $R$ and the Lagrangian density of matter induces a non-vanishing covariant derivative of the energy-momentum tensor, implying non-geodesic motion and, consequently, leading to the appearance of an extra force \cite{Bertolami:2007gv}. The extra force is always orthogonal to the four-velocity of the massive particles moving in the gravitational field, and the corresponding acceleration law was obtained in the weak field limit. The theory also points towards interesting connections with MOND, and with the Pioneer anomaly, which were further discussed.
The curvature-matter coupling theories in which the matter Lagrangian couples to the curvature scalar include the more evolved generalizations of $f(R,L_m)$  theory \cite{Harko:2010mv}.  An alternative approach is represented by the $f(R,T)$ gravities \cite{Harko:2011kv}, in which the trace of the matter Lagrangian is coupled nonminimally to the scalar curvature. Amongst other features, these models allow for an explicit breaking of the equivalence principle (EP), which is highly constrained by Solar System experimental tests \cite{Faraoni:2004pi}, by imposing a matter-dependent deviation from geodesic motion.

Note that the weak equivalence principle is considered one of the pillars of GR and, in fact, even a sizable part of the modified-gravity community considers this principle as truly fundamental \cite{Will:2014kxa}. This fact has to be stressed because it demonstrates somehow a limitation for this class of theories. However, it has been recently reported, from observational data of the Abell Cluster A586, that interaction of dark matter
and dark energy does imply the violation of the EP \cite{Bertolami:2007zm}. Notice that the violation of the EP is also found as a low-energy feature of some compactified version of higher-dimensional theories.
Indeed, as emphasized by Thibault Damour, it is important to note that the EP is not one of the ``universal'' principles of physics \cite{Damour:2001fn}. It is a heuristic hypothesis, which was introduced by Einstein in 1907, and used by him to construct his theory of GR. In modern language, the (Einsteinian) EP consists in assuming that the only long-range field with gravitational-strength couplings to matter is a massless spin-2 field. Modern unification theories, and notably String Theory, suggest the existence of new fields (in particular, scalar fields: ``dilaton'' and ``moduli'') with gravitational-strength couplings. In most cases the couplings of these new fields violate the EP. If the field is long-ranged, these EP violations lead to many observable consequences, such as the variation of fundamental ``constants'', the non-universality of free fall, and the relative drift of atomic clocks, amongst others. The best experimental probe of a possible violation of the EP is to compare the free-fall acceleration of different materials. Further tests of this principle remain important and relevant for the discovery of new physics, and do indeed strongly restrict the parameters of the considered theory \cite{Damour:1996xt, Damour:2010rp}. However, it is important to note that, in this context, the violations of the EP does not in principle rule out the specific theory. Moreover, it is important to note that up to now all experimental/observational tests of the equivalence principle have been performed on Earth, or in the Earth-Moon system, and they involve only weak gravitational fields.

Besides the major theoretical difficulties posed by present day cosmological observations, a primal problem in physics is that of the unification of quantum mechanics and gravitation. The investigation of the early Universe as a whole by using the methods and the theoretical framework of quantum mechanics is the topic of quantum cosmology \cite{Bojowald:2011zzb,Bojowald:2015iga}. Indeed, quantum cosmology is based on the fundamental idea that quantum mechanics must be able to describe all microscopic physical processes, including the Universe itself in its earliest evolutionary phases. In fact, the standard model of particle physics dictates that the gravitational force is difficult ({or perhaps even impossible) to unify with the other fundamental forces. This is a direct consequence of the fact that when trying to quantize gravitation, we must include in the quantization procedure not only ordinary matter, but also space and time, as quantum physical objects. In GR, space and time abide to specific dynamical laws that have their own excitations, such as the gravitational waves that interact with each other. These physical aspects makes the quantization of the gravitational field and of the Universe understood as a quantum system extremely nontrivial, due to the nonlinear interactions between the quanta of the field. On the other hand, the formation of cosmic structures is governed by the force of gravity, which is determined by the structure of the spacetime interaction. Hence quantum cosmology, dealing with the description of the observed properties of the Universe, is closely correlated to quantum gravity, whose object of study is the quantum theory of spacetime, and of the gravitational force \cite{Mukhanov:2007zz}.

Quantum cosmology has a long history \cite{DeWitt:1967yk,DeWitt:1967ub,DeWitt:1967uc}, during which it has tried (and still tries) to overcome several difficult conceptual problems. But it still remains today a speculative and controversial subject, and, even so, it continues to attract a large attention.  Various competing attempts have been proposed to quantize the gravitational field, for instance canonical quantum gravity, loop quantum gravity, and string theory \cite{Bojowald:2011zzb,Bojowald:2015iga,DeWitt:1967yk,DeWitt:1967ub,DeWitt:1967uc,Banerjee:2011qu,MartinBenito:2008ej}. In the context of $f(R)$ gravity, quantum cosmology was discussed in \cite{Vakili:2010rf,Vakili:2009he}, and, in this review, we consider these issues within $f(R,T)$ gravity \cite{Xu:2016rdf,Harko:2018ayt}.

%%%%%%%%%%%%%%%%%%%%%%%%%%%%%%%%%%%%%%%%%%%%%%%%%%%%%%%%%%%%%%%%%%%%%%%%%%
\subsection{Outline of article}
%%%%%%%%%%%%%%%%%%%%%%%%%%%%%%%%%%%%%%%%%%%%%%%%%%%%%%%%%%%%%%%%%%%%%%%%%%

This review article is dedicated to exploring a plethora of topics related to some of the theories mentioned above. For outlining the essential basics and for a more detailed treatment, however, we refer the interested reader to \cite{Harko:2018ayt}. The present paper is organized in the following manner:

Part \ref{Part:HMPLG} is dedicated to the hybrid metric-Palatini gravitational theory, where in Section \ref{II:formalism}, we outline the general formalism and explore the Newtonian and post-Newtonian limits is some detail. In Section \ref{II:cosmology}, we investigate the cosmological applications of the theory, ranging from FLRW cosmological models, Einstein's static universe, to a dynamical system analysis and cosmological perturbations. In Section \ref{II:galactic}, we consider dark matter as a geometric effect of modified gravity, in the context of the galactic rotation curves and the mass discrepancy in galaxy clusters. In Section \ref{II:compact}, we briefly analyze static and spherically symmetric compact astrophysical objects, and in Section \ref{II:branesGWs}, we refer to several research topics of the hybrid theory, such as thermodynamics, thick branes, screening mechanisms and gravitational waves. Finally, in Section \ref{II:beyond}, we venture beyond the linear hybrid metric-Palatini gravitational theory.

In Part \ref{PartIII:CMcoup}, we consider in detail modified theories of gravity involving specific curvature-matter couplings. In Section \ref{III:linearNCM}, we present the action and the gravitational field equations of the linear nonminimal curvature-matter coupling, the respective scalar-tensor representation and we further consider generalizations of the theory, by assuming a maximal extension of the Hilbert-Einstein action. In Section \ref{III:fRTgravity}, we analyst in some detail an interesting extension of curvature-matter couplings that has been given considerable attention recently, namely, $f(R,T)$ gravity, and consider a specific cosmological application.
In Section \ref{1d_Sec:VI}, we generalize the $f(R,T)$ theory, by considering an explicit first order coupling between the matter energy-momentum $T_{\mu \nu}$ and the Ricci tensor, and analyze several aspects of the theory, such as the Dolgov-Kawasaki instability and a specific cosmological solution to illustrate the subtleties of the theory.
In Section \ref{III:mattcreation}, we consider irreversible cosmological thermodynamical processes through a nonminimal curvature-matter coupling, which seems to be a natural scenario where the matter creation corresponds to an irreversible energy flow from the gravitational field to the created matter fluid.
Section \ref{1g_Part1:new2} presents a brief introduction to the study of the quantum cosmology of $f(R,T)$ gravity, and of some of its physical and theoretical implications.

Finally, in Part \ref{Conclusions} we conclude and build bridges for future work.

\newpage

%%%%%%%%%%%%%%%%%%%%%%%%%%%%%%%%%%%%%%%%%%%%%%%%%%%%%%%%%%%%%%%%%%%%%%%%%%
\part{Hybrid metric-Palatini gravity}\label{Part:HMPLG}
%%%%%%%%%%%%%%%%%%%%%%%%%%%%%%%%%%%%%%%%%%%%%%%%%%%%%%%%%%%%%%%%%%%%%%%%%%

%%%%%%%%%%%%%%%%%%%%%%%%%%%%%%%%%%%%%%%%%%%%%%%%%%%%%%%%%%%%%%%%%%%%%%%%%%
\section{General formalism}\label{II:formalism}
%%%%%%%%%%%%%%%%%%%%%%%%%%%%%%%%%%%%%%%%%%%%%%%%%%%%%%%%%%%%%%%%%%%%%%%%%%

The basic idea of the hybrid metric-Palatini gravity (HMPG) is to add to the  usual Einstein-Hilbert action, $S[g]=\int {\left(R/2\kappa+ L_m\right)\sqrt{-g}d^4x}$, where $\kappa $ is the gravitational coupling constant, defined below, and $L_m$ is the matter Lagrangian density, a supplementary term,  containing a Palatini type correction of the form $f(\R)$, where $\R$ is the Palatini curvature. From a physical point of view  such hybrid theories may be motivated by previous theoretical results obtained  in the context of the perturbative quantization methods, or in the framework of quantum geometries (for a detailed discussion see \cite{Harko:2011nh}, and references therein).

%%%%%%%%%%%%%%%%%%%%%%%%%%%%%%%%%%%%%%%%%%%%%%%%%%%%%%%%%%%%%%%%%%%%%%%%%%
\subsection{Action and field equations}
%%%%%%%%%%%%%%%%%%%%%%%%%%%%%%%%%%%%%%%%%%%%%%%%%%%%%%%%%%%%%%%%%%%%%%%%%%

The action of HMPG is constructed as the sum of two terms according to the prescription
\begin{equation} \label{eq:S_hybrid}
S= \frac{1}{2\kappa^2}\int d^4 x \sqrt{-g} \left[ R + f(\R)\right] +S_m \ ,
\end{equation}
where  $\kappa^2\equiv 8\pi G$, $R$ is
the standard Ricci scalar, and $\R  \equiv g^{\mu\nu}\R_{\mu\nu} $ is
the Palatini curvature, respectively. In the following we use a system of units with $c=1$.  $\R_{\mu\nu}$ is defined with the help of
an independent connection $\hat{\Gamma}^\alpha_{\mu\nu}$  in the form
\begin{equation}
\R_{\mu\nu} \equiv \hat{\Gamma}^\alpha_{\mu\nu ,\alpha} -
\hat{\Gamma}^\alpha_{\mu\alpha , \nu} +
\hat{\Gamma}^\alpha_{\alpha\lambda}\hat{\Gamma}^\lambda_{\mu\nu}
-\hat{\Gamma}^\alpha_{\mu\lambda}\hat{\Gamma}^\lambda_{\alpha\nu}\,.
\end{equation}

Moreover, $S_m$ denotes the matter action. It is more advantageous from both theoretical and computational points of view to transform the action (\ref{eq:S_hybrid}) into the equivalent form  of a scalar-tensor theory by introducing a new auxiliary field $A$, which allows to write the action as
\begin{equation} \label{eq:S_scalar0}
S= \frac{1}{2\kappa^2}\int d^4 x \sqrt{-g} \left[ R + f(A)+f_A(\R-A)\right] +S_m \ ,
\end{equation}
where $f_A\equiv df/dA$. Now, by defining $\phi\equiv f_A$, $V(\phi)=A f_A-
f(A)$, the action (\ref{eq:S_scalar0}) takes the mathematically equivalent form
\begin{equation} \label{eq:S_scalar1}
S= \frac{1}{2\kappa^2}\int d^4 x \sqrt{-g} \left[ R + \phi\R-V(\phi)\right] +S_m \ .
\end{equation}

After varying the action with respect to the metric, the connection and the scalar $\phi$, we obtain the full set of field equations of the HMPG theory as
\begin{eqnarray}
R_{\mu\nu}+\phi \R_{\mu\nu}-\frac{1}{2}\left(R+\phi\R-V\right)g_{\mu\nu}&=&\kappa^2 T_{\mu\nu}
\ ,
\label{eq:var-gab}\\
\R-V_\phi&=&0 \,,\label{eq:var-phi}\\
\hat{\nabla}_\alpha\left(\sqrt{-g}\phi g^{\mu\nu}\right)&=&0 \,,\label{eq:connection}\
\end{eqnarray}
respectively.

Equation (\ref{eq:connection}) can be immediately solved, and it shows that the independent connection is nothing but the Levi-Civita connection of the metric $t_{\mu\nu}=\phi g_{\mu\nu}$. This result implies the following relation between $\R_{\mu\nu}$ and
$R_{\mu\nu}$,
\begin{equation}
\R_{\mu\nu}=R_{\mu\nu}+\frac{3}{2\phi^2}\partial_\mu \phi \partial_\nu \phi-\frac{1}
{\phi}\left(\nabla_\mu \nabla_\nu \phi+\frac{1}{2}g_{\mu\nu}\square \phi\right) \ .
\end{equation}
 By using the above relation we can reformulate the action principle (\ref{eq:S_scalar1}) in the form of the following scalar-tensor theory,
\begin{equation} \label{eq:S_scalar2}
S=\int \frac{d^4 x \sqrt{-g} }{2\kappa^2}\left[ (1+\phi)R +\frac{3}{2\phi}\partial_\mu \phi
\partial^\mu \phi -V(\phi)\right]+S_m .
\end{equation}

This action is similar to the action of the $w=-3/2$ Brans-Dicke theory. But an important and subtle difference appears in the coupling of the scalar field to the curvature, which in the $w=-3/2$ Brans-Dicke theory is of the form $\phi R$ \cite{Harko:2011nh}. By using the expression of $\R_{\mu\nu}$
and Eq.~(\ref{eq:var-phi}), Eq.~(\ref{eq:var-gab}) can be rewritten as
\begin{eqnarray}
(1+\phi) R_{\mu\nu}&=&\kappa^2\left(T_{\mu\nu}-\frac{1}{2}g_{\mu\nu} T\right)+\frac{1}
{2}g_{\mu\nu}\left(V+\square \phi\right)
+
\nabla_\mu\nabla_\nu\phi-\frac{3}
{2\phi}\partial_\mu \phi \partial_\nu \phi \ \label{eq:evol-gab} .
\end{eqnarray}
By contracting Eq.
(\ref{eq:var-gab}) with $g^{\mu\nu}$ and with the use of Eq. (\ref{eq:var-phi}) we obtain the scalar equation
\begin{equation}\label{eq:phi(X)}
2V-\phi V_\phi=\kappa^2T+R \ .
\end{equation}

Equation (\ref{eq:phi(X)}) has the important implication that $\phi$ is
an algebraic function of the scalar $X\equiv \kappa^2T+R$, that is,
$\phi=\phi(X)$. Thus, from the physical and mathematical points of view, the HMPG theory can be interpreted as a higher order derivative theory in both the metric and the matter fields. But an alternative interpretation, in which the field equations do not contain higher-order derivatives terms,
is also possible. To arrive at it we replace in Eq.~(\ref{eq:phi(X)}) the curvature scalar $R$, which is given by
\be
R=\R+\frac{3}{\phi}\square \phi-\frac{3}{2\phi^2}\partial_\mu \phi \partial^\mu \phi,
\ee
and we also use the relation $\R=V_\phi$. Then for the scalar field we find the following second-order differential equation,
\begin{equation}\label{eq:evol-phi}
-\square \phi+\frac{1}{2\phi}\partial_\mu \phi \partial^\mu \phi +\frac{\phi[2V-(1+\phi)V_\phi]}
{3}=\frac{\phi\kappa^2}{3}T \ .
\end{equation}
The above equation shows that in HMPG the scalar field is dynamical. This represents an important and interesting difference with respect to the standard Palatini case \cite{Olmo:2011uz}.

We can reformulate the field equations of HMPG in a form more closer to standard GR as
\be\label{einstein_phi}
G_{\mu\nu}=\kappa^2\left(\frac{1}{1+\phi}T_{\mu\nu} +T_{\mu \nu}^{(\phi)}\right),
\ee
where $T_{\mu \nu}$ is the ordinary matter energy-momentum tensor, and
\bea\label{tens_perfect}
T_{\mu \nu}^{(\phi)}&=&\frac{1}{\kappa ^2}\frac{1}{1+\phi }\Bigg[\nabla_\mu\nabla_\nu\phi -\frac{3}{2\phi}\nabla_\mu\phi\nabla_\nu\phi +\Bigg(\frac{3}{4\phi}\nabla_\lambda\phi\nabla^\lambda\phi - \square \phi- \frac{1}{2}V\Bigg)g_{\mu \nu}\Bigg],
\eea
is the energy-momentum tensor of the  scalar field of the theory.

The well-formulation and the well-posedness of the Cauchy problem in HMPG was studied in \cite{Capozziello:2013gza}. By adopting generalized harmonic coordinates, it was shown that the initial value problem can always be {\it well-formulated} and, furthermore, can be {\it well-posed}, depending on the adopted matter sources.

%\subsubsection{The fluid form of the energy-momentum tensor}

Note at this moment that $T_{\mu \nu}^{(\phi)}$ can be represented in an effective form similar to the energy-momentum tensor of an ordinary fluid
\be
T_{\mu \nu}=\left(\rho +p\right)u_{\mu }u_{\nu}+pg_{\mu \nu }+q_{\mu }u_{\nu }+q_{\nu }u_{\mu }+S_{\mu \nu},
\ee
where $u_{\mu }$ is the four-velocity, defined in the standard way, $\rho $ and $p$ are the thermodynamic parameters of the fluid (energy density and isotropic pressure), $q_{\mu }$ is the heat flux, while  $S_{\mu \nu}$ is the tensor of the anisotropic dissipative stresses. For the heat flux four-vector and for the anisotropic stress tensor we impose the standard conditions $u^{\mu }q_{\mu }=0$, $S_{\mu }^{\mu }=0$, and $S_{\mu }^{\nu }u_{\nu }=0$, respectively, indicating that the heat flux and the anisotropic stresses are perpendicular to the four-velocity. By introducing the projection tensor $h^{\mu \nu}=g^{\mu \nu}+u^{\mu \nu}$, with the properties  $g_{\mu \nu}h^{\mu \nu}=3$, $h^{\mu \nu}u_{\nu }=0$, and $h_{\mu \nu}u^{\mu }u^{\nu }=0$, we can obtain the thermodynamic parameters of the fluid as
$\rho =u^{\mu }u^{\nu }T_{\mu \nu }$, $ p=-\frac{1}{3}h^{\mu \nu}T_{\mu \nu}$, $q_{\mu }=u^{\alpha }h^{\beta }_{\mu}T_{\alpha \beta }$, and
$S_{\mu \nu}=h^{\alpha }_{\mu }h^{\beta }_{\nu}T_{\alpha \beta }+ph_{\mu \nu}$, respectively. Hence, by introducing the four-velocity of the scalar field according to the definition
\be
u^{\mu }_{(\phi)} =\frac{\nabla ^{\mu }\phi }{\sqrt{\nabla _{\alpha }\phi \nabla ^{\alpha }\phi }},
\ee
 satisfying the relation $u^{\mu }_{(\phi)}u_{(\phi)\mu}=-1$,  we obtain the effective energy density $\rho _{\phi }$, pressure $p_{\phi }$, the heat flux and the anisotropic stress tensor in the scalar field description of HMPG as
\bea
\rho _{\phi} &=&\frac{1}{\kappa ^2(1+\phi)}\left\{\frac{1}{2}\nabla ^{\mu }\phi \nabla _{\mu }\ln \left(\nabla _{\alpha }\phi \nabla ^{\alpha }\phi \right)-\frac{5}{4\phi}\nabla _{\alpha }\phi \nabla ^{\alpha }\phi
-\frac{1}{3}\left[2V(\phi)-(1+\phi)V'(\phi)\right]+\frac{\kappa ^2}{3}\phi T-\frac{1}{2}V(\phi)\right\},
\eea
\bea
p_{\phi}&=&\frac{1}{\kappa ^2(1+\phi)}\left\{-\sqrt{\nabla _{\alpha }\phi \nabla ^{\alpha }\phi}\;\frac{\Theta }{3}-\frac{1}{4\phi }\nabla _{\alpha }\phi \nabla ^{\alpha }\phi
+\frac{1}{3}\left[2V(\phi)-(1+\phi)V'(\phi)\right]-\frac{\kappa ^2}{3}\phi T+\frac{1}{2}V(\phi)\right\},
\eea
\be
q^{(\phi )\mu }=\frac{1}{\kappa ^2(1+\phi)}\sqrt{\nabla _{\alpha }\phi \nabla ^{\alpha }\phi}\; u^{\alpha }_{(\phi) }\,\nabla _{\alpha }u^{\mu }_{(\phi) },
\ee
and
\bea
S^{(\phi )\mu \nu}&=&\frac{1}{\kappa ^2(1+\phi)}
 \left[\sqrt{\nabla _{\alpha }\phi \nabla ^{\alpha }\phi }\left(\nabla ^{\mu }u^{\nu }_{(\phi )} -u_{(\phi) }^{\alpha}u^{\mu }_{\phi }\nabla _{\alpha }u^{\nu }_{(\phi ) }\right)
 +\left(-\sqrt{\nabla _{\alpha }\phi \nabla ^{\alpha }\phi }\;\frac{\Theta }{3}+\frac{1}{2\phi }\nabla _{\alpha }\phi \nabla ^{\alpha }\phi\right) h^{\mu \nu}\right],
\eea
respectively, where by $\Theta =\nabla _{\alpha }u^{\alpha }_{(\phi )}$ we have denoted the expansion of the fluid. The equivalent fluid type representation of the  energy-momentum tensor in HMPG is not of a perfect fluid form, since it contains ``heat transfer'' components, together with  an anisotropic  dissipative constituent.

Hence, finally the energy-momentum tensor of the HMPG scalar field can be rewritten in a representation mathematically equivalent to an ordinary dissipative matter fluid as
\bea
T^{(\phi)\mu \nu}&=&\left(\rho _{\phi }+p_{\phi}\right)u^{\mu }_{(\phi )}u^{\nu }_{(\phi)}-p_{\phi }g^{\mu \nu}+q^{(\phi )\mu }u^{\nu }_{(\phi)}+q^{(\phi)\nu }u^{\mu }_{(\phi ) }+S^{(\phi )\mu \nu}\,.
\eea

%%%%%%%%%%%%%%%%%%%%%%%%%%%%%%%%%%%%%%%%%%%%%%%%%%%%%%%%%%%%%%%%%%%%%%%%%%
\subsection{The Newtonian limit}
%%%%%%%%%%%%%%%%%%%%%%%%%%%%%%%%%%%%%%%%%%%%%%%%%%%%%%%%%%%%%%%%%%%%%%%%%%

One of the major drawbacks in the scalar-tensor representation of the $f(R)$ theory is related to the difficulty of the theory to pass the important, and fundamental, Solar System tests, which generally allow to strongly constrain the physical parameters of any gravitational theory. In order to consider the classical tests of GR in HMPG one must first consider the weak-field and slow-motion limits of the gravitational field equations (\ref{eq:evol-gab}), and of the generalized Klein-Gordon equation (\ref{eq:evol-phi}). In order to do this we follow the standard approach, well known from GR, and we introduce first a quasi-Minkowskian coordinate system, in which the metric can be represented as $g_{\mu\nu}\approx \eta_{\mu\nu}+h_{\mu\nu}$, with $|h_{\mu\nu}|\ll 1$.  Then we expand both the metric and the scalar field with respect to an appropriately chosen background cosmological solution, which fully determines the asymptotic boundary expressions of the metric. In the following, we denote the asymptotic value of $\phi$ by  $\phi_0$, while the local perturbation is denoted as $\varphi(x)$. Then, in the first order of approximation Equation~(\ref{eq:evol-phi}) takes the form
\begin{equation}\label{eq:linear-phi}
(\vec{\nabla}^2-m_\varphi^2)\varphi=\frac{\phi_0\kappa^2}{3}\rho \ ,
\end{equation}
where we have introduced the effective mass $m_\varphi^2$ of the scalar field, defined as
\be
m_\varphi^2 \equiv \left. [2V-V_{\phi}-\phi(1+\phi)V_{\phi\phi}]/3\right|
_{\phi=\phi_0}.
\ee
The derivatives of $\varphi$ with respect to time have also been neglected, under the assumption of the slow motion.

The metric perturbations $h_{\mu\nu}=g_{\mu\nu}-\eta_{\mu\nu}$ can be obtained from the equation
\begin{equation}\label{eq:linear-gab}
-\frac{1}{2}\vec{\nabla}^2h_{\mu\nu}=\frac{1}{1+\phi_0}\left(T_{\mu\nu}-\frac{1}{2}T
\eta_{\mu\nu}\right)+\frac{V_0+\vec{\nabla}^2\varphi}{2(1+\phi_0)}\eta_{\mu\nu} \ .
\end{equation}

In the first order of approximation we are considering at this moment the components of the energy-momentum tensor are given by $T_{00}=\rho$, $T_{ij}=0$, $T=-\rho$, where $\rho$ is the ordinary matter density. We also define the mass of the gravitating system as $M=\int d^3x \rho(x)$. In spherical symmetry, an assumption valid at least approximately for most of the astrophysical systems in slow motion,  and far away from the matter sources, we obtain
the solutions of Eqs.~(\ref{eq:linear-phi}) and (\ref{eq:linear-gab}) as
\begin{eqnarray}
\varphi(r)&=&\frac{2G}{3}\frac{\phi_0 M}{r}e^{-m_\varphi r}, \label{cor1}\\
h_{00}^{(2)}(r)&=& \frac{2G_{\rm eff} M}{r} +\frac{V_0}{1+\phi_0}\frac{r^2}{6}, \label{cor2} \\
h_{ij}^{(2)}(r)&=& \left(\frac{2\gamma G_{\rm eff} M}{r} -\frac{V_0}{1+\phi_0}\frac{r^2}
{6}\right)\delta_{ij}
\label{cor3}\ ,
\end{eqnarray}
where we have introduced the effective Newtonian gravitational constant $G_{\rm eff}$ and the post-Newtonian
parameter $\gamma$, defined as
\begin{eqnarray}
G_{\rm eff}&\equiv & \frac{G}{1+\phi_0}\left[1-\left(\phi_0/3\right)e^{-m_\varphi r}\right]\,,
\label{g_eff}\\
\gamma &\equiv & \frac{1+\left(\phi_0/3\right)e^{-m_\varphi r}}{1-\left(\phi_0/3\right)e^{-
m_\varphi r}} \,. \label{gamma}
\end{eqnarray}

As one can be seen easily from Eqs.~(\ref{g_eff}) and (\ref{gamma}), the coupling of the scalar field to a local astrophysical system
depends on the magnitude of the background value $\phi_0$ of the scalar field. For a small  $\phi_0$, it follows that $G_{\rm
eff}\approx G$ and $\gamma\approx 1$. It is important to note at this moment that these results are independent on the value of the effective field mass
$m_\varphi^2$. If $m_\varphi^2<0$, the exponential terms in the above expressions will become, due to its imaginary nature, some rapidly oscillatory functions.

The previously obtained results are very different from the outcomes obtained in the metric version of $f(R)$ gravity theories \cite{Olmo:2005zr,Olmo:2005hc},  for which
\be
\varphi=\frac{2G}{3}\left(\frac{M}{r}\right)e^{-m_f r}, \qquad G_{\rm eff}\equiv  \frac{G}{\phi _0}
\left(1+\frac{e^{-m_f r}}{3}\right) ,
\ee
and
\be
\gamma \equiv \frac{\left(1-e^{-m_f r}/
3\right)}{\left(1+e^{-m_f r}/3\right)} ,
\ee
respectively. Hence in order for the Yukawa-type corrections to become insignificant at the level of the Solar System  in $f(R)$ gravity theories a large mass $m_f^2\equiv (\phi V_{\phi\phi}-V_\phi)/3$ of the scalar field is required, which is indeed not the case in the HMPG theory.

%%%%%%%%%%%%%%%%%%%%%%%%%%%%%%%%%%%%%%%%%%%%%%%%%%%%%%%%%%%%%%%%%%%%%%%%%%
\subsection{The Post-Newtonian limit}
%%%%%%%%%%%%%%%%%%%%%%%%%%%%%%%%%%%%%%%%%%%%%%%%%%%%%%%%%%%%%%%%%%%%%%%%%%

The post-Newtonian limit of the HMPG theory was investigated in \cite{Dyadina:2019yon}, and in the following we will summarize the results of this study. The starting point of the approach used in \cite{Dyadina:2019yon} is the standard expansion of the scalar field $\phi$ and of the metric tensor $g_{\mu\nu}$ according to the prescription
	\begin{equation}
	\phi=\phi_0+\varphi, \qquad  g_{\mu\nu}= \eta_{\mu\nu}+h_{\mu\nu},
	 \end{equation}
where $\phi_0$ is taken as the asymptotic background value of the scalar field far away from the ordinary matter sources, $\eta_{\mu\nu}$ denotes, as usual, the Minkowski metric, while $h_{\mu\nu}$ and $\varphi$ are the small perturbations of the metric tensor, and of the scalar field, respectively. Even though $\phi_0$ is generally a function of time $\phi =\phi(t)$, in the considered investigation $\phi_0$ was taken as a constant.
	
The post-Newtonian limits are obtained  by evaluating the perturbations of the metric and of the scalar field, respectively, to the orders $h_{00}\sim O(2) + O(4), h_{0j}\sim O(3), h_{ij}\sim O(2)$ and $\varphi\sim O(2) + O(4)$. To achieve these goals the scalar field potential $V(\phi)$ is expanded as follows \cite{Dyadina:2019yon}
	\begin{equation}
	V(\phi)=V_0\left(\phi_0\right)+V'\left(\phi_0\right)\varphi+\frac{V''\left(\phi_0\right)\varphi^2}{2!}+\frac{V'''\left(\phi_0\right)\varphi^3}{3!}...
	 \end{equation}
	
For a point-mass gravitational system the energy-momentum tensor is defined in a standard way as \cite{Dyadina:2019yon}
	\begin{equation}
	T^{\mu\nu}=\frac{c}{\sqrt{-g}}\sum_k m_k\frac{u^{\mu}u^{\nu}}{u^0} \delta^3( \mathbf{r}- \mathbf{r}_k),
	 \end{equation}
where $m_k$, $k=1,2,...,n$ are the masses of the particles in the system, $\mathbf{r}_k$ is the position vector of the $k$-th particle, $\delta^3(\mathbf{r}-\mathbf{r}_k(t))$ is the three-dimensional Dirac delta function, while the other quantities have their usual meanings. We also denote by  $v_k$ the three-dimensional velocity of the $k$-th particle.
	
Note that the energy-momentum tensor and its trace  have, in the post-Newtonian approximation, the following components \cite{Dyadina:2019yon}
	\begin{eqnarray}
		T_{00}&=&c^2\sum_k m_k \delta^3( \mathbf{r}- \mathbf{r_k})\left[1-\frac{3}{2}h_{00}+\frac{1}{2}\frac{v^2_k}{c^2}-\frac{1}{2}h\right],\\
	T_{0i}&=&-c\sum_km_kv_a^i \delta^3( \mathbf{r}- \mathbf{r_k}),\\
	T_{ij}&=&\sum_km_kv_a^iv_a^j \delta^3( \mathbf{r}- \mathbf{r_k}),\\
	T&=&-c^2\sum_km_k \delta^3( \mathbf{r}- \mathbf{r_k})\left[1-\frac{1}{2}h_{00}-\frac{1}{2}\frac{v_k^2}{c^2}-\frac{1}{2}h\right].
	 \end{eqnarray}
Moreover, the following gauge condition is also imposed on the perturbations $h^\alpha_{\beta}$,
	\begin{equation}\label{gauge}
	h^\alpha_{\beta,\alpha}-\frac{1}{2} \delta^\alpha_\beta h^\mu_{\mu,\alpha}=\frac{\varphi_{,\beta}}{1+\phi_0}.
	 \end{equation}
	
	In the leading perturbation order ($O(2)$) the generalized Klein-Gordon equation for the scalar field becomes \cite{Dyadina:2019yon}
	\begin{equation}
	\left(\nabla^2-m_\varphi^2\right)\varphi^{(2)}=\frac{k^2\phi_0}{3c^2}\sum_km_k \delta^3( \mathbf{r}- \mathbf{r_k}),
	 \end{equation}
with the general solution for the second order perturbation of the gravitational field given by
\begin{equation}
	\varphi^{(2)}=-\frac{k^2\phi_0}{12\pi c^2}\sum_km_k\frac{ e^{-m_\varphi r_k}}{r_k},
	 \end{equation}
where we have also introduced the notation $r_k=|\mathbf{r}-\mathbf{r}_k|$.
	
On the other hand, the $h_{00}^{(2)}$ perturbation term satisfies the equation
	\begin{eqnarray}
	\nabla^2\left(h_{00}^{(2)}-\frac{\varphi^{(2)}}{1+\phi_0}\right)=-\frac{k^2}{c^2(1+\phi_0)}\sum_am_a \delta^3( \mathbf{r}- \mathbf{r_a})
	+\frac{V_0}{1+\phi_0},
	 \end{eqnarray}
and it is obtained as	
	\begin{equation}
	h_{00}^{(2)}=\frac{k^2}{4\pi(1+\phi_0)c^2}\frac{M_{\odot}}{r}\left(1-\frac{\phi_0}{3}   e^{-m_\varphi r}\right)+\frac{V_0}{1+\phi_0}\frac{r^2}{6},
	 \end{equation}
where $M_{\odot}$ is, as usual,  the mass of the Sun.
	The $h_{ij}^{(2)}$-component of the metric tensor perturbation satisfies the equation
	\begin{eqnarray}
	\nabla^2\left(h_{ij}^{(2)}+ \delta_{ij}\frac{\varphi^{(2)}}{1+\phi_0}\right)
	=-\left(\frac{k^2}{(1+\phi_0)c^2} \sum_am_a \delta^3( \mathbf{r}- \mathbf{r_a})
	- \frac{V_0}{1+\phi_0}\right)\delta_{ij},
	 \end{eqnarray}
and can be obtained as \cite{Dyadina:2019yon}
\begin{equation}
	h_{ij}^{(2)}=\frac{ \delta_{ij}k^2}{4\pi(1+\phi_0)c^2}\frac{M}{r}\left(1+\frac{\phi_0}{3}   e^{-m_\varphi r}\right)- \delta_{ij}\frac{V_0}{1+\phi_0}\frac{r^2}{6}.
	 \end{equation}
	We refer the reader to \cite{Dyadina:2019yon} for the fourth order corrections of the scalar field and of the metric.

Observational constraints on $\phi _0$  and on $m_{\phi}$ can be obtained from the MESSENGER data that provides the values of $\gamma $ and $\beta$. By using the observational values $\gamma^{\rm exp}=1-0.3\times10^{-5}\pm2.5\times10^{-5}$ and $\beta^{\rm exp}=1+0.2\times10^{-5}\pm 2.5\times10^{-5}$ \cite{Will:2018mcj}, and by considering that in the HMPG theory the scalar field is very light, one can obtain for $\phi _0$ the constraints $-8\times 10^{-5}<\phi _0<7\times 10^{-5}$ and $-9\times 10^{-4}<\phi _0<9\times 10^{-4}$ \cite{Dyadina:2019yon}. Thus, these observational limits indicate that indeed $\phi _0$ takes  very small values, as assumed in the HMPG theory. It is important to mention that in this case the mass of the scalar field remains arbitrary.

Stronger constraints on the HMPG theory parameters can be obtained from the observational data from PSR J0737-3039, which give the following limits for $\phi _0$ and $m_{\phi}$ \cite{Dyadina:2018ryl},
\be
0.975\leq \frac{1}{\left(1+\phi _0\right)^{5/3}}\left[1-\frac{5\phi _0}{18\left(1-2\times 10^{26}m_{\phi}^2\right)}\right]\leq 1.
\ee
The mixed binary system PSR J1738+0333 provides the following limits \cite{Dyadina:2018ryl}
\be
0.67 \leq \frac{1}{\left(1+\phi _0\right)^{5/3}}\left[1-\frac{5\phi _0}{18\left(1-3\times 10^{27}m_{\phi}^2\right)}\right]\leq 1.
\ee
From these two bounds combined we obtain the following important restrictions on $\phi$ and $m_{\phi}^2$ \cite{Dyadina:2018ryl},
\be
\phi _0\leq 0.00004, \qquad m_{\phi}^2\leq 1.4\times 10^{-14}\;({\rm cm}^{-1} ).
\ee

%%%%%%%%%%%%%%%%%%%%%%%%%%%%%%%%%%%%%%%%%%%%%%%%%%%%%%%%%%%%%%%%%%%%%%%%%%
\subsection{$f(X)$ representation}
%%%%%%%%%%%%%%%%%%%%%%%%%%%%%%%%%%%%%%%%%%%%%%%%%%%%%%%%%%%%%%%%%%%%%%%%%%

The HMPG theory can also be directly formulated in terms of the scalar quantity $X\equiv \kappa^2 T+R$ \cite{Capozziello:2012ny}. To obtain this representation of the theory we begin with the $D$ dimensional action of the HMPG, given by
 \be \label{action}
 S= \int d^D x \sqrt{-g} \left(
R + f(\R) + 2\kappa ^2 \mathcal{L}_m \right)\,,
\ee
where, as usual, $\mathcal{L}_m$ is the matter Lagrangian.

After solving the equation of motion for the connection, it follows that, as expected,  it is compatible with the metric $F(\R)^{\frac{2}{D-2}}g_{\mu\nu}$. In other words, it turns out that it is conformally related  to the metric $g_{\mu\nu}$, with the conformal factor given by
\be
F(\R)
\equiv \frac{df(\R)}{d\R}\,.
\ee

Hence the $D$-dimensional Palatini Ricci tensor can be obtained as
\be \label{ricci1}
 \R_{\mu\nu}  =  R_{\mu\nu} +
\frac{D-1}{D-2}\frac{1}{F^2(\R)}F(\R)_{,\mu}F(\R)_{,\nu}
  - \frac{1}{F(\R)}\nabla_\mu F(\R)_{,\nu} -
\frac{1}{(D-2)}\frac{1}{F(\R)}g_{\mu\nu}\square F(\R)\,.
\ee

After varying the higher dimensional action (\ref{action}) with respect to the metric, we find
\be \label{efe1} G_{\mu\nu} +
F(\R)\R_{\mu\nu}-\frac{1}{2}f(\R)g_{\mu\nu} = \kappa ^2 T_{\mu\nu}\,,
\ee
where we have defined the matter energy-momentum tensor in the standard way,
 \be \label{memt}
 T_{\mu\nu} \equiv -\frac{2}{\sqrt{-g}} \frac{\delta
 (\sqrt{-g}\mathcal{L}_m)}{\delta(g^{\mu\nu})}.
 \ee
Solving the trace of the field equation for $\R$ yields the higher dimensional scalar relation
\be \label{trace1}
\frac{D}{2}f(\R)-F(\R)\R =- \kappa ^2 T + \left(
\frac{D}{2}-1\right) R \equiv X\,.
\ee

In the following we make the mathematical assumption that the above equation has a solution for the function $f(\R)$. If this is indeed the case, it turns out that we can obtain $\R$ in terms of $X$. The scalar quantity $X$ thus describes the deviations of the HMPG theory from the general relativistic trace equation, given, for $D=4$, by $R=-\kappa ^2 T$. Therefore we obtain the important result that the field equation (\ref{efe1}) can be expressed in terms of the metric and $X$ as
\bea \label{efex} G_{\mu\nu}
& = &
\frac{1}{2}f(X)g_{\mu\nu}- F(X)R_{\mu\nu}  +  F'(X)
\nabla_{\mu}X_{,\nu}
     + \frac{1}{D-2}\lb F'(X)\square X + F''(X)\left( \partial X\rp^2 \right)
g_{\mu\nu}
    \nonumber  \\
&&+ \left[ F''(X)-\frac{D-1}{D-2}\frac{\left(
F'(X)\right)^2}{F(X)}\right] X_{,\mu}X_{,\nu} + \kappa ^2 T_{\mu\nu}\,,
\eea
where $(\partial X)^2=X_{,\mu}X^{,\mu}$, $F(X)$ denotes
 $F(\R(X))$, $F'(X)$ denotes $F'(X)=\partial
F(\R(X))/\partial X$ etc.

Note that the presence of the second order derivatives acting on $X$ makes the HMPG theory in the $f(X)$ representation fourth order in the metric derivatives.
The trace of the field equations can be obtained as
\be \label{trace21}
F'(X)\square X + \left[
F''(X)-\frac{1}{2}\frac{\left( F'(X)\right)^2}{F(X)}\right] \left( \partial
X\right)^2
+ \frac{D-2}{2(D-1)}\left [ X + \frac{D}{2}f(X)-F(X)R\right]= 0 \,.
\ee

The Ricci scalar $R$ is related to the Palatini $\R$ by the equation
\be \label{ricciscalar1} \R(X) =
R+\frac{D-1}{D-2}\left[ \left(\frac{F'(X)}{F(X)}\right)^2-2\frac{\square
F(X)}{F(X)}\right]\,.
\ee

Hence it follows that Eqs.~(\ref{trace21}) and (\ref{ricciscalar1}) are redundant with respect to
Eq.~(\ref{trace1}).\\

%\subsubsection{The four-dimensional case}

{\bf The four-dimensional case:}

In four dimensions the action (\ref{action}) takes the form given by (\ref{eq:S_hybrid}).
After varying with respect to the metric the action given by Eq. (\ref{eq:S_hybrid}) we obtain the following gravitational field equations of the HMPG theory in the $f(X)$ representation
\be
\label{efe} G_{\mu\nu} +
F(\R)\R_{\mu\nu}-\frac{1}{2}f(\R)g_{\mu\nu} = \kappa ^2 T_{\mu\nu}\,,
\ee
where as usual the matter energy-momentum tensor is defined according to the prescription of Eq. (\ref{memt}). As we have already pointed out, also in the four-dimensional case we have a  compatibility relation between the independent connection, and the metric $F(\R)g_{\mu\nu}$, which is conformal to $g_{\mu\nu}$, with the conformal factor defined as $F(\R) \equiv df(\R)/d\R$. Hence we obtain
\bea
\label{ricci} \R_{\mu\nu} & = & R_{\mu\nu} +
\frac{3}{2}\frac{1}{F^2(\R)}F(\R)_{,\mu}F(\R)_{,\nu}
  - \frac{1}{F(\R)}\nabla_\mu F(\R)_{,\nu} -
\frac{1}{2}\frac{1}{F(\R)}g_{\mu\nu}\square F(\R)\,. \eea
For the Palatini curvature, $\R$, which can be
obtained from the trace of the field equation (\ref{efe}), we find
\be \label{trace}
F(\R)\R -2f(\R)= \kappa ^2 T +  R \equiv X\,.
\ee
If the functional form of $f(\R)$ allows to obtain some analytic representations, we can express $\R$ in
terms of $X$. As already  mentioned, the variable $X$ has {\it the important physical interpretation} as describing the differences of the HMPG theory with respect to the general relativistic trace equation $R=-\kappa ^2 T$.

In the four-dimensional case the gravitational field equations (\ref{efe}) can be reformulated  as
\be
G_{\mu\nu}=\kappa ^2T^{\rm eff}_{\mu\nu},
\ee
where
\be
T^{\rm eff}_{\mu\nu}= T^{\rm
X}_{\mu\nu}+T_{\mu\nu},
\ee
with the effective energy-momentum tensor $T^{\rm X}_{\mu\nu}$ given by
\bea\label{Tefex}
T^{\rm X}_{\mu\nu}& = &\frac{1}{\kappa ^2} \Bigg\{
\frac{1}{2}f(X)g_{\mu\nu}- F(X)R_{\mu\nu}  +  F'(X)
\nabla_{\mu}X_{,\nu}
     + \frac{1}{2}\left[ F'(X)\square X + F''(X)\left( \partial X\right)^2 \right]
g_{\mu\nu}
    \nonumber  \\
&&+ \left[ F''(X)-\frac{3}{2}\frac{\left( F'(X)\right)^2}{F(X)}\right]
X_{,\mu}X_{,\nu} \Bigg\}\,.
\eea

 For the trace of the field equations we find the expression
\bea \label{trace2} F'(X)\square X + \left[
F''(X)-\frac{1}{2}\frac{\left( F'(X)\right)^2}{F(X)}\right] \left( \partial
X\right)^2+
 \frac{1}{3}\left[ X + 2f(X)-F(X)R\right]= 0 \,.
\eea

By contracting Eq.~(\ref{ricci}) we obtain the following relation between the metric  Ricci scalar curvature  $R$ and
the Palatini scalar curvature $\R$ as
\be \label{ricciscalar} \R(X) =
R+\frac{3}{2}\lb \lp\frac{F'(X)}{F(X)}\rp^2-2\frac{\square
F(X)}{F(X)}\rb\,.
 \ee

The action given by Eq.~(\ref{eq:S_hybrid}) for the $f(X)$ representation of the HMPG theory can be transformed into the equivalent action of a scalar-tensor theory after introducing an auxiliary field $A$, which reduces it to the standard form of Eq.~(\ref{eq:S_scalar0}). In the following, when discussing the different astrophysical and cosmological applications, we will concentrate mostly on the scalar-tensor form of the HMPG theory.

%%%%%%%%%%%%%%%%%%%%%%%%%%%%%%%%%%%%%%%%%%%%%%%%%%%%%%%%%%%%%%%%%%%%%%%%%%
\section{Cosmological applications}\label{II:cosmology}
%%%%%%%%%%%%%%%%%%%%%%%%%%%%%%%%%%%%%%%%%%%%%%%%%%%%%%%%%%%%%%%%%%%%%%%%%%

In the present Section, we investigate the cosmological consequences of the HMPG theory. We will consider only the case of an isotropic, homogeneous and spatially flat geometry, with the metric given by the
Friedman-Lemaitre-Robertson-Walker (FLRW) form,
\begin{equation} \label{metric}
ds^2=-dt^2+a^2(t) d{\bf x}^2 \,,
\end{equation}
where $a(t)$ is the time dependent only scale factor. We also introduce the Hubble function, an important cosmological parameter, defined as $\dot{a}\equiv da/dt$, where a dot denotes the derivative with respect to the time.

%%%%%%%%%%%%%%%%%%%%%%%%%%%%%%%%%%%%%%%%%%%%%%%%%%%%%%%%%%%%%%%%%%%%%%%%%%
\subsection{The generalized Friedmann equations, and the deceleration parameter}
%%%%%%%%%%%%%%%%%%%%%%%%%%%%%%%%%%%%%%%%%%%%%%%%%%%%%%%%%%%%%%%%%%%%%%%%%%

For the FLRW metric the Ricci scalar is given by $R=6(2H^2+\dot{H})$. Hence, for the metric (\ref{metric}), Eq. (\ref{eq:evol-gab}) gives the following cosmological evolution equations \cite{Harko:2011nh}
\begin{eqnarray}
3H^2&=& \frac{1}{1+\phi }\left[\kappa^2\rho +\frac{V}{2}-3\dot{\phi}\left(H+\frac{\dot{\phi}}
{4\phi}\right)\right] \ ,\label{field1} \\
2\dot{H}&=&\frac{1}{1+\phi }\left[ -\kappa^2(\rho+P)+H\dot{\phi}+\frac{3}
{2}\frac{\dot{\phi}^2}{\phi}-\ddot{\phi}\right] \ . \label{field2}
\end{eqnarray}
The above equations represent the two generalized Friedmann equations, describing the cosmological dynamics in HMPG, with $\rho $ and $P$ denoting the baryonic matter energy-density, and thermodynamic pressure, respectively.

The generalized cosmological Klein-Gordon Eq.~(\ref{eq:evol-phi}) takes the form
\begin{equation}
\ddot{\phi}+3H\dot{\phi}-\frac{\dot{\phi}^2}{2\phi}+\frac{\phi}{3}
[2V-(1+\phi)V_\phi]=-\frac{\phi\kappa^2}{3}(\rho-3P) \,, \label{3}
\end{equation}
which can be reformulated as follows
\begin{equation}
\ddot{\phi}+3H\dot{\phi}-\frac{\dot{\phi}^2}{2\phi}+M^2_\phi(T)\phi=0 \ ,  \label{3a}
\end{equation}
where $T=-(\rho-3P)$ is the trace of the matter energy-momentum tensor, and $M^2_\phi(T)$ is given by
\begin{equation}\label{eq:mass}
M^2_\phi(T)\equiv m_\phi^2-\frac{1}{3}\kappa^2T=\frac{1}{3}[2V-(1+\phi)V_\phi-\kappa^2T].
\end{equation}
Equation (\ref{3a}) describes the evolution of a massive scalar field on an FLRW cosmological background, and as compared to the Klein-Gordon equation of standard GR for massive scalar fields, it contains an extra term $\dot{\phi}^2/\phi$. This new term makes a significant contribution to the evolution of the cosmological models in HMPG only when the scalar field is changing rapidly.

For a vacuum Universe, the generalized Friedmann equations can be rewritten in the form
\begin{eqnarray}
H^{2}&=&\frac{\kappa ^{2}}{3}\rho _{\rm eff},\\
\dot{H}&=&-\frac{\kappa ^{2}}{2}\left(
\rho _{\rm eff}+p_{\rm eff}\right) ,
\end{eqnarray}
where we have denoted
\begin{eqnarray}
\left( 1+\phi \right) \kappa ^{2}\rho _{\rm eff}&=&-\frac{3}{4\phi }\dot{\phi}%
^{2}+\kappa ^{2}V\left( \phi \right) -3H\dot{\phi},  \label{1a}
\\
\left( 1+\phi \right) \kappa ^{2}p_{\rm eff}&=&-\frac{3}{4\phi }\dot{\phi}%
^{2}-\kappa ^{2}V\left( \phi \right) +2H\dot{\phi}+\ddot{\phi}.
\label{2a}
\end{eqnarray}

In the absence of matter the scalar field  satisfies the generalized Klein-Gordon equation,
\begin{equation}
\ddot{\phi}+3H\dot{\phi}-\frac{1}{2\phi }\dot{\phi}^{2}+\frac{2\kappa ^{2}}{3%
}\phi \left[ 2V\left( \phi \right) -\left( 1+\phi \right)
V^{\prime }\left( \phi \right) \right] =0.  \label{3a2}
\end{equation}
In order to describe the accelerating or decelerating nature of the cosmological expansion we introduce, as usual in cosmology, the deceleration parameter, defined according to
\begin{equation}
q=\frac{d}{dt}\frac{1}{H}-1=-\frac{\dot{H}}{H^{2}}-1=\frac{3}{2}\frac{\rho
_{\rm eff}+p_{\rm eff}}{\rho _{\rm eff}}-1, \label{4a}
\end{equation}
where negative values of $q$ correspond to an accelerated expansion.

From Eqs.~(\ref{1a}) and (\ref{2a}) we can immediately find the relation
\begin{equation}
\left( 1+\phi \right) \kappa ^{2}\left( \rho _{\rm eff}+p_{\rm eff}\right) =-\frac{3%
}{2\phi }\dot{\phi}^{2}-H\dot{\phi}+\ddot{\phi}.  \label{5a}
\end{equation}
Eliminating $\ddot{\phi}$ with the help of the generalized Klein-Gordon (\ref{3a2}), we obtain
\begin{equation}
\left( 1+\phi \right) \kappa ^{2}\left( \rho _{\rm eff}+p_{\rm eff}\right) =-\frac{1}{\phi }\dot{\phi}^{2}-4H\dot{\phi}-\frac{2\kappa ^{2}}{3}\phi
\left[
2V\left( \phi \right) -\left( 1+\phi \right) V^{\prime }\left( \phi \right) %
\right] .  \label{6a}
\end{equation}
Hence, in the vacuum the generalized Friedmann equations take the simpler form
\begin{eqnarray}
3H^{2}&=&\frac{1}{1+\phi }\left[ -\frac{3}{4\phi }\dot{\phi}^{2}-3H\dot{\phi}%
+\kappa ^{2}V\left( \phi \right) \right] ,  \label{6b}
 \\
2\dot{H}&=&\frac{1}{1+\phi }\left[ \frac{1}{\phi }\dot{\phi}^{2}+4H\dot{\phi}+%
\frac{2\kappa ^{2}}{3}\phi \left[ 2V\left( \phi \right) -\left(
1+\phi \right) V^{\prime }\left( \phi \right) \right] \right] .
\label{6c}
\end{eqnarray}
We can represent now the deceleration parameter $q$ as a function of the field in the form
\begin{equation}\label{7a}
q=-\frac{3}{2}\left\{\frac{\dot{\phi}^{2}/\phi +4H\dot{\phi}+2\kappa
^{2}\phi \left[
2V\left( \phi \right) -\left( 1+\phi \right) V^{\prime }\left( \phi \right) %
\right] /3}{-3\dot{\phi}^{2}/4\phi +\kappa ^{2}V\left( \phi \right) -3H\dot{%
\phi}}\right\}-1\,.
\end{equation}
 Therefore it follows that based on the physical properties of the scalar field, and of its potential, a large number of cosmological evolutionary scenarios can be constructed, generally giving both accelerated and decelerated evolutions. In \cite{Kausar:2019iwu}, the cosmological inflation in HMPG was explored by investigating the cosmic parameters such as the slow-roll parameters, scalar-to-tensor ratio and spectral index.

%%%%%%%%%%%%%%%%%%%%%%%%%%%%%%%%%%%%%%%%%%%%%%%%%%%%%%%%%%%%%%%%%%%%%%%%%%
\subsection{Alternative form of the generalized Friedmann equations}
%%%%%%%%%%%%%%%%%%%%%%%%%%%%%%%%%%%%%%%%%%%%%%%%%%%%%%%%%%%%%%%%%%%%%%%%%%

The generalized cosmological Friedmann equations of the HMPG theory can be simplified if we introduce a new independent variable $x=\ln a$, so that $dx/dt=\dot{a}/a=H$. Hence we obtain first
\begin{equation}
\frac{d}{dt}=\frac{d}{dx}\frac{dx}{dt}=\frac{d}{dx}\frac{\dot{a}}{a}=H\frac{d%
}{dx},
\end{equation}
and
\begin{equation}
\frac{d^{2}}{dt^{2}}=\dot{H}\frac{d}{dx}+H\frac{d^{2}}{dx^{2}}\frac{dx}{dt}=%
\dot{H}\frac{d}{dx}+H^{2}\frac{d^{2}}{dx^{2}},
\end{equation}
respectively.
In the new variables, the first modified Friedmann equation (\ref{6b}) takes the form
\begin{equation}
H^{2}=\frac{\kappa ^{2}}{3}\left[\frac{V\left( \phi \right) }{1+\phi
+\phi _{,x}+\phi _{,x}^{2}/4\phi }\right].\label{pippo}
\end{equation}
The second Friedmann equation (\ref{6c}) becomes
\begin{equation}
\frac{\dot{H}}{H^{2}}=\frac{1}{2(1+\phi) }\left\{ \frac{1}{\phi }%
\phi _{,x}^{2}+4\phi _{,x}+\frac{2\kappa ^{2}}{3}\frac{\phi \left[
2V\left(
\phi \right) -\left( 1+\phi \right) V^{\prime }\left( \phi \right) \right] }{%
H^{2}}\right\} ,
\end{equation}
%\begin{equation}
%q=-\frac{1}{2}\frac{1}{1+\phi }\left\{ \frac{1}{\phi }\phi
%_{,x}^{2}+4\phi _{,x}+\frac{2\kappa ^{2}}{3}\frac{\phi \left[
%2V\left( \phi \right) -\left( 1+\phi \right) V^{\prime }\left(
%\phi \right) \right] }{H^{2}}\right\} -1,
%\end{equation}
while the deceleration parameter is given by
\begin{equation}
q=-\frac{1}{2(1+\phi) }\left\{ \frac{1}{\phi }\phi
_{,x}^{2}+4\phi _{,x}+2\frac{\phi \left[ 2V\left( \phi \right)
-\left( 1+\phi \right) V^{\prime }\left( \phi \right) \right]
\left( 1+\phi +\phi _{,x}+\phi _{,x}^{2}/4\phi \right) }{V(\phi
)}\right\} -1.
\end{equation}%
The generalized Klein-Gordon equation (\ref{3a2}) can be formulated as
\begin{equation}
\phi _{,xx}+3\phi _{,x}-\frac{1}{2\phi }\phi _{,x}^{2}+\frac{\dot{H}}{H^{2}}%
\phi _{,x}+\frac{2\kappa ^{2}}{3}\frac{\phi \left[ 2V\left( \phi
\right) -\left( 1+\phi \right) V^{\prime }\left( \phi \right)
\right] }{H^{2}}=0. \label{10}
\end{equation}

The mathematical as well as the physical condition for the existence of an accelerating cosmological phase can be formulated as
\begin{equation}
\frac{V_{,x}}{V}<\frac{\left[ 2\left( 1+\phi \right) +\phi
_{,x}^{2}/\phi +4\phi _{,x}\right] \phi _{,x}}{2\phi \left( 1+\phi
\right) \left( 1+\phi +\phi _{,x}+\phi _{,x}^{2}/4\phi \right)
}+\frac{2\phi _{,x}}{1+\phi },
\end{equation}%
where we have used the mathematical relations $dV/d\phi =\left( dV/dx\right) \left( dx/d\phi
\right) =\left( dV/dx\right) \left( 1/\phi _{,x}\right) $.\\

It is interesting to note that the generalized Klein-Gordon equation Eq.~(\ref{10}) can be expressed as a first order  Abel equation. To this effect, Eq.~(\ref{10}) can be rewritten as \cite{Harko:2011nh}
\begin{eqnarray}\label{12a}
\phi _{,xx}+3\phi _{,x}-\frac{1}{2\phi }\phi _{,x}^{2}&+&\frac{1}{2(1+\phi) }\left\{ \frac{1}{\phi }\phi _{,x}^{2}+4\phi _{,x}+f(\phi
)\left( 1+\phi +\phi _{,x}+\phi _{,x}^{2}/4\phi \right) \right\}
\phi _{,x}
\nonumber\\
&&+ f(\phi )\left( 1+\phi +\phi _{,x}+\phi _{,x}^{2}/4\phi \right)
=0,
\end{eqnarray}
where we have denoted $U(\phi )=V^{\prime }/V$, and $f\left( \phi
\right) =2\phi \left[ 2-\left( 1+\phi \right) U(\phi )\right] $.

After a few simple algebraic transformations Eq. (\ref{12a}) becomes
\begin{equation}
\phi _{,xx}+\frac{3}{2}(f+2)\phi _{,x}+\frac{3\phi \left( f+2\right) +f-2}{%
4\phi \left( 1+\phi \right) }\phi _{,x}^{2}+\frac{4+f}{8\phi
\left( 1+\phi \right) }\phi _{,x}^{3}+\left( 1+\phi \right) f=0.
\end{equation}
By introducing a new variable $u=d\phi/dx$, by dividing the resulting equation by $u^{3}$, and by denoting $u=1/v$ it turns out that the generalized second order Klein-Gordon equation can be reduced to the mathematical form of a first order Abel equation,  given by
\begin{equation}
\frac{dv}{d\phi }-\frac{3\phi \left( f+2\right) +f-2}{4\phi \left(
1+\phi
\right) }v-\frac{3}{2}(f+2)v^{2}-\left( 1+\phi \right) fv^{3}-\frac{4+f}{%
8\phi \left( 1+\phi \right) }=0\,.
\label{abelHMPG}
\end{equation}

%%%%%%%%%%%%%%%%%%%%%%%%%%%%%%%%%%%%%%%%%%%%%%%%%%%%%%%%%%%%%%%%%%%%%%%%%%
\subsection{Specific FLRW cosmological models}
%%%%%%%%%%%%%%%%%%%%%%%%%%%%%%%%%%%%%%%%%%%%%%%%%%%%%%%%%%%%%%%%%%%%%%%%%%

In the HMPG theory, the cosmological expansion of the Universe essentially depends on the analytical form of the scalar field potential $V(\phi)$.  In the following, we will briefly introduce several cosmological solutions of the generalized Friedmann equations in the HMPG theory.

%%%%%%%%%%%%%%%%%%%%%%%%%%%%%%%%%%%%%%%%%%%%%%%%%%%%%%%%%%%%%%%%%%%%%%%%%%
\subsubsection{de Sitter type solutions}
%%%%%%%%%%%%%%%%%%%%%%%%%%%%%%%%%%%%%%%%%%%%%%%%%%%%%%%%%%%%%%%%%%%%%%%%%%

A first example of a scalar field potential leading to interesting cosmological and astrophysical consequences is \cite{Harko:2011nh}
\begin{equation} \label{pot1}
V(\phi)=V_0+V_1\phi^2\,.
\end{equation}
Then from Eq.~(\ref{eq:phi(X)}) we obtain first the scalar relation $R=-\kappa^2T+2V_0$. As $T\to 0$ in the late Universe, this choice of the potential generates a de Sitter type accelerating expansion of the Universe,  with $V_0\sim \Lambda$, where $\Lambda$ is the cosmological constant.
If $V_1>0$, then the de Sitter phase corresponds to the minimum of the HMPG scalar field potential.  The effective mass $m_\varphi^2=2(V_0-2 V_1 \phi)/3>0$, and it takes a very small value for
$\phi<V_0/V_1$. Moreover, if $V_1$ is enough large, the field amplitude can be made
enough small, and thus it can satisfy all the important Solar System tests.

Now, Eq. (\ref{field1}) can be rewritten as
\begin{equation}\label{field1a}
\left(H+\frac{\dot{\phi}}{2(1+\phi)}\right)^2=\frac{\kappa^2\rho+V/2}
{3(1+\phi)}-\frac{\dot{\phi}^2}{4\phi(1+\phi)^2}.
\end{equation}
We can find another class of cosmological solutions by imposing the constraint
\begin{equation}
H+\frac{\dot{\phi}}{2(1+\phi)}=\tilde{H}_0= {\rm constant},
\end{equation}
so that one immediately obtains
\begin{equation}
  a (t)=\frac{a_0 e^{\tilde{H}_0t}}{\sqrt{1+\phi}}.
\end{equation}

In the limit $\rho\to 0$, from the generalized field equations and the condition $\dot{\tilde{H}}_0=0$ one can show that
\begin{equation}
\left(\tilde{H}_0^2-\frac{V}{6}\right)^2=9\tilde{H}_0^2\phi\left[\frac{V}
{6(1+\phi)}-\tilde{H}_0^2\right] \ ,
\end{equation}
giving for the scalar field potential the complicated expression
\begin{equation} \label{pot2}
V(\phi)=\frac{3\tilde{H}_0^2}{(1+\phi)}\left[2+11\phi\pm 3\phi \sqrt{5-4\phi}\right] \ .
\end{equation}
In this model the only free parameter is the amplitude $\tilde{H}_0^2$ of the generalized Hubble parameter, which, in order to satisfy the observational constraints, must have the same order of magnitude as the cosmological
constant $\Lambda$. It is a matter of further research to investigate the relevance of the potential (\ref{pot2}) from the elementary particle physics point of view.

%%%%%%%%%%%%%%%%%%%%%%%%%%%%%%%%%%%%%%%%%%%%%%%%%%%%%%%%%%%%%%%%%%%%%%%%%%
\subsubsection{Marginally accelerating solutions}
%%%%%%%%%%%%%%%%%%%%%%%%%%%%%%%%%%%%%%%%%%%%%%%%%%%%%%%%%%%%%%%%%%%%%%%%%%

We call marginally accelerating solutions of the Friedmann equations the cosmological solutions satisfying the condition $q=0$. From Eq.~(\ref{7a}) it follows that in HMPG theory the scalar field corresponding to a marginally accelerating solution must satisfy the equation \cite{Capozziello:2012ny}
\begin{equation}
\frac{\dot{\phi}^{2}}{2\phi }+2H\dot{\phi}+\frac{2}{3}\kappa
^{2}\phi \left[
3V\left( \phi \right) -\left( 1+\phi \right) V^{\prime }\left( \phi \right) %
\right] =0.
\end{equation}

A simple marginally accelerating solution can be obtained by
assuming that the scalar field  potential satisfies the condition
\begin{equation}
3V\left( \phi \right) -\left( 1+\phi \right) V^{\prime }\left(
\phi \right) =0,
\end{equation}
which immediately fixes the analytical form of the potential as
\begin{equation}
V(\phi )=V_{0}\left( 1+\phi \right) ^{3},
\end{equation}%
with $V_{0}$ an arbitrary constant of integration. The scalar field and the Hubble function are given by
$\phi =\phi _{0}/a^{4}$, and  $H=-\dot{\phi}/4\phi$. For the marginally accelerating case the general solutions of the Friedmann equations in HMPG theory are given by \cite{Capozziello:2012ny}
\begin{equation}
\phi \left( t\right) =\frac{\exp \left[ -4\kappa \sqrt{\frac{V_{0}}{3}}%
\left( t-t_{0}\right) \right] }{1-\exp \left[ -4\kappa \sqrt{\frac{V_{0}}{3}}%
\left( t-t_{0}\right) \right] },
\end{equation}
and
\begin{equation}
a(t)=a_{0}\left\{ \exp \left[ 4\kappa \sqrt{\frac{V_{0}}{3}}\left(
t-t_{0}\right) \right] -1\right\} ^{1/4},  \label{9a}
\end{equation}
respectively, where $t_0$ is an arbitrary constant of integration.
Marginally accelerating models are important in cosmology since all models
that reach the value $q=0$ of the deceleration parameter, expanding from phases with $q>0$, once they reach the marginally accelerating era, will necessarily enter into an accelerating cosmological regime, with $q<0$. Moreover,  these solutions represent the important transition point between the period of structure formation of the early Universe, and the dark energy dominated phase that occurred in the later evolutionary phases.

%%%%%%%%%%%%%%%%%%%%%%%%%%%%%%%%%%%%%%%%%%%%%%%%%%%%%%%%%%%%%%%%%%%%%%%%%%
\subsubsection{Accelerating solutions of the field equations}
%%%%%%%%%%%%%%%%%%%%%%%%%%%%%%%%%%%%%%%%%%%%%%%%%%%%%%%%%%%%%%%%%%%%%%%%%%

%%%%%%%%%%%%%%%%%%%%%%%%%%%%%%%%%%%%%%%%%%%%%%%%%%%%%%%%%%%%%%%%%%%%%%%%%%
\paragraph{Power-law accelerating models.}
%%%%%%%%%%%%%%%%%%%%%%%%%%%%%%%%%%%%%%%%%%%%%%%%%%%%%%%%%%%%%%%%%%%%%%%%%%

A particular solution of the cosmological field equations of the HMPG theory can be obtained by assuming that the deceleration parameter $q=-\dot{H}/H^{2}-1<0$ is a constant, $q=q_0={\rm constant}$ (during the de Sitter phase the deceleration parameter is $q=-1$).
We may further assume that the scalar field potential satisfies the simple condition
\begin{equation}
2V\left( \phi \right) -\left( 1+\phi \right) V^{\prime }\left(
\phi \right) =0,
\end{equation}
a condition that fixes the functional form of the potential as
\begin{equation}
V(\phi )=V_{0}\left( 1+\phi \right) ^{2}.
\end{equation}
Then it follows that the generalized Friedmann equations of the HMPG theory admit the solutions
\begin{equation}
H=\frac{\dot{a}}{a}=\frac{1}{\left( 1-q_{0}\right) t},
\end{equation}
and
\begin{equation}
a(t)=a_0t^{1-q_{0}},
\end{equation}
respectively, describing a power-law accelerating phase, whose nature depends on the numerical value of $q_0$.\\

%%%%%%%%%%%%%%%%%%%%%%%%%%%%%%%%%%%%%%%%%%%%%%%%%%%%%%%%%%%%%%%%%%%%%%%%%%
\paragraph{The case $f=0$, $V\left( \phi \right)=V_{0}\left( 1+\phi \right) ^{2}$.}
%%%%%%%%%%%%%%%%%%%%%%%%%%%%%%%%%%%%%%%%%%%%%%%%%%%%%%%%%%%%%%%%%%%%%%%%%%

In the case  $f=0$, $V\left( \phi \right)
=V_{0}\left( 1+\phi \right) ^{2}$, and $\phi \ll 1$, the solutions of the Abel equations (\ref{abelHMPG}) equivalent to the generalized Klein-Gordon equation can be approximated as
\begin{equation}
v=1+\frac{C}{\sqrt{6\phi }},\qquad u=\phi _{,x}=\frac{\sqrt{6\phi }}{\sqrt{6\phi }+C%
}.  \label{13}
\end{equation}

For small values of $\phi$, $\phi \ll 1$, the potential can be approximated as $V\approx
V_{0}={\rm constant}$. The condition for accelerating expansion becomes
\begin{equation}
\frac{\left[ 2+\phi _{,x}^{2}/\phi +4\phi _{,x}\right] \phi
_{,x}}{2\phi \left( 1+\phi _{,x}+\phi _{,x}^{2}/4\phi \right)
}+2\phi _{,x}>0,
\end{equation}%
and this condition is always satisfied, since if $C>0$, then $\phi _{,x}>0$. We also assume $\phi >0$.\\

%%%%%%%%%%%%%%%%%%%%%%%%%%%%%%%%%%%%%%%%%%%%%%%%%%%%%%%%%%%%%%%%%%%%%%%%%%
\paragraph{Solutions parameterized by $q$.}
%%%%%%%%%%%%%%%%%%%%%%%%%%%%%%%%%%%%%%%%%%%%%%%%%%%%%%%%%%%%%%%%%%%%%%%%%%

The generalized Friedmann equations take a very simple form if they are reformulated in terms of the deceleration parameter $q$ \cite{Capozziello:2012ny}.  Hence Eqs.~(\ref{6b}) and (\ref{6c}) can be rewritten in the following transparent form
\begin{eqnarray}
3H^{2}&=&\frac{\kappa ^{2}}{\left( 1+q\right) }\left[ 2V\left( \phi
\right) -\phi V^{\prime }\left( \phi \right) \right] ,  \label{8d}
 \\
\dot{H}&=&-\frac{\kappa ^{2}\left( 1-q\right) }{3\left( 1+q\right)
}\left[ 2V\left( \phi \right) -\phi V^{\prime }\left( \phi \right)
\right]. \label{8e}
\end{eqnarray}

 In order to completely solve the problem, and to close the system of cosmological equations, the generalized Klein-Gordon equation (\ref{3a2}), must also be considered in the same variables. This representation in terms of the deceleration parameter gives a powerful and extremely efficient method to obtain some solutions of the Friedmann equations.

In the case of the exponential expansion $q=-1$, $H=H_{0}$ = constant, $\dot{H%
}=0$, and $a(t)=a_{0}\exp \left[ H_{0}\left( t-t_{0}\right)
\right] $,
respectively. For this case, Eq.~(\ref{8e}) is automatically satisfied, while Eq.~(\ref{8d}%
) gives for the potential the expression
\begin{equation}
V\left( \phi \right) =\frac{3}{\kappa ^{2}}H_{0}^{2}+\frac{V_{0}}{\kappa ^{2}%
}\phi ^{2},
\end{equation}%
where $V_{0}$ is an arbitrary constant of integration. The scalar field is obtained by solving the generalized Klein-Gordon equation, and it is found as
\begin{equation}
\phi \left( t\right) =\left[ \frac{3H_{0}}{\exp \left[ H_{0}\left(
t-t_0\right) \right] \pm \sqrt{3V_{0}}}\right] ^{2}.
\end{equation}
In order to obtain a nonsingular expression of the scalar field we must chose the positive sign in the above expression of $\phi$.

In the case of  the accelerated expansion with a power law scale factor,  $q=q_{0}={\rm constant}$ and
$H(t)=1/\left( 1-q_{0}\right) t$, $\dot{H}=-\left[1\left(1-q_{0}\right)\right]
t^{-2}$, where $q_{0}\neq 1$. The generalized Friedmann equations reduce to a single condition for the
scalar field potential, given by
\begin{equation}
2V\left( \phi \right) -\phi V^{\prime }\left( \phi \right)
=\frac{3\left( 1+q_{0}\right) }{\left( 1-q_{0}\right)
^{2}}\frac{1}{t^{2}}.  \label{11}
\end{equation}
 Hence from the above differential equation we can determine the forms of the scalar field potential compatible with accelerated power law expansion, with the scale factor given by $a(t)=a_0 t^{1-q_0}$.

%%%%%%%%%%%%%%%%%%%%%%%%%%%%%%%%%%%%%%%%%%%%%%%%%%%%%%%%%%%%%%%%%%%%%%%%%%
\subsection{Einstein's static Universe}
%%%%%%%%%%%%%%%%%%%%%%%%%%%%%%%%%%%%%%%%%%%%%%%%%%%%%%%%%%%%%%%%%%%%%%%%%%

Einstein's static Universe, representing the first cosmological solution of GR, has always attracted great interest, despite its static nature. Moreover, the interesting possibility that the Universe began in an asymptotically Einstein static state, followed by an inflationary expansion, has also been considered. The properties of the Einstein static Universe in the HMPG theory were considered in \cite{Boehmer:2013oxa}.

The Einstein static Universe is characterized by a constant value of the scale factor $a = a_0 = {\rm const}$ in the FLRW metric, giving $H=\dot{H}=0$, and $R = 6K/a^2_0$, where $K$ is a constant. In \cite{Boehmer:2013oxa} it was assumed that the ordinary matter satisfies the linear barotropic equation of state $p_m = w \rho_m$. Then the generalized Friedmann equations of the HMPG theory take the very simple form
\begin{align}
  \kappa^2\rho_{\rm eff} &= \frac{3K}{a_0^2}\,,
  \label{bg1}\\
  \frac{\kappa^2}{2}\lp\rho_{\rm eff}+p_{\rm eff}\rp &= \frac{K}{a_0^2}\,.
  \label{bg2}
\end{align}
The above equations lead immediately to the following constraint on the thermodynamic parameters of ordinary matter,
\begin{align}
  \rho_{\rm eff} + 3 p_{\rm eff} = 0\,.
  \label{bg3}
\end{align}

Next, after assuming  $\phi = \phi_0 = {\rm const}$, and by taking into account the definitions of the effective density and pressure, the following relation between the potential and matter density is easily obtained,
\begin{align}
  \kappa^2\rho_m(1+3w) =  V(\phi_0) \,.
  \label{bg4}
\end{align}
In the HMPG theory's Einstein static Universe the generalized Klein-Gordon equation (\ref{3a2}) takes the simple form
\begin{align}
  \frac{6K}{a^2_0} = V'(\phi_0) \,.
  \label{bg5}
\end{align}

Hence it follows that one can obtain the matter density  $\rho_m$ and the scale factor $a_0$ of the Einstein Universe in terms of $\phi_0$ and of the scalar field potential $V$. Now the generalized Friedmann equations give the following  equation, from which  the value of the field $\phi_0$ can be obtained as
\begin{align}
  1+ \phi_0 = \frac{V(\phi_0)}{V'(\phi_0)}\frac{3(1 + w)}{(1+3w)} \,.
  \label{bg6}
\end{align}
Once the potential is given from cosmological or physical considerations, one can obtain $\phi_0$ explicitly. It is interesting to point out that in the HMPG theory $K=-1$ is a physically acceptable parameter for Einstein's static Universe. If $V'(\phi_0) < 0$, the constant scale factor $a_0$ can be found from  Eq.~(\ref{bg5}). The linear homogeneous perturbations of Einstein's Universe were also considered in \cite{Boehmer:2013oxa}, and it was explicitly shown that a large class of stable solutions does exist.

%%%%%%%%%%%%%%%%%%%%%%%%%%%%%%%%%%%%%%%%%%%%%%%%%%%%%%%%%%%%%%%%%%%%%%%%%%
\subsection{G\"{o}del type solutions}
%%%%%%%%%%%%%%%%%%%%%%%%%%%%%%%%%%%%%%%%%%%%%%%%%%%%%%%%%%%%%%%%%%%%%%%%%%

G\"{o}del type solutions of the gravitational field equations in the HMPG theory have been obtained in \cite{Santos:2016tds}, for a metric of the form
\begin{equation}
ds^2 = [dt + H(r)d\phi]^2 - D^2(r)d\phi^2 - dr^2 - dz^2\,,
\end{equation}
written in cylindrical coordinates, $(r, \phi, z)$. The necessary and sufficient conditions for the G\"{o}del-type metric
to be space-time homogeneous (ST-homogeneous) are given by (see \cite{Santos:2016tds} and references therein)
\begin{equation}
\frac{H'}{D}  =  2\omega \qquad \; \text{and} \qquad \;
\frac{D''}{D}  =  m^{2},
\end{equation}
where the prime denotes the derivative with respect to $r$, while the parameters $(\Omega,m)$ are constants satisfying the conditions $\Omega^{2}>0$ and $-\infty\leq m^{2}\leq\infty$, respectively \cite{Santos:2016tds}.  The general G\"{o}del-type solutions, whose
matter source is a combination of a scalar with an electromagnetic fields plus a perfect fluid, were also determined.  This general solution contains all previously known G\"{o}del type solutions as special cases. The important result of the existence of the  G\"{o}del-type solutions, found in \cite{Santos:2016tds}, shows that the HMPG theory does not solve the fundamental causal inconsistency that appears in the form of closed timelike curves in the spacetime geometry, which is also present in standard GR.

%%%%%%%%%%%%%%%%%%%%%%%%%%%%%%%%%%%%%%%%%%%%%%%%%%%%%%%%%%%%%%%%%%%%%%%%%%
\subsection{Dynamical system analysis}
%%%%%%%%%%%%%%%%%%%%%%%%%%%%%%%%%%%%%%%%%%%%%%%%%%%%%%%%%%%%%%%%%%%%%%%%%%

The cosmological equations describing the dynamics of a homogeneous and isotropic Universe in both general relativity and modified gravity theories are systems of ordinary strongly nonlinear differential equations.  One important and elegant method allowing their investigation is by formulating them as dynamical systems. This formulation permits the use of powerful analytical and numerical tools to obtain a qualitative as well as quantitative comprehension of the cosmological dynamics. In particular, using dynamical systems theory one can study the fixed points,  Lyapunov stability, centre manifold, linear stability, as well as the global structure of the cosmological solutions \cite{Coley:2003mj}. Moreover,  the stability conditions allow to tightly constrain and even rule out on purely theoretical grounds different models, as well as a comparison with observational data.

The cosmology of the HMPG theory was studied by using methods from the theory of the qualitative analysis of the dynamical systems in \cite{Carloni:2015bua}. As a first step in this study, the propagation equations were formulated in terms of some appropriately defined variables, forming an autonomous system. The independent cosmological variables adopted in \cite{Carloni:2015bua} are
\begin{align}\label{100}
&X=\frac{\h}{H}\,,\qquad Y=\frac{\R}{6H^2}\,,\qquad Z=\frac{f}{6H^2}\,, \qquad\Omega= \frac{\mu}{3H^2}\,,\qquad K=\frac{k}{a^2H^2}\,,
\end{align}
and they must be considered together with the  dimensionless time variable $N=\ln a$.

In the $f(X)$ representation of the HMPG theory, the cosmological equations can be formulated in terms of the above variables as a dynamical system given by \cite{Carloni:2015bua},
\begin{align}
\begin{split}
& X_{,N}=\frac{1}{2} X \left[2 Z-X-2\mathcal{F}\left(K+X^2\right)+(1+3 w) \Omega \right]-K+Y\,,\\
& Y_{,N}=Y \left[2(1+Z)-2\mathcal{F}\left(K+X^2\right)+(1+3 w) \Omega +2
   (X-1) \mathcal{Q}\right]\,,\\
& Z_{,N}=Z \left[2(1+Z)-2\mathcal{F}\left(K+X^2\right)+(1+3 w) \Omega\right]+2  Y(X-1)
  \mathcal{F}\mathcal{Q}\,,\\
&\Omega_{,N}=\Omega  \left[2 Z-2\mathcal{F}\left(K+X^2\right)+(1+3 w) ( \Omega-1)\right]\,, \\
&K_{,N}=K \left[2 Z-2\mathcal{F}\left(K+X^2\right)+(1+3 w) \Omega  \right]\,.
\end{split}
\end{align}
together with the constraint
 \begin{equation}\label{101}
 K [1+\mathcal{F}]+X^2 \mathcal{F}-Y \mathcal{F}-\Omega +Z+1=0\,.
 \end{equation}
Note that the quantities $\mathcal{F}= F(\R)$ and $\mathcal{Q}(\R)=F/(\R F')$ are only functions of $\R$. In order to close the above system of equations, all geometrical and physical quantities have to be represented in terms of the variables introduced in Eq.~(\ref{100}). Once the functional form of $f$ is known, we can do this by taking into account that $Z/Y$ is a function of $\R$ only, that is, we have the relation
 \begin{equation}\label{KeyEq}
 \frac{Z}{Y}=\frac{f(\R)}{\R}.
 \end{equation}

 From the above relation and for $Y\neq0$, we obtain $\R=\R(Z/Y)$. Therefore the dynamical system can be closed.
By using the constraint equation (\ref{101}) we can easily eliminate the variable $\Omega$.  Hence, finally the  independent equations of the dynamical system describing the cosmological evolution in the HMPG theory are
 \begin{align}
 \begin{split}\label{DynSysRed}
& X_{,N}=\frac{1}{2} X \left\{\mathcal{F} \left[(3 w-1) \left(K+X^2\right)-(3 w+1) Y\right]+(1+3 w)(K+1)+3 (w+1) Z-2X\right\}-K+Y\,,\\
& Y_{,N}=Y\left\{\mathcal{F} \left[(3 w-1) \left(K+X^2\right)-(3 w+1) Y\right]+(1+3 w)K+3 (w+1) (Z+1)+2 \mathcal{Q} (X-1)+2\right\}\,,\\
&Z_{,N}=Z \left\{\mathcal{F} \left[(3 w-1) \left(K+X^2\right)-(3 w+1) Y\right]+(1+3  w)K+3(w+1) (Z+1)+2\right\}+2 \mathcal{F}\mathcal{Q}Y(X-1) \,,\\
&K_{,N}=K \left\{\mathcal{F} \left[(3 w-1) \left(K+X^2\right)-(3 w+1) Y\right]+(1+3  w)(K+1)+3 (w+1) Z\right\}\,.
\end{split}
\end{align}

The fixed (critical) points of the above system of strongly nonlinear system of differential equations represent exact cosmological solutions of the HMPG  theory, corresponding to either power-law or de Sitter type cosmological evolutions.  In \cite{Carloni:2015bua} two specific cases, corresponding to the choices of the function $f$ as $f=\chi \R^n$, where $\chi $ and $n$ are constants, and $f=\alpha \R/\left(1+\beta \R\right)$, with $\alpha $ and $\beta $ constants, were investigated in detail. For the first case the important result was found that the cosmological evolution always possesses an attractor, corresponding to an exponential, de Sitter type, expansion. For the second model,  some interesting properties of the cosmological model were found, together with some bounds on the model parameters. It also turns out that the critical points are not global attractors for this cosmological model. The thought-provoking problem of the stability of the Einstein static Universe, and of the bouncing solutions, was also investigated.

It is instructive to compare the dynamical system approach to HMPG cosmology with similar investigations in other modified gravity theories.  An investigation of the cosmology of $f(R)$ gravity from the perspective of the dynamical systems, in the absence of matter, was performed in \cite{Odintsov:2017tbc}. By introducing the new variables
$x_1=-\frac{\dot{F}(R)}{F(R)H}$, $x_2=-\frac{f(R)}{6F(R)H^2}$,$x_3=
\frac{R}{6H^2}$, the dynamical system takes the form
\begin{align}
& \frac{\mathrm{d}x_1}{\mathrm{d}N}=-4-3x_1+2x_3-x_1x_3+x_1^2\, ,
\\ \notag &
\frac{\mathrm{d}x_2}{\mathrm{d}N}=8+m-4x_3+x_2x_1-2x_2x_3+4x_2 \, ,\\
\notag & \frac{\mathrm{d}x_3}{\mathrm{d}N}=-8-m+8x_3-2x_3^2 \, ,
\end{align}
where the $e$-foldings number $N$ was used instead of the cosmic time, $d/dN=(1/H)d/dt$, while the parameter $m$ is given by $
m=-\ddot{H}/H^3$. If the parameter $m$ is a constant, the dynamical system is becomes autonomous, with all explicit time dependence removed from it.  Two cases were investigated, with the first leading to  a stable de Sitter attractor fixed point and to an unstable de Sitter fixed point, respectively, while the second is related to a matter dominated era. A detailed numerical analysis of the dynamical system was performed, together with a detailed analysis of the stability of all fixed points. The occurrence of future cosmological finite-time singularities in the dynamical system corresponding to two cosmological theories, namely that of vacuum $f(R)$ gravity and that of three fluids was investigated in \cite{Odintsov:2018uaw}. In the case of $f(R)$ gravity a non-autonomous dynamical system was introduced. However, this system is still attracted asymptotically to an accelerating attractor solution. In the case of the multifluid cosmology, consisting of three fluids, the interacting dark matter and dark energy fluids, and the baryonic fluid, respectively, by appropriately choosing the variables, one can obtain an autonomous polynomial dynamical system. By using a dominant balance analysis, one can investigate the occurrence of finite-time singularities.

%%%%%%%%%%%%%%%%%%%%%%%%%%%%%%%%%%%%%%%%%%%%%%%%%%%%%%%%%%%%%%%%%%%%%%%%%%
\subsection{Cosmological perturbations}
%%%%%%%%%%%%%%%%%%%%%%%%%%%%%%%%%%%%%%%%%%%%%%%%%%%%%%%%%%%%%%%%%%%%%%%%%%

The important problem of the analysis of perturbations of the cosmological models in the HMPG theory was considered in \cite{Harko:2011nh,Capozziello:2012ny} and \cite{Lima:2014aza}, respectively. The full set of the perturbation equations was obtained in the Jordan frame and in the Newtonian and synchronous gauges in \cite{Lima:2014aza}. The Poisson equation, giving the Newtonian limit of the theory, was also derived, and the evolution of the gravitational lensing potential for a cosmological model having a background evolution identical to $\Lambda$CDM was also considered.

In the following we briefly present the approach to the cosmological perturbations developed in \cite{Capozziello:2012ny}. For the sake of generality, the action considered in our presentation is taken as
\be
S=\frac{1}{2\kappa^2}\int d^4 x \sqrt{-g}\lb (\OA+\phi) R + \frac{3}{2\phi}\lp \partial\phi\rp^2 - 2\kappa^2V(\phi)\rb + S_m\,,
\ee
where
\be
\kappa^2V(\phi)=\frac{1}{2}\lb r(\phi)\phi-f(r(\phi))\rb, \quad r(\phi) \equiv {f'}^{-1}(\phi) \,.
\ee
and $\Omega_A$ is an arbitrary (constant) parameter. The Newtonian gauge, in which the perturbations are considered, can be parameterized by the two gravitational potentials $\Phi$ and $\Psi$, so that
\be
ds^2= -\lp 1+2\Psi \rp dt^2+a^2(t)\lp 1+2\Phi\rp d\vec{x}^2\,.
\ee
The full set of the perturbations equations is obtained as \cite{Capozziello:2012ny}
\bea
&&\frac{k^2}{a^2}\Phi
+ 3\lp H-\frac{\dot{\phi}}{2\Fp}\rp\dot{\Phi}
- 3\lp H^2 +\frac{H\dot{\phi}}{\F}-\frac{\dot{\phi}^2}{4\phi\Fp}\rp\Psi
   \nonumber \\
&&\qquad =\frac{1}{2\F}\lb \kappa^2\delta\rho_m + \lp \frac{3}{4\phi^2}\dot{\phi}^2+V'(\phi)-3H^2-\frac{k^2}{a^2}\rp\varphi - 3\lp H + \frac{\dot{\phi}}{2\phi}\rp
\dot{\varphi}\rb\,,
\eea
\bea
&&\lb 6\lp H^2+2\dot{H}\rp - 2\frac{k^2}{a^2} + \frac{6}{\F}\lp
\ddot{\phi}-\frac{\dot{\phi}^2}{\phi^2}+H\dot{\phi}\rp \rb \Psi
+ 3\lp 2H - \frac{\dot{\phi}}{\F}\rp \lp \dot{\Phi} - \dot{\Psi}\rp
-6\ddot{\Phi}
  \nonumber \\
&&\qquad
= \frac{1}{\F}\lb \kappa^2\lp \delta\rho_m+3\delta p_m\rp
+\lp 6H^2+6\dot{H} + 3\frac{\ddot{\phi}}{\phi^2}-2V'(\phi)+\frac{k^2}{a^2}\rp\varphi
+3\lp H-\frac{2\dot{\phi}}{\phi}\rp \dot{\varphi}
+ 3\ddot{\varphi}
\rb\,.
\eea
\be
- \lp H+\frac{\dot{\phi}}{2\Fp}\rp\Phi + \dot{\Phi}
= \frac{1}{2\Fp}\lb \kappa^2\lp\rho_m+p_m\rp a v_m + \lp H+\frac{3\dot{\phi}}{2\phi}\rp\varphi+\dot{\varphi}\rb\,
\ee
where we have denoted $\varphi=\delta\phi$. To the above set of perturbations equations one must also add the off-diagonal spatial part,
\be
\Psi+\Phi=-\frac{\varphi}{\F}\,.
\ee

For the case of a perfect cosmological fluid, the continuity and Euler equations for the ordinary baryonic matter components are
\bea
\dot{\delta} + 3H\lp c_s^2-w\rp\delta & = & -\lp 1+w \rp \lp 3\dot{\Phi}-\frac{k^2}{a}v\rp\,, \\
\ddot{v}+\lp 1-3c_a^2\rp Hv & = &  \frac{1}{a}\lp \Psi + \frac{c_s^2}{1+w}\delta\rp\,,
\eea
respectively. In the HMPG theory the linearly perturbed component of the generalized Klein-Gordon equation is given by \cite{Capozziello:2012ny}
\bea
\ddot{\varphi}+\lp 3H+\frac{1}{\phi}\rp \dot{\varphi} + \lp \frac{k^2}{a^2}+\frac{\dot{\phi}^2}{2\phi^2}- \frac{2}{3}V''(\phi)\rp\varphi &= & \nonumber
\lp 2\ddot{\phi}+6H\dot{\phi}-\frac{3}{2\phi}\dot{\phi}^2\rp\Psi + \dot{\phi}\lp \dot{\Psi}-3\dot{\Phi}\rp - \frac{\phi}{3}\delta R\,.
\eea

%%%%%%%%%%%%%%%%%%%%%%%%%%%%%%%%%%%%%%%%%%%%%%%%%%%%%%%%%%%%%%%%%%%%%%%%%%
\subsubsection{Perturbations in the matter dominated cosmology}
%%%%%%%%%%%%%%%%%%%%%%%%%%%%%%%%%%%%%%%%%%%%%%%%%%%%%%%%%%%%%%%%%%%%%%%%%%

In the case of a matter-dominated Universe we have $w_m=c_s^2=0$, where $c_s$ denotes the speed of the sound. Moreover, for the matter dominated epoch in the perturbation equations the spatial gradients are much bigger than the time derivatives. Therefore, in this regime the perturbations of the
matter density are more important than the gravitational potentials. Hence, from the continuity and the Euler equation we immediately obtain
\be
\ddot{\delta}=-2H\dot{\delta}-\frac{k^2}{a^2}\Psi\,.
\ee
After defining $\Pi = a^2\rho_m\delta_m/k^2$, we obtain the field equations and  the generalized Klein-Gordon equation as
\bea
\Fp\Phi &=& \Pi-\varphi\,, \\
\Fp\lp\Psi+\Phi\rp & = & -\varphi\,, \\
-2\Fp\Psi & = & \Pi + \varphi\,, \\
3\varphi & = & -2\phi\lp \Psi + 2\Phi\rp\,.
\eea

Finally, one can obtain the perturbation equation of the baryonic matter density as
\be \label{delta_evol}
\ddot{\delta} + 2H\dot{\delta} = 4\pi G_{\rm eff}\rho_m\delta\,,
\ee
with
\be \label{delta_evol2}
 G_{\rm eff} \equiv \frac{\OA-\frac{1}{3}\phi}{\OA\Fp}G\,.
\ee
The above results indicate that in the HMPG theory there are no instabilities in the cosmological evolution of the matter inhomogeneities, in contrast to the Palatini-$f(\R)$ type theories \cite{Capozziello:2012ny}.

%%%%%%%%%%%%%%%%%%%%%%%%%%%%%%%%%%%%%%%%%%%%%%%%%%%%%%%%%%%%%%%%%%%%%%%%%%
\subsubsection{Perturbation propagations in the vacuum}
%%%%%%%%%%%%%%%%%%%%%%%%%%%%%%%%%%%%%%%%%%%%%%%%%%%%%%%%%%%%%%%%%%%%%%%%%%

Let's consider next the vacuum case, with  $\rho_m=0$, and investigate the curvature perturbations in the uniform-field gauge $\zeta$, given by
\be
\zeta= \Phi-\frac{H}{\dot{\phi}}\varphi\,.
\ee

The exact linear evolution equation for $\zeta $ can be obtained now as \cite{Capozziello:2012ny}
\be \label{eta_evol}
\ddot{\zeta}+\lb 3H
-2\frac{\ddot{\phi}+2\dot{H}\Fp -\frac{\dot{\phi}^2}{\F}}{\dot{\phi}+2H\Fp}
+ \frac{\phi\Fp}{\dot{\phi}^2}\lp\frac{2\ddot{\phi}\dot{\phi}}{\phi\Fp}
 + \frac{\dot{\phi}^3\Fp^2\phi}{1-\phi^3\Fp^3}\rp\rb\dot{\zeta} = - \frac{k^2}{a^2}\zeta\,.
\ee
From the above equation the important result follows  that small scale perturbations propagate with the speed of light, similarly to the canonical scalar field
theory. Hence, it turns out again that there are no instabilities in the scalar field perturbations.

%%%%%%%%%%%%%%%%%%%%%%%%%%%%%%%%%%%%%%%%%%%%%%%%%%%%%%%%%%%%%%%%%%%%%%%%%%
\subsection{Constraining the cosmological evolution}
%%%%%%%%%%%%%%%%%%%%%%%%%%%%%%%%%%%%%%%%%%%%%%%%%%%%%%%%%%%%%%%%%%%%%%%%%%

Two models of the HMPG theory were introduced in \cite{Lima:2015nma}, corresponding to two particular choices of the function $f(\R)$. The first adopted form is the exponential form $f(\R)=\Lambda _*\left(1+e^{-\R/\R_s}\right)$, where $\Lambda _*$ and $\R_s$ are the model parameters. The second choice is given by $f(\R)=\Lambda _*\left(1+\R^2/\R_s^2\right)$.  In both cases the background cosmological evolution was investigated, and it was shown that in the HMPG theory
 standard general relativity in the presence of a cosmological constant is recovered in the large time limit. The main reason behind this behavior is the evolution of the Palatini Ricci scalar towards the minimum of the effective scalar field potential. By using a combination of cosmological background data obtained from the observations of  the Cosmic Microwave Background, of supernovae
and of baryonic acoustic oscillations, the free parameters of the considered models were constrained. It turns out that for both HMPG models the maximum departure from the present day value of the gravitational constant $G$ is of the order of 1\%.

The cosmological evolution of a specific HMPG model, with $f(\R)\propto  \R^2$ was investigated in \cite{Leanizbarrutia:2017xyd}. In order to obtain some observational constraints on the free parameters of the model, and to test its  viability, the
conditions imposed by the Supernovae Ia and Baryonic Acoustic Oscillations data were analyzed. The results
were compared with the local constraints from Solar System tests. For the present day background value of the scalar field one obtains the constraint $-3.4\times 10^{-5}<\phi _0<3.4\times 10^{-5}$, while  the mass of the scalar field must have the value of the order of $m_{\phi} = 1 - 3H_0$ at the present cosmological
time. This leads to a very small field mass of the scalar field, and $m_{\phi}r \ll 1$ at the local scales, like the Earth or the Solar System. Hence, if the constraint on $\phi _0$ is satisfied, one could  successfully avoid corrections on the gravitational constant and on the parameter $\gamma$. On the other hand, the present day value of the scalar field $\phi _0$ can be set as small as required by the theoretical of observational constraints.
Moreover, for the considered redshift range $0<z<1$,  the evolution of the scalar field can be restricted to small values, in order  to satisfy both the cosmological and the local requirement.

%%%%%%%%%%%%%%%%%%%%%%%%%%%%%%%%%%%%%%%%%%%%%%%%%%%%%%%%%%%%%%%%%%%%%%%%%%
\section{Galactic and extragalactic dynamics}\label{II:galactic}
%%%%%%%%%%%%%%%%%%%%%%%%%%%%%%%%%%%%%%%%%%%%%%%%%%%%%%%%%%%%%%%%%%%%%%%%%%

The unsolved problem of the dark matter still continues to represent one of the biggest mysteries in present day science \cite{Boehmer:2007kx,Bohmer:2007fh,Harko:2018ayt,Borka:2015vqa}. Indeed, two important observations, namely, the behavior of the galactic rotation curves, and the mass discrepancy in clusters of galaxies, suggest the necessity of considering the existence of a special form of matter, called dark matter, mostly residing at galactic and extra-galactic scales. Physically dark matter exists in the form of a zero pressure (dust) gas of particles. The determination of the cosmological matter composition of the Universe from many cosmological observations also indicates the presence of a hypothetical dark matter component, fully determining, together with the dark energy, the cosmological dynamics.

%%%%%%%%%%%%%%%%%%%%%%%%%%%%%%%%%%%%%%%%%%%%%%%%%%%%%%%%%%%%%%%%%%%%%%%%%%
\subsection{Galactic rotation curves}
%%%%%%%%%%%%%%%%%%%%%%%%%%%%%%%%%%%%%%%%%%%%%%%%%%%%%%%%%%%%%%%%%%%%%%%%%%

One of the most important astrophysical observations backing the existence of dark matter is related to the behavior of the galactic rotation curves, which shows the unexpected existence of a constant velocity region, situated far away from the luminous galactic baryonic matter distribution. However, it is also possible to explain the strange behavior of the rotation curves by assuming that the nature of the gravitational force itself alters at galactic and extragalactic scales. Hence at scales much larger than that of the Solar System a new theory of gravity may be necessary to describe the observational phenomena related to the gravitational interactions. The possibility that the behavior of the rotation curves of massive test particles (hydrogen clouds) gravitating in circular orbits around galactic structures can be explained within the framework of the HMPG theory was investigated in \cite{Capozziello:2013yha}.

As a starting point in this analysis it was assumed that the metric outside the galactic baryonic mass distribution is static and spherically symmetric, given by
\begin{equation}
ds^{2}=-e^{\nu \left( r\right) }c^2dt^{2}+e^{\lambda \left( r\right)
}dr^{2}+r^{2}\left( d\theta ^{2}+\sin ^{2}\theta d\phi ^{2}\right) ,
\label{line}
\end{equation}
with the metric tensor components $\nu (r)$ and $\lambda (r)$ assumed to be functions of the radial coordinate $r$ only. The three-dimensional  velocity of a particle with nonzero mass is generally given by the expression
\begin{equation}
v^{2}(r)=e^{-\nu }\left[ e^{\lambda }\left( \frac{dr}{dt}\right)
^{2}+r^{2}\left( \frac{d\Omega }{dt}\right) ^{2}\right] ,
\end{equation}
where $d\Omega ^{2}=d\theta ^{2}+\sin ^{2}\theta d\phi ^{2}$. For a stable circular orbit $dr/dt=0$, and the tangential velocity is immediately obtained as
\begin{equation}
v_{tg}^{2}(r)=e^{-\nu }r^{2}\left( \frac{d\Omega }{dt}\right) ^{2}=e^{-\nu }r^{2}\left( \frac{d\Omega }{ds}\right) ^{2}\left(\frac{ds}{dt}^2\right).
\end{equation}
Along the galactic plane with $\theta =\pi /2$, the angular velocity is given by
\begin{equation}
\frac{v_{tg}^{2}(r)}{c^2}=\frac{r \nu ^{\prime }}{2}.  \label{vtg}
\end{equation}

Hence, in order to obtain the tangential velocity of a particle orbiting in the galactic plane we need to find the metric tensor component $\nu$. In the weak field approximation the static spherically symmetric field equations of the HMPG theory reduce to \cite{Capozziello:2013yha}
\bea\label{f1}
\frac{1}{r^2}\frac{d}{dr}\left(r\lambda\right)=\alpha \varphi +\beta &=&\rho ^{(\rm eff)},
  \\
\label{f2}
-\frac{\nu '}{r}+\frac{\lambda }{r^2}=\varphi ''+\alpha \varphi +\beta &=&-p_r^{(\rm eff)},
   \\
\label{f3}
-\frac{1}{2}\left(\nu ''+\frac{\nu '-\lambda '}{r}\right)=\alpha \varphi +\beta &=&-p_{\perp}^{(\rm eff)},
\eea
where $\alpha$ and $\beta$ are defined according to
\be
\alpha =\frac{1}{r_{\varphi }^2}- \frac{1}{2}V'\left(\phi _0\right), \qquad  \beta =-\frac{1}{2}V\left(\phi _0\right),
\ee
while $\varphi$ is related to the scalar field $\phi$ by $\phi =\phi _0+\varphi$.

The field equations can be solved to give for the tangential velocity the expression \cite{Capozziello:2013yha}
\bea\label{vfin}
\frac{v_{tg}^2}{c^2}&=&\frac{V_0}{6} r^2 + \frac{GM_B}{c^2r}\left\{1+\frac{2\phi_0}{3}e^{\frac{GM_B/c^2-r}{r_\varphi}}\left[\left(1+\frac{r}{r_\varphi}\right) (2+\alpha r_\varphi^2)+\frac{r^2}{r_\varphi^2}(1+\alpha r_\varphi^2)\right]\right\},
\eea
where $r_{\phi}=1/m_{\phi}$, and $M_B$ is the mass of the baryonic matter in the galaxy. In the HMPG theory we can also introduce formally an equivalent ``dark matter'' mass profile $M_{DM}(r)$, associated to the tangential velocity profile, and which can be defined as
\begin{equation}\label{darkmass}
\frac{2GM_{DM}(r)}{c^2}=\int_{R_B}^r{\left(\alpha \varphi +\beta \right)r^2dr}.
\end{equation}
Thus, the metric tensor component $\lambda(r)$ can be obtained as
\be
\lambda(r)=\frac{2GM_{DM}(r)}{r}.
\ee
We can also obtain the effective ``dark matter'' density profile $\rho _{DM}$ as given by the relation
\be
 \rho _{DM}(r)=\frac{1}{4\pi r^2}\frac{dM_{DM}(r)}{dr}=\frac{c^2}{2G}\left(\alpha \varphi+\beta \right).
 \ee

The above results show that the galactic rotation curves and the baryonic mass deficit in galaxies can be naturally explained in the HMPG theory. First of all, The extra-terms that appear in the gravitational field equations, due to the presence in the gravitational action of the Palatini Ricci scalar, influence, through the metric tensor, the equations of the particles, moving in the galactic plane. Secondly, the theory also induces a supplementary gravitational type universal interaction, which can explain the observational data  related to the unusual behavior of the galactic rotation curves. The scalar field also contributes with an effective energy density and mass to the total mass balance of the galaxy, and this effective supplementary and nonbaryonic mass, of gravitational (geometric) origin, can be interpreted physically as representing the effective mass of the "dark matter" in the galactic halo \cite{Capozziello:2013yha}.

%%%%%%%%%%%%%%%%%%%%%%%%%%%%%%%%%%%%%%%%%%%%%%%%%%%%%%%%%%%%%%%%%%%%%%%%%%
\subsection{Clusters of galaxies}
%%%%%%%%%%%%%%%%%%%%%%%%%%%%%%%%%%%%%%%%%%%%%%%%%%%%%%%%%%%%%%%%%%%%%%%%%%

A second fundamental astrophysical phenomena that strongly points towards the existence of dark matter is related to the well known mass inconsistency in galaxy clusters, large scale astrophysical systems bounded by gravity, and which consists from a few hundred to thousands of galaxies. There are two basic methods for the determination of the total gravitational mass of a cluster of galaxies. Firstly, one can compute an estimate of the mass $M_V$ of the cluster with the use of the virial theorem, and by using the observational data related to the dynamical evolutions of the galaxies in the cluster, giving their kinetic energy. On the other hand, in a second method, the total baryonic mass $M_B$ of the cluster can be determined through the simple algebraic addition of the masses of the individual galaxies of the cluster. The mass inconsistency  problem for the clusters of galaxies appears because many high quality astrophysical observations show that $M_V$ is always much greater than $M_B$, so that  $M_V /M_B \sim 20 - 30$.

The properties of the clusters of galaxies in the HMPG theory were investigated in \cite{Capozziello:2012qt}, under the basic assumption that dark matter is just a modification of the gravitational force at extragalactic scales. The starting point in the analysis of the properties of the clusters of galaxies in the HMPG theory is the fundamental Boltzmann equation of kinetic physics in an arbitrary Riemann geometry for the galactic distribution function $f_B$,
\be  \label{distr}
\left( p^{\alpha }\frac{\partial }{\partial x^{\alpha }}-p^{\alpha }p^{\beta
}\Gamma _{\alpha \beta }^{i}\frac{\partial }{\partial p^{i}}\right) f_B=0,
\ee
where $\Gamma _{\alpha \beta }^{i}$ are the Christoffel symbols associated to the metric, while $p^{\alpha }$ is the four-momentum of the particle (the galaxy, in the present investigation). In the case of a static spherically symmetric geometry, with a metric given by Eq.~(\ref{line}), the Boltzmann equation takes the general form
\bea
&&r\frac{\partial}{\partial r}\left[\rho\left\langle u_{1}^{2}\right\rangle%
\right]+ \frac{1}{2}\rho \left[ \left\langle u_{0}^{2}\right\rangle +
\left\langle u_{1}^{2}\right\rangle\right] r\frac{\partial \nu }{\partial r}
-\rho \left[ \left\langle u_{2}^{2}\right\rangle +\left\langle
u_{3}^{2}\right\rangle -2\left\langle u_{1}^{2}\right\rangle \right] =0,
\label{tetr2}
\eea
where $\rho $ is the mass density of the ordinary matter in the cluster, while $\left\langle u_{i}^{2}\right\rangle$, $i=0,1,2,3$  denotes the averages of the four-velocities of the galaxies.

After multiplying the above equation by $4\pi r^{2}$, and integrating over the cluster we obtain the important relation
\bea
&&\int_{0}^{R}4\pi \rho \left[ \left\langle u_{1}^{2}\right\rangle
+\left\langle u_{2}^{2}\right\rangle +\left\langle u_{3}^{2}\right\rangle%
\right] r^{2}dr -\frac{1}{2}\int_{0}^{R}4\pi r^{3}\rho \left[ \left\langle
u_{0}^{2}\right\rangle +\left\langle u_{1}^{2}\right\rangle\right] \frac{%
\partial \nu }{\partial r}dr=0.  \label{kin}
\eea
In the limit of small  velocities $\langle u_{1}^{2}\rangle \approx \langle u_{2}^{2}\rangle \approx
\langle u_{3}^{2}\rangle \ll \langle u_{0}^{2}\rangle \approx 1$, and the gravitational field equations of HMPG reduce to
\be  \label{fin1}
\frac{1}{2r^{2}}\frac{\partial }{\partial r}\left(r^{2} \frac{\partial \nu }{%
\partial r}\right) = 4\pi G\rho + 4\pi G\rho_{\phi}^{(eff)},
\ee
where \be\label{approx}
4\pi G\rho _{\phi }^{(eff)} \simeq  V\left( \phi \right) +\left( 2\nabla
_{t}\nabla ^{t}+\square \right) \phi -\frac{3}{\phi }\nabla _t\phi \nabla ^t\phi ,
\ee
represents the effective energy of the scalar field. Equation (\ref{kin}) becomes
\be
2K-\frac{1}{2}\int_{0}^{R}4\pi r^{3}\rho \frac{\partial \nu }{\partial r}%
dr=0,  \label{cond1}
\ee
where by
\begin{align}
K=\int_{0}^{R}2\pi \rho \left[ \left\langle u_{1}^{2}\right\rangle
+\left\langle u_{2}^{2}\right\rangle +\left\langle u_{3}^{2}\right\rangle %
\right] r^{2}dr,
\end{align}
we have denoted the total kinetic energy of the galaxies in the cluster.

In the following we introduce the total baryonic mass of the cluster $M_B$,
defined as $M_B=\int_{0}^{R}dM(r)=\int_{0}^{R} 4\pi \rho r^{2}dr$, the geometric mass of the cluster
$
M_{\phi }^{(\rm eff)}\left( r\right) =4\pi \int_{0}^{r}\rho _{\phi}^{(\rm eff)}(r')r'^{2}
dr'$,  as well as the baryonic $\Omega _B$ and the geometric $\Omega _{\phi }^{(\rm eff)}$ potential energies of the cluster, defined as
\begin{eqnarray}
\Omega _B=&-\int_{0}^{R}\frac{GM_B(r)}{r}\,dM_B(r),
\qquad \qquad
\Omega _{\phi }^{(\rm eff)}=\int_{0}^{R}\frac{GM_{\phi }^{(\rm eff)}(r)}{r}\,dM_B(r),
\end{eqnarray}
where $R$ is the cluster radius. We also define the radii $R_{V}$ and $%
R_{\phi }$ as
\be\label{RU3}
R_{V}=M_B^{2}\Bigl/\int_{0}^{R}\frac{M_B(r)}{r}\,dM_B(r),\Bigr.
\qquad
R_{\phi }^{(\rm eff)}=\left[M_{\phi }^{(\rm eff)}\right]^{2}\Bigl/\int_{0}^{R}\frac{M_{\phi }^{(\rm eff)}(r)}{r}\,dM_B(r).\Bigr.
\ee

All these quantities are related through the virial theorem, which follows immediately from Eq.~(\ref{kin}), and which can be formulated as
\be
2K+ \Omega _B- \Omega_{\phi }^{(\rm eff)} =0.
\ee
 The baryonic potential energy  $\Omega _B$ and the geometric potential energy $\Omega _{\phi}^{(\rm eff)}$ can also be represented in the mathematically equivalent form
\begin{eqnarray}
\Omega _B =-\frac{GM_B^{2}}{R_{V}}, \qquad
 \Omega _{\phi}^{(\rm eff)} =\frac{G\left[M_{\phi }^{(\rm eff)}\right]^{2}}{R_{\phi }^{(\rm eff)}}.
\end{eqnarray}
By defining the virial mass $M_{V}$ of the cluster of galaxies as $2K=GM_BM_{V}/R_{V}$,
we obtain the following relation between the virial and the baryonic masses of the cluster of galaxies,
\be  \label{fin6}
\frac{M_{V}}{M_B}=1+\frac{\left[M_{\phi }^{(\rm eff)}\right]^{2}R_{V}}{M_B^{2}R_{\phi }^{(\rm eff)}},
\ee
a relation which for  $M_{V}/M_B>3$, can be approximated very well as
\be
M_{V}\approx \frac{\left[M_{\phi}^{(\rm eff)}\right]^2}{M_B}\frac{R_{V}}{R_{\phi }^{(\rm eff)}}. \label{virial}
\ee

Thus, due to the presence of the scalar field of the HMPG theory, and of the effective energy associated to it, one can convincingly explain the observed mass inconsistency in the clusters of galaxies \cite{Capozziello:2012qt}. The geometric quantities introduced in the theoretical analysis of the cluster dynamics can also be determined effectively from observations at the level of galaxy clusters, by using the astrophysical properties of the clusters, including the high precision data on the gas temperature and the core radius.

%%%%%%%%%%%%%%%%%%%%%%%%%%%%%%%%%%%%%%%%%%%%%%%%%%%%%%%%%%%%%%%%%%%%%%%%%%
\section{Stellar type objects, black holes and wormholes}\label{II:compact}
%%%%%%%%%%%%%%%%%%%%%%%%%%%%%%%%%%%%%%%%%%%%%%%%%%%%%%%%%%%%%%%%%%%%%%%%%%

As compared to Newtonian gravity, GR has a deep impact on the structure and properties of high density compact astrophysical objects, such as, for example, neutron stars. In particular, it predicts a maximum mass of stars consisting of baryonic matter, of the order of $3-3.5M_{\odot}$, and the existence of strange objects, characterized by the existence of a membrane type event horizon, called black holes. Exotic objects called wormholes, which may allow time travel, could also exist, and they are consistent with the mathematical structure of GR. It is an interesting topic of research to consider the impact of the modifications to the gravitational dynamics that appear in HMPG on the properties of stars, black holes, and wormholes.

For a static and spherically symmetric geometry, with the metric given by Eq.~(\ref{line}),
the Einstein gravitational field equations in HMPG take the form \cite{Capozziello:2012qt}
\begin{eqnarray}
\kappa ^{2}\rho (r)c^{2} =\frac{1}{r^{2}}\left[ 1-e^{-\lambda }\left(
1-r\lambda ^{\prime }\right) \right] (1+\phi )-e^{-\lambda }\left[ \phi
^{\prime \prime }-\frac{3\phi ^{\prime 2}}{4\phi }\right]
+\frac{\phi ^{\prime }}{2r}e^{-\lambda }\left( r\lambda ^{\prime
}-4\right) -\frac{V}{2}\,,
\label{hybrid_rho}
\end{eqnarray}%
\begin{equation}
\kappa ^{2}p_{r}(r)=\left[ \frac{1}{r^{2}}(e^{-\lambda }-1)+\frac{\nu
^{\prime }}{r}e^{-\lambda }\right] (1+\phi )+\phi ^{\prime }\left( \frac{\nu
^{\prime }}{2}+\frac{2}{r}+\frac{3\phi ^{\prime }}{4\phi }\right)
e^{-\lambda }+\frac{V}{2}\,,  \label{hybrid_pr}
\end{equation}%
\begin{eqnarray}
\kappa ^{2}p_{t}(r) &=&\Bigg[\left( \frac{\nu ^{\prime \prime }}{2}+\left(
\frac{\nu ^{\prime }}{2}\right) ^{2}+\frac{\nu ^{\prime }}{2r}\right)
e^{-\lambda }-\frac{1}{2}\frac{\lambda ^{\prime }e^{-\lambda }}{r}\left( 1+r%
\frac{\nu ^{\prime }}{2}\right) \Bigg](1+\phi )  \nonumber \\
&&+\left[ \phi ^{\prime \prime }+\frac{\phi ^{\prime }\nu ^{\prime }}{2}+%
\frac{3\phi ^{\prime 2}}{4\phi }\right] e^{-\lambda }+\frac{\phi ^{\prime }}{%
r}e^{-\lambda }\left( 1-\frac{r\lambda ^{\prime }}{2}\right) +\frac{V}{2}\,,
\end{eqnarray}
where a prime denotes the derivative with respect to the radial coordinate $r$, and $\rho $, $p_r$ and $p_t$ denote the matter energy density, and the radial and transversal pressures, respectively. The generalized Klein-Gordon equation describing the scalar field variation inside the compact object is given by
\begin{equation}
-\left[ \phi ^{\prime \prime }+\frac{\phi ^{\prime }\nu ^{\prime }}{2}-\frac{%
\phi ^{\prime 2}}{2\phi }+\frac{2\phi ^{\prime }}{r}\right] e^{-\lambda }+%
\frac{\phi ^{\prime }\lambda ^{\prime }}{2}e^{-\lambda }+\frac{\phi }{3}%
\left[ 2V-(1+\phi )V_{\phi }\right] =\frac{\phi \kappa ^{2}}{3}T\,,
\label{modKGeq}
\end{equation}
while the conservation of the matter energy-momentum tensor gives the following
relation between the thermodynamic quantities appearing in the energy-momentum tensor, and the
metric tensor component $\nu $,
\begin{equation}
\nu ^{\prime }=-\frac{2p_{r}^{\prime }}{\rho c^2+p_{r}}+\frac{2\left(
p_{t}-p_{r}\right) }{r}.  \label{eqcons}
\end{equation}

The above system of structure equations describing the properties of  compact objects in HMPG theory can be reformulated in a form somehow similar to standard GR by introducing the effective mass $m_{\rm eff}(r)$ of the compact object that allows  to represent the metric tensor coefficient $e^{-\lambda }$\ as
\begin{equation}
e^{-\lambda }=1-\frac{2G_{0}m_{\rm eff}(r)}{c^{2}r}.
\end{equation}%
Then the effective mass satisfies the mass continuity equation
\begin{equation}\label{26}
\frac{dm_{\rm eff}}{dr}=4\pi \rho_{\rm eff}r^{2},
\end{equation}%
where $\rho_{\rm eff}$ is the effective density of the star, defined as
\begin{equation}
\rho_{\rm eff}=-\frac{rf(\Phi )+3\Phi ^{\prime }/2}{4\pi r^{2}\left[ 1+\Phi
^{\prime }r/2\right] }m_{\rm eff}+\frac{1}{\kappa ^{2}c^{2}\left[ 1+\Phi
^{\prime }r/2\right] }\left[ 2\frac{\Phi ^{\prime }}{r}+\frac{U\left( \Phi
\right) }{2}+f(\Phi )+\kappa^{2}\rho c^{2}\right] .
\end{equation}

The generalized Tolman-Oppenheimer-Volkoff equation is obtained in the
HMPG theory as
\begin{equation}
\frac{dp}{dr}=-\frac{\left( \rho c^{2}+p\right) \left\{ \left( \kappa
^{2}pe^{-\Phi }-U/2\right) r^{2}-\left( 1-2G_{0}m_{\rm eff}/c^{2}r\right)
\left\{ 1+r\left[ 2+rh(\Phi )\right] \Phi ^{\prime }\right\} +1\right\} }{%
r\left( 1-2G_{0}m_{\rm eff}/c^{2}r\right) \left( 2+\Phi ^{\prime }r\right) }.
\label{peq}
\end{equation}

The metric tensor coefficient  $\nu ^{\prime }$ can be determined from the equation
\begin{equation}
\mathbf{\nu }^{\prime }=\frac{\left( \kappa ^{2}pe^{-\Phi }-U/2\right)
r^{2}-\left( 1-2G_{0}m_{\rm eff}/c^{2}r\right) \left\{ 1+r\left[ 2+rh(\Phi )%
\right] \Phi ^{\prime }\right\} +1}{r\left( 1-2G_{0}m_{\rm eff}/c^{2}r\right)
\left( 1+\Phi ^{\prime }r/2\right) } \,, \label{nuprime}
\end{equation}
where we have defined
\begin{equation}
h(\Phi )=\frac{3e^{\Phi }\Phi ^{\prime }}{4\left( e^{\Phi }-1\right) }.
\end{equation}

Finally, the generalized Klein-Gordon equation becomes
\begin{eqnarray}\label{31}
-\Phi ^{\prime \prime }+\frac{1}{2}\frac{2-e^{\Phi }}{e^{\Phi }-1}\Phi
^{\prime 2}-\Phi ^{\prime }\left[ -\frac{p^{\prime }}{\rho c^{2}+p}+\frac{2}{%
r}-\frac{G_{0}}{c^{2}}\frac{4\pi \rho_{\rm eff}r^{3}-m_{\rm eff}}{r^{2}\left(
1-2G_{0}m_{\rm eff}/c^{2}r\right) }\right]   \nonumber \\
+\frac{ e^{\Phi }-1 }{3\left( 1-2G_{0}m_{\rm eff}/c^{2}r\right) }%
\left[ U\left( \Phi \right) -\frac{dU\left( \Phi \right) }{d\Phi }-\kappa_{\rm eff}^{2}T\right] =0.
\end{eqnarray}

In the case of stellar type objects the system of equations  (\ref{26}), (\ref{peq}) and (\ref{modKGeq}) must be closed by specifying the equation of state for the matter (baryonic or of other type) inside the
star, $p=p(\rho)$, and by imposing the boundary conditions $m_{\rm eff}(0)=0$, $\rho
(0)=\rho _c$, $\Phi (0)=\Phi _0$, $\Phi ^{\prime }(0)=\Phi ^{\prime }_0 (0)$, and $p(R)=0$, respectively, where $\rho _c$ denotes the central density, with $R$ being the radius of the star.

%%%%%%%%%%%%%%%%%%%%%%%%%%%%%%%%%%%%%%%%%%%%%%%%%%%%%%%%%%%%%%%%%%%%%%%%%%
\subsubsection{Stars and black holes}
%%%%%%%%%%%%%%%%%%%%%%%%%%%%%%%%%%%%%%%%%%%%%%%%%%%%%%%%%%%%%%%%%%%%%%%%%%

\paragraph{Stellar properties in HMPG.} The structure and properties of the stellar type objects in HMPG were studied in \cite{Danila:2016lqx}. For the equations of state of the dense matter several cases were considered, corresponding to the stiff fluid, radiation-like fluid, MIT quark bag model, and to the Bose-Einstein Condensate (BEC) superfluid, respectively. The gravitational field equations were solved numerically, and the stellar models were studied in both GR and HMPG, respectively. An in-depth comparison between the predictions of these two gravitational theories was also performed. As a general result obtained from the investigation of the stellar properties it follows  that for all the  equations of state mentioned above, HMPG stars are, as expected, more massive than their general relativistic counterparts, with the mass of the stable stellar type configurations showing a significant increase of the gravitational mass. A comparison of the maximum masses of compact stellar type objects in GR and HMPG is presented in Table~\ref{TableI} \cite{Danila:2016lqx}.
The much higher masses of the stellar objects in HMPG may suggest the possibility that the so-called  stellar mass black holes, massive objects with masses in the range of $3.8M_{\odot}$ and $6M_{\odot}$,  may  be in fact HMPG stars \cite{Danila:2016lqx}.

  \begin{table}
\begin{center}
 \begin{tabular}{|c | c| c|}
 \hline
 EOS & $M_{\rm max}^{\rm GR}/M_{\odot}$ & $M_{\rm max}^{\rm HMP}/M_{\odot}$ \\
 \hline\hline
 MIT Bag Model & 2.025 & 4.359 \\
 \hline
 Stiff fluid  & 3.279 & 3.968 \\
 \hline
 Radiation fluid  & 2.256 & 3.660 \\
 \hline
 BEC  & 2.230 & 4.971 \\
 \hline
\end{tabular}
\caption{Maximum masses of stellar objects in GR and HMPG.}\label{TableI}
\end{center}
\end{table}

Recently, several massive neutron stars have been observed, and their masses measured, which are given by  $2.01\pm 0:04 M_{\odot}$ \cite{Antoniadis:2013pzd},
$2.27\pm 0.17M_{\odot}$ \cite{Linares:2018ppq}, and $2.14^{+0.20}_{-0.18}M_{\odot}$ \cite{Cromartie:2019kug}, all consistent
with a mass range of around $2.10M_{\odot}$. These masses, as well as possible higher masses can be easily explained in the framework of HMPG gravity, without resorting to exotic equations of state, but considering only the effects of the modified gravity on the stellar structure. Hence supermassive stellar models can be constructed without the need of introducing physics beyond the Standard Model of particle physics.

Similar results can be obtained in the framework of the $f(R)$ gravity theory. In \cite{Capozziello:2015yza} the stellar structure was investigated, for four different equations of state, for two distinct models, with the Lagrangians given by $f(R)=R + \alpha R^2\left(1 + \gamma R\right)$, and $f(R)\sim R^{1+\epsilon}$, respectively.  It was found that generally the weight of the gravitational corrections to the Lagrangian reshapes the structure of the neutron stars, due to the supplementary  curvature pressure-density terms present in the structure equations of the star. The case of the $R^{1+\epsilon}$ is particularly interesting, with different values of the parameter $\epsilon$ allowing the formation of larger stable stellar structures, in terms of both mass and radius. Extreme objects with masses as high as $3.1M_{\odot}$ are also possible. Hence modified gravity theories may offer the possibility of explaining the higher observed masses of neutron stars without any need for exotic equations of state, or new non-gravitational physics.

\paragraph{Black holes.} Black hole solutions in HMPG were obtained numerically in \cite{Danila:2018xya}.  From a numerical and mathematical point of view it is more advantageous to reformulate the field equations in a dimensionless form, and to introduce as an independent coordinate the quantity $\xi =1/r$. The formation of a black hole in the HMPG theory  is indicated by the  presence of a Killing horizon for the time-like Killing vector in the metric tensor components. Several black hole
models can be obtained, corresponding to different functional forms of the scalar field potential, which has a strong influence on the black hole properties. The location of the event horizon of the black holes is significantly different in HMPG as compared to GR, with  deviations as high as 25\% in the value of the Schwarzschild radius. The thermodynamic properties of the HMPG black holes (horizon temperature, specific heat, entropy and evaporation time due to Hawking luminosity) also show important variations when compared to their general relativistic counterparts.

 An in depth investigation of the vacuum solutions in HMPG was performed in \cite{Bronnikov:2019ugl,Bronnikov:2020vgg}, where it was pointed out  that a scalar-tensor theory of the HMPG type is identical with GR with a phantom conformally coupled scalar field representing the source of gravity. As a general result it was shown that asymptotically flat solutions of the field equations in HMPG theory either contain naked central singularities, or they describe traversable wormholes. Moreover, it was shown that a special two-parameter family of globally regular black hole solutions with extremal horizons does in fact exist in this type of theories. Another interesting class of solutions found in \cite{Bronnikov:2019ugl} is represented by a one-parameter family of solutions, containing an infinite number of extremal horizons located between static regions, and a spherical radius. It is important to note that the spherical radius monotonically changes from region to region. An important and interesting property of the black hole and wormhole type solutions obtained in \cite{Bronnikov:2019ugl} is that they are unstable under monopole perturbations. Moreover, a scalar-tensor theory with $V(\phi) \equiv 0$, and which contains at least one nontrivial ($\phi ={\rm constant}$ vacuum solution with $R = 0$, always reduces to a theory containing a conformal scalar field.

%%%%%%%%%%%%%%%%%%%%%%%%%%%%%%%%%%%%%%%%%%%%%%%%%%%%%%%%%%%%%%%%%%%%%%%%%%
\subsubsection{Wormhole solutions}
%%%%%%%%%%%%%%%%%%%%%%%%%%%%%%%%%%%%%%%%%%%%%%%%%%%%%%%%%%%%%%%%%%%%%%%%%%

Wormhole solutions in modified theories of gravity are particularly interesting, as in GR these exotic geometries violate the energy conditions \cite{Capozziello:2013vna,Capozziello:2014bqa,Mimoso:2014ofa,Lobo:2009ip}. Now, it was shown in modified gravity that one may impose that the matter threading the wormhole satisfies the energy conditions, so that it is the effective energy-momentum tensor containing higher order curvature derivatives that is responsible for the null energy condition violation \cite{Harko:2013yb}. Thus, the higher order curvature terms, interpreted as a gravitational fluid, sustain these non-standard wormhole geometries, fundamentally different from their counterparts in GR \cite{Lobo:2009ip,Bohmer:2011si,Harko:2013yb,Lobo:2007qi,MontelongoGarcia:2010xd,Garcia:2010xb,Lobo:2008zu,Mehdizadeh:2015jra,Boehmer:2007rm,Zangeneh:2015jda,Lobo:2010sb,Harko:2013aya,Boehmer:2007rm,Korolev:2020ohi}.

The possibility of the existence of wormhole solutions in HMPG theory was investigated in \cite{Capozziello:2012hr}, where the general conditions for wormhole formation, according to the null energy conditions at the throat were presented. Specific examples were also investigated in detail. In the first one the redshift function, the scalar field and the scalar field potential that simplifies the modified Klein-Gordon equation were indicated. The obtained solution is not asymptotically flat, and must be matched to an external  vacuum solution. In the second case one can obtain an asymptotically flat geometry, by  specifying the metric functions and the scalar field. We refer the reader to \cite{Capozziello:2012hr} for more details.

%%%%%%%%%%%%%%%%%%%%%%%%%%%%%%%%%%%%%%%%%%%%%%%%%%%%%%%%%%%%%%%%%%%%%%%%%%
\section{Thermodynamics, branes, screening, gravitational waves, strings, and further work}\label{II:branesGWs}
%%%%%%%%%%%%%%%%%%%%%%%%%%%%%%%%%%%%%%%%%%%%%%%%%%%%%%%%%%%%%%%%%%%%%%%%%%

 The first law and the generalized second law of thermodynamics on the dynamically apparent horizon of a black hole in HMPG theory was investigated in \cite{Azizi:2015ina}. The validity of the generalized second law of thermodynamics in a universe bounded by the cosmological event horizon and filled with a perfect fluid was also discussed.

 The tensor perturbation of thick branes in HMPG were analyzed in \cite{Fu:2016szo}. Two models were considered in great detail, including the cases of thick branes constructed by a background scalar field,  and thick branes constructed by pure gravity. Note that in both models the thick branes have no inner structure. But when the parameter of the first model becomes larger than its critical value, it turns out that the graviton zero mode has inner structure. It was also found that on the brane the effective four-dimensional gravity can be effectively reproduced for both cases. Moreover, both thick brane systems are stable against the tensor perturbations.

The efficiency of screening mechanisms in HMPG was explored in \cite{VargasdosSantos:2017ggl}. The value of the gravitational field was computed around spherical bodies embedded in a material background of constant energy and matter density. A thin shell condition depending on the background field value was derived. To find out the astrophysical relevance of the thin shell effect its behavior in the neighborhood of different astrophysical objects (planets, moons or stars) was analyzed, and it was found that the condition is well satisfied with the exception of  some distinctive objects. Bounds on the model were established using data from Solar System experiments, including the spectral deviation measured by the Cassini mission, and the stability of the Earth-Moon system. The obtained bounds fix the important range of viable hybrid gravity models.

The equations describing the propagation of the gravitational wave in HMPG were presented in \cite{Kausar:2018ipo}. The solution of the propagation equations was also obtained, and the polarization states were fully determined. The obtained results on the gravitational polarized modes were compared with the existing results in the original approaches of the hybrid combination. It was found that the difference in the results is due to the coupling of the Ricci scalar with the trace of the matter energy momentum tensor.

Static and cylindrically symmetric interior string-type solutions in the scalar-tensor representation of the HMPG theory of gravity were obtained in \cite{Harko:2020oxq}. The gravitational field equations were significantly simplified through the imposition of the Lorentz invariance condition along the $t$ and $z$ axes. Interestingly enough, the general solution of the field equations can be obtained, for an arbitrary form of the scalar field potential, in an exact closed parametric form, with the scalar field $\phi$ taken as a parameter. Several exact classes of solutions of the field equations were presented, and investigated in detail, corresponding to a constant potential, and to a power-law potential of the form $V (\phi )=V_0\phi ^{3 /4}$. String models with exponential and Higgs-type scalar field potentials were also investigated by using numerical methods. Hence a large class of novel stable stringlike solutions in the of HMPG gravity can be constructed, in which the basic parameters, such as the string tension, the metric tensor components as well as the scalar field depend substantially on the initial values of the scalar field, and of its derivative, on the $r =0$ circular axis of the string.

Furthermore, it is important to mention that symmetries play a crucial role in physics and, in particular, the Noether symmetries are a useful tool both to select models motivated at a fundamental level, and to find exact solutions for specific Lagrangians. In \cite{Borowiec:2014wva}, the application of point symmetries was considered in the HMPG theory in order to select the $f(\mathcal{R})$ functional form, and to find analytical solutions of the field equations, and for the related Wheeler-DeWitt (WDW) equation. It was shown that, in order to find out integrable $f(\mathcal{R})$ models, conformal transformations in the Lagrangians are extremely useful. In this context, specific conformal transformations were analyzed, and it was shown that it is possible to transform the field equations by using normal coordinates to simplify the dynamical system, and to obtain exact solutions. Furthermore, the WDW equation was derived for the minisuperspace model, and the Lie point symmetries for the WDW equation were determined, and used to find invariant solutions.

%%%%%%%%%%%%%%%%%%%%%%%%%%%%%%%%%%%%%%%%%%%%%%%%%%%%%%%%%%%%%%%%%%%%%%%%%%
\section{Beyond the standard linear hybrid metric-Palatini theory}\label{II:beyond}
%%%%%%%%%%%%%%%%%%%%%%%%%%%%%%%%%%%%%%%%%%%%%%%%%%%%%%%%%%%%%%%%%%%%%%%%%%

A generalization of the HMPG theory was introduced in \cite{Tamanini:2013ltp}. We call this theory Generalized Hybrid Metric-Palatini Gravity (GHMPG). The action of the theory is
\be
  S_{f}=\frac{1}{2\kappa^2}\int d^4x \sqrt{-g}\, f(R,\mathcal{R}) \,,
\ee
where $f(R,\mathcal{R})$ is an arbitrary function of the Ricci and Palatini scalars. Varying with respect to the independent connection $\hat\Gamma^\lambda_{\mu\nu}$ we obtain \cite{Tamanini:2013ltp}
\be
  \hat\nabla_\lambda\left(\sqrt{-g}\frac{\partial f}{\partial \mathcal{R}} g^{\mu\nu}\right)=0 \,,
\ee
an equation that can be solved in terms of a Levi-Civita connection associated to the conformal metric $h_{\mu\nu}=\frac{\partial f}{\partial \mathcal{R}}g_{\mu\nu}$, and given by
\be
  \hat\Gamma^\lambda_{\mu\nu} = \frac{1}{2} h^{\lambda\sigma} \left(\partial_\mu h_{\nu\sigma}+\partial_\nu h_{\mu\sigma}-\partial_\sigma h_{\mu\nu}\right) \,.
\ee
The variation with respect to the metric provides the set of the gravitational field equations as
\bea
  \frac{\partial f}{\partial{R}} R_{\mu\nu} -\frac{1}{2}g_{\mu\nu}f
  -\left(\nabla_\mu\nabla_\nu-g_{\mu\nu}\square \right)\frac{\partial f}{\partial{R}}  +\frac{\partial f}{\partial \mathcal{R}}\mathcal{R}_{\mu\nu}=8\pi T_{\mu\nu} \,.
\eea

The theory admits a scalar-tensor representation, with action given by
\be
  S=\frac{1}{2\kappa^2}\int d^4x\sqrt{-g} \left[\phi\,R-\frac{3}{2\xi
    }(\partial\xi)^2 -W(\phi,\xi)\right] \,,
\ee
 which can be interpreted theoretically as corresponding to a Brans-Dicke theory with a vanishing BD parameter in the presence of a potential $W(\phi,\xi)$, and interacting with a second minimally coupled scalar field $\xi$. In the Einstein frame the full set of gravitational field equations can be formulated as
 \be
  G_{\mu\nu}=\kappa^2 \left(T_{\mu\nu}^{(\phi)} +e^{-\kappa\phi\sqrt{2/3}}\,T_{\mu\nu}^{(\xi)}-g_{\mu\nu}W\right) \,,
\ee
\bea
  \square\phi +\frac{\kappa}{\sqrt{6}} e^{-\kappa\phi\sqrt{2/3}} (\nabla\xi)^2 -W_\phi &=0& \,, \\
  \square\xi +\frac{\kappa\sqrt{2}}{\sqrt{3}} \nabla_\mu\phi\nabla^\mu\xi -e^{\kappa\phi\sqrt{2/3}}\,W_\xi &=& 0 \,,
\eea
where
\be
  T_{\mu\nu}^{(\phi)}=\nabla_\mu\phi\nabla_\nu\phi-\frac{1}{2}g_{\mu\nu} (\nabla\phi)^2 \,,
  \ee
  \be
  T_{\mu\nu}^{(\xi)}=\nabla_\mu\xi\nabla_\nu\xi-\frac{1}{2}g_{\mu\nu} (\nabla\xi)^2 \,.
\ee
and $W_\phi$ and $W_\xi$ are the derivatives of the potential with respect to the two scalar fields $\phi$ and $\xi$, respectively.

The cosmological implications of the theory were also investigated. For a flat, homogeneous and isotropic geometry described by the FLRW metric (\ref{metric}),  the generalized Friedmann equations and the evolution equations for the two scalar fields are obtained as
\be
3\frac{k}{a^2}+3H^2= \frac{\kappa^2}{2}e^{-\sqrt{2/3} \kappa  \phi} \dot\xi^2+\frac{\kappa^2}{2}\dot\phi^2+ \kappa^2\,W \,,
\ee
\be
\frac{k}{a^2}+2 \dot H+3 H^2=-\frac{\kappa^2}{2}e^{-\sqrt{2/3} \kappa  \phi} \dot\xi^2 -\frac{\kappa^2}{2}\dot\phi^2+\kappa^2\,W \,,
\ee
\be
\ddot\phi+3 H \dot\phi+\frac{\kappa}{\sqrt{6}}  e^{-\sqrt{2/3} \kappa  \phi}\, \dot\xi^2+W_\phi=0 \,,
\ee
\be
\ddot\xi+3 H \dot\xi-\frac{\kappa\sqrt{2}}{\sqrt{3}}\, \dot\xi\,\dot\phi+e^{\sqrt{2/3} \kappa  \phi}\, W_\xi=0 \,.
\ee
By using the above equations the evolution of the cosmological models can be studied using methods from the theory of the dynamical systems.

The cosmological behavior in GHMPG was investigated in \cite{Rosa:2017jld,Rosa:2019ejh},  by  adopting some natural ansatzes for the scale factor and for the effective potential $W$. Exponentially expanding solutions for flat universes can be obtained in the considered models, even when the background cosmology is not vacuum. Wormhole solutions in GHMPG theory were investigated in \cite{Rosa:2018jwp}, by using the scalar-tensor representation of the theory. One very interesting property of these wormhole solutions is that the matter fields obeys the null energy condition (NEC) everywhere, including at the throat and up to infinity. Consequently, the important result emerges that there is no need for exotic matter to support the wormhole. The wormhole geometry with its flaring out at the throat is determined by the higher-order curvature terms, or, in the equivalent scalar-tensor representation, by the two fundamental scalar fields.  However, both of these possibilities require the presence of a gravitational fluid. Thus, in GHMPG, in building up a wormhole, one can exchange the exoticity of matter with the exoticity of the gravitational sector. The specific wormhole models obtained in \cite{Rosa:2018jwp}, and specifically built to obey the matter NEC from the throat to infinity, have generally three regions, namely, an interior region containing the throat, a thin shell of matter, and a vacuum Schwarzschild anti-de Sitter (AdS) exterior.

The generalized hybrid metric-Palatini theory of gravity admits a scalar-tensor representation in terms of two interacting scalar fields. In \cite{Sa:2020qfd,Sa:2020fvn}, it was shown that upon an appropriate choice of the interaction potential, one of the scalar fields behaves like dark energy, inducing a late-time accelerated expansion of the universe, while the other scalar field behaves like pressureless dark matter that, together with ordinary baryonic matter, dominates the intermediate phases of cosmic evolution. This unified description of dark energy and dark matter gives rise to viable cosmological solutions, which reproduce the main features of the evolution of the universe.
In \cite{Borowiec:2020lfx}, a class of scalar-tensor theories (STT) including a non-metricity that unifies metric, Palatini and hybrid metric-Palatini gravitational actions with non-minimal interaction was proposed and investigated from the point of view of their consistency with generalized conformal transformations. It was shown that these theories can be represented on-shell by a purely metric STT possessing the same solutions for a metric and a scalar field. A set of generalized invariants was also proposed.

In \cite{Koivisto:2013kwa}, a method was developed to analyze the field content of the (generalized) hybrid theories, in particular to determine whether the propagating degrees of freedom are ghosts or tachyons. New types of second, fourth and sixth order derivative gravity theories were investigated and the so called linear HMPG theory was singled out as a viable class of ``hybrid'' extensions of GR.
The nature of the two additional scalar degrees of freedom in GHMPG, which contains two coupled dynamical scalar modes, was studied in \cite{Bombacigno:2019did}. The weak field limit of the theory was investigated both in the static case, and from a gravitational waves motivated perspective. In the first case, and in the lowest order of the post parameterized Newtonian approximation of the model, the Yukawa corrections to the Newtonian potential do appear. An interesting consequence of the theory is that one scalar field can have long range interactions, and hence it can be used to model dark matter effects. For the gravitational waves it is possible to obtain well-defined physical degrees of freedom, once appropriate constraints are imposed on the model parameters. The study of the geodesic deviation indicates the presence of breathing and longitudinal polarizations of the scalar waves, which can give rise to beating phenomena. In \cite{Rosa:2020uoi} it was shown that the Kerr solution exists in GHMPG theory, and that for certain choices of the function $f(R,\R)$ that describe the theory, the Kerr solution can be stable against perturbations on the scalar degree of freedom of the theory. By using the separation methods stability bounds on the effective mass of the Ricci scalar perturbation have also been obtained. Some specific forms of the function $f(R,\R)$, allowing the existence of a stable Kerr solution, were also considered.

\newpage

%%%%%%%%%%%%%%%%%%%%%%%%%%%%%%%%%%%%%%%%%%%%%%%%%%%%%%%%%%%%%%%%%%%%%%%%%%
\part{Curvature-matter couplings}\label{PartIII:CMcoup}
%%%%%%%%%%%%%%%%%%%%%%%%%%%%%%%%%%%%%%%%%%%%%%%%%%%%%%%%%%%%%%%%%%%%%%%%%%

The explanation of the recent acceleration of the Universe, whose first observational evidences were provided in \cite{Perlmutter:1998np} and \cite{Riess:1998cb}, respectively, has led to the necessity of finding a convincing and credible explanation of this intriguing observation, as well as of the closely related problem of the strange composition of the Universe. The assumption of the existence of a simple cosmological constant, despite extremely attractive as a phenomenological explanation of observations, requires an unrealistic fine-tuning, and hence it is theoretically unsatisfactory. Hence one should look at more general modifications of standard general relativity. One of an important directions of study are the generalized gravity models involving nonminimal couplings between geometry and matter. In the present Chapter we will review this class of modified gravity theories, concentrating mainly on three such theories, $f\left(R,L_m\right)$, $f(R,T)$ and $f\left(R,T, R_{\mu \nu}T^{\mu \nu}\right)$ gravities, respectively.

%%%%%%%%%%%%%%%%%%%%%%%%%%%%%%%%%%%%%%%%%%%%%%%%%%%%%%%%%%%%%%%%%%%%%%%%%%
\section{Linear nonminimal curvature-matter couplings actions and gravitational field equations}\label{III:linearNCM}
%%%%%%%%%%%%%%%%%%%%%%%%%%%%%%%%%%%%%%%%%%%%%%%%%%%%%%%%%%%%%%%%%%%%%%%%%%

%%%%%%%%%%%%%%%%%%%%%%%%%%%%%%%%%%%%%%%%%%%%%%%%%%%%%%%%%%%%%%%%%%%%%%%%%%
%\subsection{Action and gravitational field equations}
%%%%%%%%%%%%%%%%%%%%%%%%%%%%%%%%%%%%%%%%%%%%%%%%%%%%%%%%%%%%%%%%%%%%%%%%%%

In this Section, we review some of the basic and interesting features and properties of the simplest class of models involving a linear nonminimal coupling between curvature and matter, which generalizes $f(R)$ gravity. This class of theories can be roughly divided into two groups: theories with scalar
field-geometry coupling, and theories with arbitrary forms of standard matter-geometry coupling, respectively. By standard matter we usually mean either baryonic matter, or radiation. In the second class of theories there is no specific evolution equation for matter, unless the case of the first class of theories, where, for example, the scalar field must satisfy a Klein-Gordon type equation. After briefly analyzing scalar field-geometry couplings, we will concentrate on the detailed presentation of the main results in the geometry-matter coupling approaches where the matter Lagrangian is taken in a general form, which includes the case of the baryonic matter.

\subsection{Scalar field-geometry couplings}

In order to solve the cosmological constant problem a dynamical approach was proposed in \cite{Mukohyama:2003nw}. The starting point in this analysis is the nonstandard Lagrangian,
\be\label{Ran}
 L  =  \int d^4x\sqrt{-g}
  \left[ \frac{R}{2\kappa^2} + \alpha R^2 + L_{kin} - V(\phi) \right],
  \ee
  where
  \be
 L_{kin}  =  \frac{\kappa^{-4}K^q}{2qf^{2q-1}},
\ee
where $f$ is an arbitrary  function of the Ricci scalar $R$ which vanishes at $R=0$, given by
$f(R) \sim (\kappa^4R^2)^m$. Here $\kappa$ denotes the Planck length, $\alpha$ and $q$ are constants, and
$K \equiv -\kappa^4\partial^{\mu}\phi\partial_{\mu}\phi$. It turns out that $f(R)$ is radiatively stable, leads to a stable dynamics, despite the  singular kinetic term. This approach reduces fine-tuning  by $60$  orders of magnitude, and can lead to a new mechanism for sampling possible cosmological constants.

The action (\ref{Ran}) can be written in a more general form as
\begin{equation}
  L=\sqrt{g}\left[ \frac{R}{2 \kappa^2}+ \alpha R^2+  f \left( R \right)
L_{kin}
- V (\phi) \right],
\end{equation}
where $L_{kin}= -\frac{1}{2}g^{\mu \nu }\partial _{\mu }\phi $ $\partial_{\nu }\phi $ is the kinetic term of the Lagrangian of a scalar field, and $f(R)$ is an arbitrary function of the Ricci scalar, which is considered to be divergent at $R=0$. Hence these approaches introduce an arbitrary coupling between gravity and the kinetic term of the scalar field.

In order to obtain a dark energy model that grows due to the decrease of the curvature, in \cite{Nojiri:2004bi} a direct (non-linear) gravitational
coupling between a particular form of a geometric term, constructed with the help of the Ricci scalar,  and a matter-like Lagrangian was proposed.
The action of the theory is given by
\be\label{actOd}
S=\int d^4 x \sqrt{-g}\left[{1 \over \kappa^2}R + \left(\frac{R}{\mu^2}\right)^\alpha L_d \right] ,
\ee
where $L_d$ is considered as a matter-like action (dark energy), $\alpha $ is a constant, and $\mu$ is a parameter that may eliminate the instabilities  occurring in higher derivative theories. The adopted form of the coupling may be justified from requirements of renormalizability, or it may be induced by quantum effects as, for example,  non-local effective action. The action (\ref{actOd}) can be rewritten in a scalar-tensor form by introducing two scalar fields $\zeta=R$ and $\eta=1/\kappa ^2+\alpha \zeta ^{\alpha -1}L_d=e^{-\sigma}$, respectively, and after rescaling the metric according to  $g_{\mu \nu}\rightarrow e^{\sigma} g_{\mu \nu}$, as
\bea
S&=&\int d^4 x \sqrt{-g}\left\{R - \frac{3}{2}g^{\mu\nu}\partial_\mu\sigma \partial_\nu\sigma
+ \left({1 \over \alpha}-1\right) \right.
 \left.
\left(\e^{-\sigma} -
{1 \over \kappa^2}\right)^{1 \over 1 - \alpha}
\left[\alpha L_d\left(\e^\sigma g_{\mu\nu}, \phi\right) \right]^{1 \over 1-\alpha}\right\}\ .
\eea

The variation with respect to the metric of Eq.~(\ref{actOd}) gives the gravitational field equations as
\be
 {1 \over \kappa^2}\left({1 \over 2}g^{\mu\nu}R - R^{\mu\nu}\right) + \tilde T^{\mu\nu}=0,
\ee
where $\tilde T_{\mu\nu}$ is the effective energy-momentum tensor obtained as
\bea
\tilde T^{\mu\nu}&\equiv& \frac{1}{\mu^{2\alpha}}\left[ - \alpha R^{\alpha - 1} R^{\mu\nu} L_d \right.
 \left. + \alpha\left(\nabla^\mu \nabla^\nu
 - g^{\mu\nu}\nabla^2 \right)\left(R^{\alpha -1 } L_d\right) + R^\alpha T^{\mu\nu}\right] ,
 \eea
and
\be
T^{\mu\nu}\equiv {1 \over \sqrt{-g}}{\delta \over \delta g_{\mu\nu}}
\left(\int d^4x\sqrt{-g} L_d\right).
\ee
As for $L_d$, it is assumed that it is the Lagrangian of a free massless scalar field $\phi$,
\be
L_d = - {1 \over 2}g^{\mu\nu}\partial_\mu \phi \partial_\nu \phi\ .
\ee

In the case of a FLRW universe it turns out that an accelerating solution of the form
\be\label{207}
a=a_0 t^{\alpha + 1 \over 3},\qquad H={\alpha + 1 \over 3t}\,
\ee
where $a_0$ is a constant, exists. The stability of this model was also studied.

A generalization of the action (\ref{actOd}) of the form
\be
S=\int d^4 x \sqrt{-g}\left[{1 \over \kappa^2}R + f(R) L_d \right],
\ee
giving the field equations
\bea
{1 \over \kappa^2}\left({1 \over 2}g^{\mu\nu}R - R^{\mu\nu}\right)
 - f'(R) R^{\mu\nu} L_d
 + \left(\nabla^\mu \nabla^\nu
 - g^{\mu\nu}\nabla^2 \right)\left[f'(R) L_d\right] + f(R) T^{\mu\nu}=0,
\eea
 was also considered. The stability of the solution (\ref{207}) was studied, by considering its perturbation in the form $a=at^{\frac{\alpha+1}{3}}\left(1+\delta\right)$, where $\left(|\delta|\ll 1\right)$.
If $\delta$  depends only on the time variable $t$, the perturbations equation takes the form
\be
\frac{1}{t}\frac{d^3\delta}{dt^3} + \frac{A_1}{t^2}\frac{d^2\delta}{dt^2}
+ \frac{A_2}{t^3}\frac{d\delta}{dt} + \frac{A_3}{t^4}\delta =0,
\ee
where $A_1$, $A_2$, $A_3$, and $A_4$ are some constants. If $\delta$ also depends on the spatial coordinates, the perturbation equation becomes
\be
\frac{1}{t}\frac{d^3\delta}{dt^3} + \frac{A_1}{t^2}\frac{d^2\delta}{dt^2}
+ \frac{A_2}{t^3}\frac{d\delta}{dt} + \frac{A_3}{t^4}\delta
+ B_1\frac{k^2}{t^2} \frac{d^2\delta}{dt^2} + \frac{B_2k^2 }{t}\frac{d\delta}{dt} =0,
\ee
where $B_1$ and $B_2$ are constants. From the analysis of the perturbation equations it turns out that the inhomogeneity of the Universe does not grow for the equation of state of the dark energy $w<-1$.

 The class of nonlinear matter-gravity theories described by an action of the type (\ref{actOd}) were  studied in the
Palatini approach in \cite{Allemandi:2005qs}. The considered action is
\begin{equation}
 S=\int{d^4x\sqrt{-g}\left[ F(R)+f\left( R\right) L_{d} + \kappa L_{mat},
(\Psi)\right]},
\end{equation}
  where  $f(R)$ and $F(R)$ are some analytic functions of the
scalar field $R=g^{\alpha\beta}R_{\alpha \beta}(\Gamma )$ and
 $L_{d}=-\frac{1}{2}g^{\mu \nu }\partial _{\mu }\phi $ $\partial_{\nu
}\phi +V \left( \phi \right)$ is the scalar field Lagrangian, or, generally any Lagrangian
describing spinors, vector fields, etc.  Note that the ordinary (baryonic) matter Lagrangian $L_{mat}$ is minimally coupled to gravity. After obtaining the field equations in the first order formalism the FLRW cosmology was studied, which leads to different results as compared to the purely metric approach. The emerging FLRW cosmology may lead either to an effective quintessence phase (cosmic acceleration), or to an effective phantom phase. A dark energy dominated phase also occurs in this model in the first order formalism. Finally, it was shown that a dynamical theory that can solve the cosmological
constant problem also exists in the Palatini formalism.

For the sake of completeness, in this context, we also mention the possibility of scalar field-matter Lagrangian coupling, which is specific to the so-called chameleon models, which contain a light scalar field whose mass is strongly dependent on the ambient cosmological matter density.  The scalar field can be non-minimally coupled to the matter sector, with the Lagrangian of the model given by \cite{Sheikhahmadi:2019gzs}
\be
S = \int d^4x \sqrt{-g} \; \left[\frac{R}{2} - {1 \over 2}\; \partial_\mu\phi \partial^\mu\phi - V(\phi) + f(\phi) L_m \right]\;,
\ee
where $L_m$ is the Lagrangian of the baryonic (standard) matter fields, $V(\phi)$ is the potential of the chameleon-like scalar field $\phi$, and $f(\phi)$ is  an arbitrary function of the scalar field.

For a FLRW metric, the Friedmann equations of the model are given by
\begin{equation}\label{Friedmann}
3H^2 = \rho_\phi + f(\phi) \rho\;, \qquad 2\dot{H} + 3 H^2 = - p_{\phi} - f(\phi) p\;,
\end{equation}
where $\rho_\phi$ and $p_\phi$ are the energy density and pressure of the scalar field, respectively.
The equation of motion of the scalar field (the generalized Klein-Gordon equation) is given by
\begin{equation}\label{EoM}
\ddot{\phi} + 3H\dot{\phi} + V'(\phi) = f'(\phi) L_m\;.
\end{equation}
For a discussion of the cosmological applications of  scalar field-matter couplings see \cite{Sheikhahmadi:2019gzs} and references therein.

 \subsection{Theories with standard curvature-geometry coupling}

The action of this modified theory of gravity is given by \cite{Bertolami:2007gv}
\begin{equation}
S=\int \left\{\frac{1}{2}f_1(R)+\left[1+\lambda f_2(R)\right]{
L}_{m}\right\} \sqrt{-g}\;d^{4}x~,
   \label{1b_actionlinear}
\end{equation}
where $f_i(R)$ (with $i=1,2$) are arbitrary functions of the Ricci scalar $R$, and the coupling constant $\lambda$ determines the strength of the interaction between $f_2(R)$ and the matter Lagrangian, denoted by ${L}_{m}$.

The field equations are provided by varying the action with respect to the metric $g_{\mu\nu}$, and are given by the following system
\bea
F_1(R)R_{\mu \nu }-\frac{1}{2}f_1(R)g_{\mu \nu }-\nabla_\mu
\nabla_\nu \,F_1(R)+g_{\mu\nu}\square\, F_1(R)
=-2\lambda F_2(R){L}_m R_{\mu\nu}
  \nonumber   \\
+2\lambda(\nabla_\mu
\nabla_\nu-g_{\mu\nu}\square \,){ L}_m F_2(R) +[1+\lambda f_2(R)]T_{\mu \nu }^{(m)}~,
\label{1b_field1a}
\eea
where $F_i(R)=f'_i(R)$, with the prime representing the derivative with respect to the scalar curvature. The matter energy-momentum tensor is defined as usual by
\begin{equation}
T_{\mu \nu
}^{(m)}=-\frac{2}{\sqrt{-g}}\frac{\delta(\sqrt{-g}\,{
L}_m)}{\delta(g^{\mu\nu})} ~. \label{1b_EMTdef}
\end{equation}
We will only consider the metric formalism, and we refer the reader \cite{Harko:2010hw} for the Palatini approach of the linear nonminimal curvature-matter coupling.

A general property of these nonminimal curvature-matter coupling theories is the non-conservation of the energy-momentum tensor. This can be easily verified by taking into account the covariant derivative of the field equation (\ref{1b_field1a}),
the Bianchi identity, $\nabla^\mu G_{\mu\nu}=0$, and the following identity $(\square\,\nabla_\nu -\nabla_\nu\square\,)F_i=R_{\mu\nu}\,\nabla^\mu F_i$, which then imply the following relationship
\begin{equation}
\nabla^\mu T_{\mu \nu }^{(m)}=\frac{\lambda F_2}{1+\lambda
f_2}\left[g_{\mu\nu}{L}_m- T_{\mu \nu
}^{(m)}\right]\nabla^\mu R ~. \label{1b_cons1}
\end{equation}

Thus, one may interpret the curvature-matter coupling as an exchange of energy and momentum
between both. It is rather important to mention that analogous couplings also arise after a conformal
transformation in the context of scalar-tensor theories of gravity, and in string theory. In the absence of the coupling, one verifies the conservation of the energy-momentum tensor \cite{Koivisto:2005yk}, which can also be confirmed from the diffeomorphism invariance of the matter part of the action.
Note that, from Eq. (\ref{1b_cons1}), one can easily check that the conservation of the matter energy-momentum tensor is also established if $f_2(R)$ is a constant, or if the matter Lagrangian is not an explicit function of the metric.

It is rather pedagogical to test the motion of massive test particles in this theory. To this effect, consider a perfect fluid, whose physical properties are described by the energy-momentum tensor introduced in its standard form: $T_{\mu \nu }^{(m)}=\left( \rho +p\right) U_{\mu }U_{\nu
}+pg_{\mu \nu }$, where $\rho$ is the energy density and $p$ is the isotropic pressure, respectively.
Note that the four-velocity, $U_{\mu }$, satisfies the standard conditions $U_{\mu }U^{\mu }=-1$ and $\nabla _{\nu }U^{\mu }U_{\mu  }=0$. It is also useful to introduce the projection operator $h_{\mu \lambda }=g_{\mu \lambda }+U_{\mu }U_{\lambda }$, which obeys the conditions $h_{\mu\lambda }U^{\mu }=0$.
The following step consists in contracting Eq. (\ref{1b_cons1}) with the projection operator
$h_{\mu \lambda }$, which yields the following expression
\bea
\left( \rho +p\right) g_{\mu \lambda }U^{\nu }\nabla_\nu
U^{\mu} & - &(\nabla_\nu p)(\delta_\lambda^\nu-U^\nu U_\lambda)
   \nonumber \\
&-&\frac{\lambda F_2}{1+\lambda f_2}\left({
L}_m-p\right)(\nabla_\nu R)\,(\delta_\lambda^\nu-U^\nu
U_\lambda)=0 ~.
\eea
Finally, by contracting this relation with $g^{\alpha \lambda }$ provides the equation of motion for a fluid element, given by
\begin{equation}
\frac{D U^{\alpha }}{ds} \equiv \frac{dU^{\alpha }}{ds}+\Gamma _{\mu
\nu }^{\alpha }U^{\mu }U^{\nu }=f^{\alpha }~, \label{1b_eq1}
\end{equation}
where the extra force is defined by
\begin{eqnarray}
\label{1b_force}
f^{\alpha }&=&\frac{1}{\rho +p}\Bigg[\frac{\lambda
F_2}{1+\lambda f_2}\left({L}_m -p\right)\nabla_\nu
R+\nabla_\nu p \Bigg] h^{\alpha \nu }\,.
\end{eqnarray}
Note that the extra force $f^{\alpha }$ is orthogonal to the four-velocity of the particle, $f^{\alpha }U_{\alpha }=0$, which can be seen directly from the properties of the projection operator. This result is consistent with the usual interpretation of the four-force in relativity, according to which only the component of the four-force that is orthogonal to the particle's four-velocity can influence its motion, and trajectory.

A particularly intriguing feature of these curvature-matter coupling theories is that the extra force depends on the functional form of the Lagrangian density. More specifically, by considering the Lagrangian density $L_m = p$, where $p$ is the pressure, the extra force vanishes \cite{Bertolami:2008ab}. On the other hand, it has been argued that this is not the unique choice for the matter Lagrangian density, and that more natural forms for $L_m$, such as $L_m = -\rho $, do not imply the vanishing of the extra force. Indeed, in the presence of the nonminimal coupling they give rise to two distinct theories with different predictions \cite{Faraoni:2009rk}, and this important issue has been further investigated in different physical and mathematical contexts \cite{Bertolami:2013raa,Minazzoli:2013bva}.

However, one can also argue that the energy-momentum tensor in modified theories of gravity with a nonminimal coupling is more general than the usual general-relativistic energy-momentum tensor for perfect fluids \cite{Harko:2010zi}, and for this reason it contains a supplementary equation of state dependent term, which could be related physically to the elastic stresses in the body, or to other forms of internal energy. Therefore, the extra force induced by the coupling between curvature and matter does not automatically vanishes as a consequence of the thermodynamic properties of the system, or for a specific choice of the matter Lagrangian, and it is non-zero in the case of a fluid of dust particles.

In an attempt to clarify this issue, consider the definition of the energy-momentum tensor, given by Eq. (\ref{1b_EMTdef}), and assuming that the Lagrangian density $L_{m}$ of matter depends only
on the metric tensor components $g_{\mu \nu }$, and not on its derivatives, so we have
\begin{equation}
T_{\mu \nu }=L_{m}g_{\mu \nu }-2\frac{\partial L_{m}}{\partial g^{\mu
\nu }}. \label{1b_defTab}
\end{equation}
By taking into account the explicit form of the field equations (\ref{1b_field1a}), one obtains for the covariant divergence of the energy-momentum tensor the
equation, given by Eq. (\ref{1b_cons1}), which can be rewritten as
\begin{equation}
\nabla ^{\mu }T_{\mu \nu }=2\left\{\nabla ^{\mu }\ln \left[ 1+\lambda f_{2}(R)%
\right] \right\}\frac{\partial L_{m}}{\partial g^{\mu \nu }}.  \label{1b_cons1b}
\end{equation}

Consider now the case in which matter is assumed to be a thermodynamic perfect fluid, obeys a barotropic equation of state, with the thermodynamic pressure $p$ being a function of the energy density $\rho$ only, so that $p=p\left( \rho \right) $. In this case, the matter Lagrangian
density, which in the general case could be a function of both density and
pressure, $L_{m}=L_{m}\left( \rho ,p\right) $, or of only one of the
thermodynamic parameters, becomes an arbitrary function of the matter density $\rho $ only, so that $L_{m}=L_{m}\left( \rho \right) $. Then the matter
energy-momentum tensor is obtained as \cite{Harko:2010zi}
\begin{equation}
T^{\mu \nu }=\rho \frac{dL_{m}}{d\rho }U^{\mu }U^{\nu }+\left( L_{m}-\rho
\frac{dL_{m}}{d\rho }\right) g^{\mu \nu },  \label{1b_tens}
\end{equation}
where the four-velocity $U^{\mu }=dx^{\mu }/ds$ satisfies the condition $%
g^{\mu \nu }U_{\mu }U_{\nu }=-1$.

Using Eqs.~(\ref{1b_cons1b}) and (\ref{1b_tens}) we obtain the equation of motion of a massive test particle, or of a test fluid, with a linear curvature-matter coupling as
\begin{equation}
\frac{D^{2}x^{\mu }}{ds^{2}}=U^{\nu }\nabla _{\nu }U^{\mu }=\frac{d^{2}x^{\mu }}{ds^{2}}+\Gamma _{\nu \lambda }^{\mu }U^{\nu }U^{\lambda
}=f^{\mu },
\label{1b_eqmota}
\end{equation}
where the word line parameter $s$ is taken as the proper time, $U^{\mu
}=dx^{\mu }/ds$ is the four-velocity of the particle, $\Gamma
_{\sigma \beta }^{\nu }$ are the Christoffel symbols associated to the
metric, and $f^{\mu }$ is defined as
\begin{equation}
f^{\mu }=-\nabla _{\nu }\ln \left\{ \left[ 1+\lambda f_{2}(R)\right] \frac{%
dL_{m}\left( \rho \right) }{d\rho }\right\} \left( U^{\mu }U^{\nu }+g^{\mu
\nu }\right) \,.
\end{equation}
As we have already mentioned, the extra force, $f^{\mu }$, generated due to the presence of the coupling between matter and geometry, is perpendicular in this type of models to the four-velocity, $f^{\mu}U_{\mu }=0$.

%%%%%%%%%%%%%%%%%%%%%%%%%%%%%%%%%%%%%%%%%%%%%%%%%%%%%%%%%%%%%%%%%%%%%%%%%%
\subsection{Scalar-tensor representation}
%%%%%%%%%%%%%%%%%%%%%%%%%%%%%%%%%%%%%%%%%%%%%%%%%%%%%%%%%%%%%%%%%%%%%%%%%%

The curvature-matter coupling theory can also be written in an equivalent form in a scalar-tensor representation, and this was established in \cite{Faraoni:2007sn}.
In fact, it can be shown that the action (\ref{1b_actionlinear}) is equivalent to a scalar-tensor Brans-Dicke-type theory, with a single scalar field, vanishing Brans-Dicke parameter $\omega$, and an unusual coupling of the potential $U(\psi)$ of the theory to matter.

As a first step, we introduce a new field $\phi$, so that the action~(\ref{1b_actionlinear}) takes the form
\begin{equation} \label{1b_100}
S = \int d^4x \sqrt{-g} \left\{ \frac{f_1(\phi)}{2} +\frac{1}{2} \,
\frac{df_1}{d\phi} \left( R-\phi \right) +\left[ 1+\lambda f_2(
\phi) \right] {L}_m \right\}.
\end{equation}
We then introduce the field $\psi(\phi) \equiv f_1'(\phi) $, where the prime  denotes differentiation with respect  to $\phi$, and thus we obtain the following action
\begin{equation} \label{1b_300}
S=\int d^4x \sqrt{-g} \left[ \frac{\psi R }{2} -V(\psi)\,
+U(\psi) {L}_m \right] \;,
\end{equation}
where
\begin{eqnarray}
V(\psi) = \frac{\phi(\psi) f_1' \left[ \phi (\psi ) \right]
-f_1\left[ \phi( \psi ) \right] }{2} \;,  \qquad
U( \psi) = 1+\lambda f_2\left[ \phi( \psi ) \right]
.\label{1b_500}
\end{eqnarray}
Note that the field $\phi (\psi)$ can be obtained by inverting $ \psi(\phi) \equiv
f_1'(\phi) $.

The actions~(\ref{1b_actionlinear}) and (\ref{1b_300}) are equivalent when $f_1''(R) \neq 0$ \cite{Faraoni:2007sn}. This can be established  by setting $\phi=R$, and then Eq.~(\ref{1b_100}) reduces trivially to Eq.~(\ref{1b_actionlinear}). On the other hand, the variation of Eq.~(\ref{1b_100}) with respect to $\phi$ gives
\begin{equation} \label{1b_600}
\left( R-\phi \right) f_1''(\phi)+2\lambda f_2'(\phi) {L}_m=0 .
\end{equation}
For the specific case of vacuum, namely, ${L}_m=0$, we verify that Eq.~(\ref{1b_600}) yields $\phi=R$ whenever $f_1''\neq 0$ \cite{Faraoni:2007yn,Teyssandier:1983zz,Wands:1993uu,Whitt:1984pd}. However, in the presence of
matter there seem to be other possibilities.
When ${L}_m\neq 0$, the actions~(\ref{1b_actionlinear}) and (\ref{1b_100}) are equivalent if
\begin{equation}
\left(
R-\phi \right) f_1''(\phi) +2\lambda f_2''(\phi) {L}_m \neq
0.
\end{equation}

If Eq.~(\ref{1b_600}) is satisfied, we have a pathological case, which corresponds to
\begin{equation} \label{1b_700}
\lambda f_2(\phi) {L}_m= \frac{ f_1'(\phi)}{2} \left(
\phi-R \right) -\frac{ f_1(\phi)}{2} \;.
\end{equation}
However, if Eq.~(\ref{1b_700}) holds, then the action~(\ref{1b_100}) reduces to the trivial case of pure matter without the gravity sector. Thus, for modified theories of gravity with a linear curvature-matter coupling the actions~(\ref{1b_actionlinear}) and (\ref{1b_300}) are equivalent when $f_1''(R) \neq 0$, similarly to the case of pure $f(R)$ gravity \cite{Faraoni:2007yn,Teyssandier:1983zz,Wands:1993uu,Whitt:1984pd}.

%%%%%%%%%%%%%%%%%%%%%%%%%%%%%%%%%%%%%%%%%%%%%%%%%%%%%%%%%%%%%%%%%%%%%%%%%%
\subsection{Generalized $f(R,L_m)$ curvature-matter couplings}\label{1d_Part1:5}
%%%%%%%%%%%%%%%%%%%%%%%%%%%%%%%%%%%%%%%%%%%%%%%%%%%%%%%%%%%%%%%%%%%%%%%%%%

The linear curvature-matter coupling theory, and in fact $f(R)$ gravity, can be generalized by assuming a maximal extension of the Einstein-Hilbert action, given by the following action \cite{Harko:2010mv}
\begin{equation}
S=\int f\left(R,L_m\right) \sqrt{-g}\;d^{4}x~,
\end{equation}
where $f\left(R,L_m\right)$ is an arbitrary function of the Ricci scalar $R$, and of the Lagrangian density corresponding to matter, $L_{m}$. The
energy-momentum tensor of the matter is defined as usual by Eq. (\ref{1b_EMTdef}). Thus, by assuming that the Lagrangian density $L_{m}$ of the matter depends only
on the metric tensor components $g_{\mu \nu }$, and not on its derivatives,
we obtain Eq. (\ref{1b_defTab}), which will be useful below.

The gravitational field equation is obtained by varying the action with respect to the metric, and yields the following
\begin{eqnarray}\label{1c_field2a}
&&f_{R}\left( R,L_{m}\right) R_{\mu \nu }+\left( g_{\mu \nu }\square\, -\nabla
_{\mu }\nabla _{\nu }\right) f_{R}\left( R,L_{m}\right)
\nonumber\\
&&-\frac{1}{2}\left[
f\left( R,L_{m}\right) -f_{L_{m}}\left( R,L_{m}\right)L_{m}\right] g_{\mu \nu }=
\frac{1}{2}%
f_{L_{m}}\left( R,L_{m}\right) T_{\mu \nu }.
\end{eqnarray}
Note that for the specific case of the Hilbert-Einstein Lagrangian, i.e., $f\left( R,L_{m}\right) =R/2+L_{m}$, we recover the standard Einstein field equation of GR. For the slightly more involved case $f\left( R,L_{m}\right)=f_{1}(R)+f_{2}(R)G\left( L_{m}\right) $, where $f_{1}$ and $f_{2}$ are
arbitrary functions of the Ricci scalar, and $G$ a function of the matter Lagrangian
density, respectively, we reobtain the field equations of the modified
gravity with a curvature-matter coupling introduced in \cite{Harko:2008qz}.

Taking the covariant divergence of Eq.~(\ref{1c_field2a}), and using the following mathematical identity \cite{Koivisto:2005yk}
\begin{eqnarray}
\nabla ^{\mu }\left[ f_R\left(R,L_m\right)R_{\mu \nu }-\frac{1}{2}f\left(R,L_m\right)g_{\mu \nu }+
\left(g_{\mu \nu }\square \, -\nabla _{\mu }\nabla _{\nu }\right) f_R\left(R,L_m\right)\right] \equiv 0\,,
\end{eqnarray}
we obtain the following expression for the divergence of the energy-momentum tensor $T_{\mu \nu}$
\begin{eqnarray}
\nabla ^{\mu }T_{\mu \nu } &=& \nabla ^{\mu }\ln \left[
f_{L_m}\left(R,L_m\right)\right] \left\{ L_{m}g_{\mu \nu
}-T_{\mu \nu }\right\}
   \nonumber \\
&=& 2\nabla
^{\mu }\ln \left[ f_{L_m}\left(R,L_m\right) \right] \frac{\partial L_{m}}{%
\partial g^{\mu \nu }}\,.  \label{1c_noncons1}
\end{eqnarray}
Note that if one imposes the conservation of the energy-momentum tensor, i.e., $\nabla ^{\mu }T_{\mu \nu }=0$, this provides an effective functional relation between the matter Lagrangian density and the function $f_{L_m}\left(R,L_m\right)$, given by
\begin{equation}
\nabla
^{\mu }\ln \left[ f_{L_m}\left(R,L_m\right) \right] \frac{\partial L_{m}}{%
\partial g^{\mu \nu }}=0\,.
\end{equation}
Thus, once the matter Lagrangian density is known, by an
appropriate choice of the function $f\left( R,L_{m}\right) $ one can construct, at least in principle, conservative models with arbitrary curvature-matter dependencies.

Now, assuming that the matter Lagrangian is a function of the rest mass density $\rho $ of the matter only, from Eq.~(\ref{1c_noncons1}) we obtain explicitly the equation of motion of the test particles, given by Eq. (\ref{1b_eqmota}), where in $f(R,L_m)$ gravity, the extra force $f^{\mu}$ is defined as
\begin{equation}
f^{\mu }=-\nabla _{\nu }\ln \left[  f_{L_m}\left(R,L_m\right) \frac{%
dL_{m}\left( \rho \right) }{d\rho }\right] \left( U^{\mu }U^{\nu }-g^{\mu
\nu }\right) .
\label{1c_extra}
\end{equation}
Note that from the expression $U_{\mu }\nabla _{\nu }U^{\mu }\equiv 0$, one verifies easily that the extra force $f^{\mu }$ is always perpendicular to the velocity, so that $U_{\mu }f^{\mu }=0$, as also mentioned in the previous Section. Due to the presence of the extra force $f^{\mu }$, the motion of test particles in modified theories of gravity with an arbitrary coupling between matter and curvature is non-geodesic.

Furthermore, by taking into account the effects of the extra force, the generalized geodesic deviation equation, describing the relative accelerations of nearby particles, and the Raychaudhury equation, giving the evolution of the kinematical quantities associated with deformations, was extensively considered in the framework of an arbitrary curvature-matter coupling, namely, $f\left(R,L_m\right)$ gravity, in \cite{Harko:2012ve}. In this context, as a physical application of the geodesic deviation equation, the modifications of the tidal forces due to the supplementary curvature-matter coupling, which induces an extra force,  were obtained in the weak field approximation. It was found that the tidal motion of test particles is directly influenced not only by the gradient of the extra force, which is basically determined by the gradient of the Ricci scalar, but also by an explicit coupling between the velocity and the Riemann tensor. As a specific example, the expression of the Roche limit, which is the orbital distance at which a satellite will begin to be tidally torn apart by the body it is orbiting, was also obtained for this class of models. We refer the reader to \cite{Harko:2012ve} for more details.

The energy conditions and cosmological applications were also explored \cite{Wang:2012rw}. In \cite{Huang:2013dca} the Wheeler-De Witt equation of $f(R,L_m)$ gravity was analyzed in a flat FLRW Universe, which is the first step of the study of $f(R,L_m)$ quantum cosmology. In the minisuperspace spanned by the FLRW scale factor and the Ricci scalar, the equivalence of the reduced action was examined, and the canonical quantization of $f(R,L_m)$ gravity was undertaken, and the corresponding Wheeler-De Witt equation derived. A similar analysis will be presented and discussed below, but in the  different context of the $f(R,T)$ gravity theory. The introduction of the invariant contractions of the Ricci and Riemann tensors were further considered, and applications to black hole and wormhole physics were investigated \cite{Tian:2014mta}.

%%%%%%%%%%%%%%%%%%%%%%%%%%%%%%%%%%%%%%%%%%%%%%%%%%%%%%%%%%%%%%%%%%%%%%%%%%
\section{$f(R,T)$ gravity}\label{III:fRTgravity}
%%%%%%%%%%%%%%%%%%%%%%%%%%%%%%%%%%%%%%%%%%%%%%%%%%%%%%%%%%%%%%%%%%%%%%%%%%

%%%%%%%%%%%%%%%%%%%%%%%%%%%%%%%%%%%%%%%%%%%%%%%%%%%%%%%%%%%%%%%%%%%%%%%%%%
\subsection{Action and gravitational field equations}\label{1d_sec5_1}
%%%%%%%%%%%%%%%%%%%%%%%%%%%%%%%%%%%%%%%%%%%%%%%%%%%%%%%%%%%%%%%%%%%%%%%%%%

In this Section, we consider an interesting extension of GR that has been given considerable attention recently, namely, $f(R,T)$ gravity, where $R$ is the Ricci scalar, and $T$ is the trace of the energy-momentum tensor \cite{Harko:2011kv}. Note that the $T$-dependence may be induced by several physical processes, including exotic imperfect fluids, or quantum effects (conformal anomaly).

The action of $f(R,T)$ gravity is given by
\begin{equation}
S=\frac{1}{16\pi}\int
f\left(R,T\right)\sqrt{-g}\;d^{4}x+\int{L_{m}\sqrt{-g}\;d^{4}x}\,.
\label{1d_gravaction}
\end{equation}
Relative to the matter content, we assume that it consists of a fluid that can be characterized by two thermodynamic parameters only, the energy density $\rho$ and the pressure $p$, respectively. The trace of the energy-momentum tensor $T$ can be expressed as a function of the matter Lagrangian as
\begin{equation}
T=g^{\mu \nu}T_{\mu \nu}=4L_m-2g^{\mu \nu}\frac{\delta L_m}{\delta g^{\mu \nu}}.
\end{equation}
Thus, $f(R,T)$ gravity can be interpreted as an extension of the $f\left(R,L_m\right)$ type theories, where the gravitational action depends not only on the matter Lagrangian, but also on its variation with respect to the metric.

Now, varying the action (\ref{1d_gravaction}) with respect to the metric tensor components $g^{\mu \nu }$, we obtain the gravitational field equations, given by
\begin{eqnarray}\label{1d_field}
f_{R}\left( R,T\right) R_{\mu \nu } - \frac{1}{2}
f\left( R,T\right)  g_{\mu \nu }
+\left( g_{\mu \nu }\square\, -\nabla_{\mu }\nabla _{\nu }\right)
f_{R}\left( R,T\right)
   \nonumber  \\
=8\pi T_{\mu \nu}-f_{T}\left( R,T\right)
T_{\mu \nu }-f_T\left( R,T\right)\Theta _{\mu \nu}\, .
\end{eqnarray}
A subtlety needs to be pointed out, as we have defined the variation of $T$ with respect to the metric tensor as
\begin{equation}
\frac{\delta \left(g^{\alpha \beta }T_{\alpha \beta }\right)}{\delta g^{\mu
\nu}}
=T_{\mu\nu}+\Theta _{\mu \nu}\, ,
\end{equation}
where
\begin{equation}
\Theta_{\mu \nu}\equiv g^{\alpha \beta }\frac{\delta T_{\alpha \beta
}}{\delta g^{\mu \nu}}\, .
\end{equation}
For $T=0$, we recover $f(R,T)\equiv f(R)$, and consequently Eq.~(\ref{1d_field}) reduces to the field equations of $f(R)$ gravity.

Taking into account the covariant divergence of Eq.~(\ref{1d_field}),
with the use of the following mathematical identity \cite{Koivisto:2005yk}
\begin{eqnarray}
\nabla ^{\mu }\left[ f_R\left(R,T\right)
R_{\mu\nu}-\frac{1}{2}f\left(R,T\right)g_{\mu\nu}
+\left(g_{\mu \nu }\square \, -\nabla_{\mu }\nabla_{\nu}\right)
f_R\left(R,T\right)\right]
\equiv -\frac{1}{2}g_{\mu \nu}f_T(R,T)\nabla ^{\mu }T\, ,
\end{eqnarray}
the divergence of the energy-momentum tensor $T_{\mu \nu}$ is provided by
\begin{equation}\label{1d_noncons}
\nabla ^{\mu }T_{\mu \nu }
=\frac{f_{T}\left( R,T\right) }{8\pi -f_{T}\left(R,T\right) }
\left[ \left( T_{\mu \nu }+\Theta _{\mu \nu }\right) \nabla^{\mu }
\ln f_{T}\left( R,T\right) +\nabla ^{\mu }\Theta _{\mu \nu }-\frac{1}{2}g_{\mu \nu}\nabla ^{\mu }T\right]\, .
\end{equation}

To deduce the specific form of the tensorial quantity $\Theta _{\mu \nu}$, we refer to the interested reader to \cite{Harko:2011kv,Harko:2018ayt}, however, here we shall only provide the definition, which is given by
\begin{equation}\label{1d_var}
\Theta _{\mu \nu}=-2T_{\mu \nu}+g_{\mu \nu }L_{m}-2g^{\alpha \beta }
\frac{\partial ^2L_{m}}{\partial g^{\mu \nu }\partial g^{\alpha \beta
}}\,.
\end{equation}

Note that this quantity depends on the matter Lagrangian. For instance, in the case of the electromagnetic field the matter Lagrangian is given by $L_{m}=-\frac{1}{16\pi }F_{\alpha \beta }F_{\gamma \sigma } g^{\alpha \gamma }g^{\beta \sigma }$, where $F_{\alpha \beta }$ is the electromagnetic field tensor, so that we obtain $\Theta _{\mu\nu}=-T_{\mu \nu }$. For a massless scalar field $\phi $ with Lagrangian
$L_{m}=g^{\alpha \beta }\nabla_{\alpha }\phi \nabla _{\beta
}\phi $, we obtain $\Theta _{\mu \nu}=-T_{\mu \nu}+(1/2)Tg_{\mu
\nu}$.

The problem of the perfect fluids, described by an energy density $\rho $, pressure $p$ and four-velocity $U^{\mu}$ is more subtle, since there is no unique definition of the matter
Lagrangian. For instance, assume that the matter Lagrangian can be taken as $L_{m}=p$, so that Eq.~(\ref{1d_var}), provides the expression $\Theta _{\mu \nu }=-2T_{\mu \nu }+pg_{\mu \nu }$.
The choice of the matter Lagrangian given by $L_m=-\rho $ leads to a different form for $\Theta _{\mu \nu }$, namely, $\Theta _{\mu \nu }=-2T_{\mu \nu}+\rho g_{\mu \nu}$, and consequently to different field equations, and thereupon to different cosmological and astrophysical models. In fact, the nature of the qualitative and quantitative differences between the two different versions of the $f(R,T)$ gravity, corresponding to different matter Lagrangians, is still an open question, although significant contributions were made.

%%%%%%%%%%%%%%%%%%%%%%%%%%%%%%%%%%%%%%%%%%%%%%%%%%%%%%%%%%%%%%%%%%%%%%%%%%
\subsection{The equations of motion of test particles}
%%%%%%%%%%%%%%%%%%%%%%%%%%%%%%%%%%%%%%%%%%%%%%%%%%%%%%%%%%%%%%%%%%%%%%%%%%

For the specific case of a perfect fluid, the divergence of the energy-momentum tensor is given by
\begin{equation}\label{cons2}
\nabla ^{\mu }T_{\mu \nu }=-\frac{1}{8\pi +f_{T}\left( R,T\right) }\left\{
T_{\mu \nu }\nabla ^{\mu }f_{T}\left( R,T\right) +g_{\mu \nu }\nabla ^{\mu }
\left[ f_{T}\left( R,T\right) p\right] \right\}\, ,
\end{equation}
and consequently, the equation of motion of a test fluid \cite{Harko:2011kv,Harko:2018ayt} can be obtained as
\begin{equation}
\frac{d^{2}x^{\mu }}{ds^{2}}+\Gamma _{\nu \lambda }^{\mu }u^{\nu }u^{\lambda
}=f^{\mu }\, ,  \label{eqmot}
\end{equation}
where the extra force $f^{\mu }$ is defined by
\begin{equation}
f^{\mu }=8\pi\frac{\nabla _{\nu }p}{\left(\rho +p\right)\left[8\pi
+f_{T}\left(R,T\right)\right]}\left(g^{\mu \nu }-U^{\mu }U^{\nu }\right)\, .
\end{equation}
The extra-force $f^{\mu }$  is perpendicular to the four-velocity $U^{\mu }$,
$f^{\mu}U_{\mu }=0$.
When $f_T\left(R,T\right)=0$, we re-obtain the equation of motion of perfect
fluids with pressure in standard general relativity, which follows from the conservation of the energy-momentum tensor, $\nabla _{\mu }T_{\nu }^{\mu }=0$. In the limit $p\rightarrow 0$, corresponding to a pressureless fluid (dust), in standard general relativity the motion of the test particles becomes geodesic. The same result holds in the $f(R,T)$ gravity. Even if $f_T\left(R,T\right)\neq 0$, the motion of the dust particles always follows the geodesic lines of the geometry.

%%%%%%%%%%%%%%%%%%%%%%%%%%%%%%%%%%%%%%%%%%%%%%%%%%%%%%%%%%%%%%%%%%%%%%%%%%
\subsection{Specific cosmological solution}\label{1d_sec5_3}
%%%%%%%%%%%%%%%%%%%%%%%%%%%%%%%%%%%%%%%%%%%%%%%%%%%%%%%%%%%%%%%%%%%%%%%%%%

Relative to cosmological applications, we refer the reader to \cite{Harko:2011kv,Harko:2018ayt} for an in-depth analysis. However, in order to illustrate the subtleties involved, here we consider a particular class of an $f(R,T)$ model, obtained by explicitly specifying the functional form of $f$.
In fact, generally, the field equations also depend, through the tensor $\Theta _{\mu \nu }$, on
the physical nature of the matter field. Thus, in the case of $f(R,T)$ gravity, depending on the type of the matter source, for each choice of $f$ we may obtain several theoretical models, corresponding to different choices of the matter models \cite{Harko:2012ar}.

As a specific example, we consider the case in which the function $f$
is given by $f\left(R,T\right)=f_1(R)+f_2(T)$, where $f_1(R)$ and
$f_2(T)$ are arbitrary functions of $R$ and $T$, respectively. In
this case, assuming that the matter content consists of a perfect fluid, the
gravitational field equations become
\begin{eqnarray}
f'_1(R)R_{\mu\nu}-\frac{1}{2}f_1(R)g_{\mu\nu}+(g_{\mu\nu} \square-\nabla_\mu
\nabla_\nu)f'_1(R)=
%\nonumber\\
8\pi
T_{\mu\nu}+f_2'(T)T_{\mu\nu}+\left[f_2'(T)p+\frac{1}{2}f_2(T)\right]g_{\mu\nu}\,.
\label{Ein21}
\end{eqnarray}
Furthermore, for simplicity, consider the case of dust with $p=0$, so that the field equations reduce to
\begin{eqnarray}
f'_1(R)R_{\mu\nu}-\frac{1}{2}f_1(R)g_{\mu\nu}+(g_{\mu\nu} \square-\nabla_\mu
\nabla_\nu)f'_1(R)=
%\nonumber\\
8\pi T_{\mu\nu}+f_2'(T)T_{\mu\nu}+\frac{1}{2}f_2(T)g_{\mu\nu}\, .
\label{Ein22}
\end{eqnarray}
In the case $f_2(T)\equiv 0$, we re-obtain the field equations of
standard $f(R)$ gravity.

Equation (\ref{Ein21}) can be reformulated as an effective Einstein field equations of the form
\be G_{\mu
\nu}=R_{\mu \nu }-\frac{1}{2}Rg_{\mu \nu} =8\pi
G_\mathrm{eff}T_{\mu\nu} +T_{\mu \nu }^\mathrm{eff}\, ,
\ee
where we have denoted
\be G_\mathrm{eff}=\frac{1}{f_1^{\prime }(R)}
\left[1+\frac{f_2^{\prime }(T)}{8\pi }\right]\, ,
\ee
and
\be
T_{\mu \nu }^\mathrm{eff}=\frac{1}{f_1^{\prime }(R)}
\left\{\frac{1}{2}\left[f_1(R)-Rf_1^{\prime }(R) +2f^{\prime
}_2(T)p+f_2(T)\right]g_{\mu \nu}-\left(g_{\mu\nu}
\square-\nabla_\mu \nabla_\nu \right)f'_1(R)\right\}\, .
\ee
Note that the gravitational coupling can be given by an effective, matter (and time) dependent coupling, which is proportional to $f_2^{\prime }(T)$. The field equations can be recast in a form that the higher order corrections, coming both from the geometry, and from the curvature-matter coupling, provide a energy-momentum tensor having both a geometrical and matter origin, describing an ``effective'' source term on the right hand side of the Einstein field equations. Thus, in the $f(R,T)$ scenario, the cosmic acceleration may result not only from  a geometrical contribution to the total cosmic energy density, but it is also dependent on the matter content of the universe, which provides new corrections to the Hilbert-Einstein Lagrangian via the curvature-matter coupling.

The $(t,t)$ component of Eq.~(\ref{Ein22}) has the following form:
\be
\label{tt1}
3H^2 = \frac{8\pi}{f_1^{\prime}(R)}\left[ 1+\frac{f_2^{\prime }(T)}{8\pi
}\right] \rho +\frac{1}{f_1^{\prime}(R)}\left[-\frac{1}{2}\left(f_1(R)
-6 \left(\dot H + 2 H^2 \right) f_1^\prime(R) \right)+2f_2^\prime (T)
-9 \left(\ddot H + 4 H\dot H\right) f_1^{\prime\prime}(R)\right] \, .
\ee
The curvature scalar is given by $R=6 \left( \dot H + 2 H^2 \right)$. For simplicity, consider that $T_{\mu\nu}$ corresponds to a matter content with a constant EoS parameter $w$, and defining the $e$-folding $N$ by $a=a_0 \e^N$, then $\rho$ and $T$ are given by
\be
\label{tt2}
\rho = \rho_0 \e^{-3(1+w)N}\, ,\qquad T = - (1-3w) \rho_0 \e^{-3(1+w)N}\, .
\ee

We now consider that the expansion of the Universe is governed by
\be \label{tt3}
H = h(N)\, ,
\ee
where $h(N)$ is an arbitrary function of $N$. Then Eq.~(\ref{tt1})
can be written as
\bea \label{tt4} f_2^\prime (T) &=& F_2 (N)
\nn \\
& \equiv &
\frac{3}{1+ \rho_0 \e^{-3(1+w)N}} \left\{ - \frac{8\pi}{3} \rho_0
\e^{-3(1+w)N} + \frac{1}{6} f_1 \left[ 6 \left( h(N) h^\prime(N) +
2 h(N)^2 \right) \right] \right.
\nn \\
&& - \left[h(N) h^\prime(N) +
h(N)^2 \right] f_1^\prime \left[ 6 \left( h(N) h^\prime(N) + 2
h(N)^2 \right) \right]
\nn \\
&& \left. + 3 \left[ h(N)^2
h^{\prime\prime} (N) + h(N) h'(N)^2 + 4 h(N)^2 h'(N) \right]
f_1^{\prime\prime} \left[ 6 \left( h(N) h^\prime(N) + 2 h(N)^2
\right) \right] \right\} \, . \eea
Equation (\ref{tt4}) dictates that for an arbitrary $f_1(R)$, and for
the following specific choice
\be \label{tt5}
f_2^\prime (T) = F_2
\left( - \frac{\ln \left( -
\frac{T}{\left(1-3w\right)\rho_0}\right)}{3(1+w)} \right) \, ,
\ee
an arbitrary development of the expansion in the Universe given by
(\ref{tt3}) can be realized. Hence, for viable $f(R,T)$
gravitational models, using the above reconstruction method, the
possibility arises to modify the description of the evolution of the universe by adding the
corresponding function depending on the trace of the energy-momentum tensor to the gravitational action.

We refer the reader to \cite{Harko:2011kv} for further examples of cosmological solutions in $f(R,T)$ theory.

%%%%%%%%%%%%%%%%%%%%%%%%%%%%%%%%%%%%%%%%%%%%%%%%%%%%%%%%%%%%%%%%%%%%%%%%%%
\section{$f\left(R,T,R_{\mu\nu}T^{\mu\nu}\right)$ gravity}\label{1d_Sec:VI}
%%%%%%%%%%%%%%%%%%%%%%%%%%%%%%%%%%%%%%%%%%%%%%%%%%%%%%%%%%%%%%%%%%%%%%%%%%

Now one may further build on the $f(R,T)$ theory, considered in the previous Section, by noting that for the specific case of a traceless energy-momentum tensor, $T=0$, for instance, when the electromagnetic field is involved, the gravitational field equations for the $f(R,T)$ gravity \cite{Harko:2011kv} reduce to that of the field equations for $f(R)$ gravity, and the nonminimal couplings of gravity to the matter field identically vanish. Thus, this motivates a new generalization of $f(R,T)$ gravity that consists in including an explicit first order coupling between the matter energy-momentum $T_{\mu \nu}$, and the Ricci tensor \cite{Haghani:2013oma, Odintsov:2013iba}.

For this case, in contrast to $f(R,T)$ gravity, for $T=0$, there still exists  a nonminimal coupling of the geometry to the electromagnetic field via the $R_{\mu\nu}T^{\mu\nu}$ coupling term in the action, which is non-zero, in general (see \cite{Harko:2014gwa} for a review). As in the previous Section, we consider that the matter content is described by a perfect fluid, characterized by only two thermodynamic parameters, $\rho $ and $p$.

The first version of $f\left(R,T,R_{\mu\nu}T^{\mu\nu}\right)$ gravity was proposed in \cite{Nojiri:2009th}. The main problem leading to this approach is related to the fact that if one considers the perturbations of a flat Minkowski geometry,  which is obviously Lorentz invariant, then in the general relativistic case non-renormalizable divergences appear, originating from the ultraviolet region in momentum space. However, if one could change the behavior of the graviton propagator in the ultraviolet region from $1/k^2$ to $1/k^4$, the theory may become renormalizable in the ultraviolet sector. In this context in \cite{Horava:2009uw}, it was proposed to modify the ultraviolet behavior of the graviton propagator in a Lorentz non-invariant way according to the scaling  $1/\left|{k}\right|^{2z}$, where ${k}$ are the spatial momenta, and $z$ could be 2, 3 or larger integers. In order to obtain a
Ho\v{r}ava type gravity model \cite{Horava:2009uw} one can adopt the action
\be
S = \int d^4 x \sqrt{-g} \left[ \frac{R}{2\kappa^2} - \alpha \left( T^{\mu\nu} R_{\mu\nu}
+ \beta T R \right)^2 \right],
\ee
where $T_{\mu\nu}$ is the energy-momentum tensor of the arbitrary type fluid, $T$ is the trace of the energy-momentum tensor, while $\alpha $ and $\beta $ are constants.   The above action is fully diffeomorphism invariant, and corresponds to a general class of modified gravity theories. More general actions of the form
\be
S = \int d^4 x \sqrt{-g} \left\{ \frac{R}{2\kappa^2} - \sum_{n=0}^N \alpha_n \left[
\left(T^{\mu\nu}\nabla_\mu \nabla_\nu + \gamma T \nabla^\rho \nabla_\rho\right)^n
\left( T^{\mu\nu} R_{\mu\nu} + \beta T R \right) \right]^2 \right\},
\ee
have also been discussed. From a cosmological point of view in the case of a FLRW geometry the generalized Friedmann equations take the form
\be
\frac{3}{\kappa^2} H^2 = \frac{27 \alpha \rho^2 (1+w)^2}{2} H^4 + \rho_\mathrm{matter},
\ee
where $\rho_\mathrm{matter}$ is the usual matter energy density (radiation, baryons etc.), different from the density $\rho $ of the fluid
appearing in the gravitational action. Since the fluid having the energy density $\rho$ has a constant EoS parameter $w$, it follows that
$\rho = \rho_0 a^{-3(1+w)}$, where $\rho_0$ is a constant.
Then the variation of the scale factor of the Universe can be obtained as
\be
a(t) \propto \left(t_0 - t\right)^{-1/3(1+w)},
\ee
which turns out to be a phantom like solution even in the case $w>-1$.

The investigations initiated in \cite{Nojiri:2009th} were continued in \cite{Nojiri:2010kx},  where covariant power-counting renormalizable gravities constrained by scalar Lagrange multiplier were studied. The actions for the $z=2n + 2$ model with $ n=0,1,2,\cdots $ are of the type
\bea
S_{2n+2} &=& \int d^4 x \sqrt{-g} \left\{ \frac{R}{2\kappa^2} - \alpha \left[
\left(\partial^\mu \phi \partial^\nu \phi \nabla_\mu \nabla_\nu
+ 2 U_0 \nabla^\rho \nabla_\rho \right)^n
\left( \partial^\mu \phi \partial^\nu \phi R_{\mu\nu} + U_0 R \right)\right]^2
\right. \nonumber\\
&& \left. - \lambda \left( \frac{1}{2} \partial_\mu \phi \partial^\mu \phi
+ U_0 \right) \right\},
\eea
where $\lambda $ is a Lagrangian multiplier. The action contains a coupling between the kinetic part of the scalar field and the Ricci tensor.
A covariant power-counting renormalizable vector gravity, invariant under $U(1)$ gauge symmetry, was also proposed. As for the vector Lagrange multiplier, several forms of it were discussed.  Power-law or de Sitter accelerating cosmological solution, which may describe the inflationary era, were found for the considered covariant scalar/vector gravity.

%%%%%%%%%%%%%%%%%%%%%%%%%%%%%%%%%%%%%%%%%%%%%%%%%%%%%%%%%%%%%%%%%%%%%%%%%%
\subsection{Action and field equations of the $f\left(R,T,R_{\mu\nu}T^{\mu\nu}\right)$ gravity}\label{1d_sec6_1}
%%%%%%%%%%%%%%%%%%%%%%%%%%%%%%%%%%%%%%%%%%%%%%%%%%%%%%%%%%%%%%%%%%%%%%%%%%

The action of the $f\left(R,T,R_{\mu\nu}T^{\mu\nu}\right)$ gravity is given by
\be\label{1d_eq200}
S=\f{1}{16\pi G}\int d^4x\sqrt{-g}f\left(R,T,R_{\mu\nu}T^{\mu\nu}\right)+\int
d^4x\sqrt{-g}L_m \,.
\ee
As before, the only requirement imposed on the function $f\left(R,T,R_{\mu\nu}T^{\mu\nu}\right)$ is that it is an arbitrary analytical function in all its arguments. The gravitational field equations takes the following form
%\begin{widetext}
\bea \label{1d_eq203}
(f_R-f_{RT}L_m)G_{\mu\nu}-\nabla_\mu\nabla_\nu f_R +\f{1}{2}\square \,\left(f_{RT}T_{\mu\nu}\right)
 +2f_{RT}R_{\alpha(\mu}T_{\nu)}^{~\alpha}
 -\nabla_\alpha\nabla_{(\mu}\left[T^\alpha_{
~\nu)}f_{RT}\right]
   \nonumber \\
+\left[\square\, f_R+\f{1}{2}Rf_R-\f{1}{2}f+f_TL_m+\f{1}{2}\nabla_\alpha\nabla_\beta\left(f_{RT}T^{\alpha\beta}\right)\right]g_{\mu\nu}
    \nonumber\\
-\left(f_T+\f{1}{2}f_{RT}R+8\pi
G\right)T_{\mu\nu}
% \nonumber\\
-2\left(f_Tg^{\alpha\beta}+f_{RT}R^{\alpha\beta}\right)\f{\partial^2
L_m}{\partial g^{\mu\nu}\partial g^{\alpha\beta}}=0.
\eea
%\end{widetext}
%
The trace of the gravitational field equation, Eq.~(\ref{1d_eq203}), which will be used below, is obtained as
\begin{eqnarray}\label{1d_trace}
3\square\, f_R+\f{1}{2}\square \,\left(f_{RT}T\right)+\nabla_\alpha\nabla_\beta\left(f_{RT}T^{\alpha\beta}\right)+Rf_R-Tf_T  -\f{1}{2}RTf_{RT}
+2R_{\alpha\beta}T^{\alpha\beta}f_{RT}
 \nonumber\\
+Rf_{RT}L_m +4f_TL_m -2f
    -8\pi GT
    -2g^{\mu\nu}\left(g^{\alpha\beta}f_T+R^{\alpha\beta}f_{RT}\right)\f{\partial^2 L_m}{\partial g^{\mu\nu}\partial g^{\alpha\beta}}=0.
\end{eqnarray}

Note that the second derivative of the matter Lagrangian with respect to the metric is non-zero if the matter Lagrangian is of second or higher order in the metric. Thus, for a perfect fluid with $L_m=-\rho$, or a scalar field with
$L_m=-\partial_\mu\phi\partial^\mu\phi /2$,
this term can be dropped. However, for instance, considering the Maxwell  field, we have
$L_m=-F_{\mu\nu}F^{\mu\nu}/4$, this term results in $\partial^2 L_m/\partial g^{\mu\nu}\partial
g^{\alpha\beta}=-F_{\mu\alpha}F_{\nu\beta}/2$,
thus giving a non-zero contribution to the field equations.

In analogy with the standard Einstein field equation, one can write the field equation (\ref{1d_eq203}) in an effective form as
\be\label{1d_eq204}
G_{\mu\nu}=8\pi G_{\rm eff}T_{\mu\nu}-\Lambda_{\rm eff}g_{\mu\nu}+T^{\rm eff}_{\mu\nu},
\ee
where we have defined the effective gravitational coupling $G_{\rm eff}$, the effective cosmological constant $\Lambda _{\rm eff}$, and an effective energy-momentum tensor $T^{\rm eff}_{\mu\nu}$ as
\be\label{1d_eq204-1}
G_{\rm eff}=\f{G+\f{1}{8\pi}\big(f_T+\f{1}{2}f_{RT}R-\f{1}{2}\square\, f_{RT}\big)}{f_R-f_{RT}L_m},
\ee
\be
\Lambda_{\rm eff}=\f{2\square\,
f_R+Rf_R-f+2f_TL_m+\nabla_\alpha\nabla_\beta(f_{RT}T^{
\alpha\beta})}{2(f_R-f_{RT}L_m)},
\ee
and
\bea \label{1d_eq204-2}
T^{\rm eff}_{\mu\nu}&=&\f{1}{f_R-f_{RT}L_m}\Bigg\{\nabla_\mu\nabla_\nu f_R-\nabla_\alpha f_{RT}\nabla^\alpha T_{\mu\nu}
    -\f{1}{2}f_{RT}\square \, T_{\mu\nu}
    -2f_{RT}R_{\alpha(\mu}T_{\nu)}^{~\alpha}
    \nonumber\\
&&+\nabla_\alpha\nabla_{(\mu}\left[T^\alpha_{
~\nu)}f_{RT}\right]+2\left(f_Tg^{\alpha\beta}+f_{RT}R^{\alpha\beta}\right)
\f{\partial^2
L_m}{\partial g^{\mu\nu}\partial g^{\alpha\beta}}\Bigg\},
\eea
respectively. Note that, in general, $G_{\rm eff}$ and $\Lambda_{\rm eff}$ are not constants, and depend on the specific model considered.

%%%%%%%%%%%%%%%%%%%%%%%%%%%%%%%%%%%%%%%%%%%%%%%%%%%%%%%%%%%%%%%%%%%%%%%%%%
\subsection{Equation of motion of massive test particles}\label{1d_sec6_2a}
%%%%%%%%%%%%%%%%%%%%%%%%%%%%%%%%%%%%%%%%%%%%%%%%%%%%%%%%%%%%%%%%%%%%%%%%%%

The covariant divergence of the energy-momentum tensor can be obtained by taking the divergence of the gravitational field equation (\ref{1d_eq203}), which takes the following form
\begin{eqnarray}\label{1d_eq401}
\nabla^\mu T_{\mu\nu}&=&\f{2}{\left(1+Rf_{TR}+2f_T\right)}\Bigg\{
\nabla_\mu\left(f_{RT}R^{\sigma\mu}T_{\sigma\nu}
\right)
    +\nabla_\nu\left(L_mf_T\right)
    -G_{\mu\nu}\nabla^\mu \left(f_{RT}L_m\right)
    \nonumber\\
&&    -\f{1}{2}\bigg(f_{RT}R_{\rho\sigma}+f_T
g_{\rho\sigma}\bigg)
\nabla_\nu T^{\rho\sigma}
  -
\f{1}{2}\left[\nabla^\mu\left(Rf_{RT}\right)+2\nabla^\mu f_T\right]T_{\mu\nu}\Bigg\},
\end{eqnarray}
where we have assumed that $\partial^2
L_m/\partial g^{\mu\nu}\partial g^{\alpha\beta}=0$, and used the mathematical identities
\bea
\nabla^\mu \left(f_R R_{\mu\nu}+\square\, f_R
g_{\mu\nu}-\f{1}{2}fg_{\mu\nu}-\nabla_\mu\nabla_\nu
f_R\right)
%\nonumber  \\
=-\f{1}{2}\left[f_T \nabla_\nu
T+f_{RT}\nabla_\nu\left(R_{\rho\sigma}T^{\rho\sigma}\right)\right],
\eea
\be
2T_{\mu\tau;\delta[;\rho;\sigma]}=T_{\mu\tau;\alpha}R^\alpha_{~\delta\rho\sigma}
+T_{\alpha\tau;\delta}R^\alpha_{~\mu\rho\sigma}+T_{\mu\alpha;\delta}R^\alpha_{
~\tau\rho\sigma},
\ee
and $\left[\square\,,\nabla_\nu\right]T=R_{\mu\nu}\nabla^\mu T$, respectively.

In order to find the equation of motion for a massive test particle, as in the previous Sections, we consider the energy-momentum tensor of a perfect fluid. Following the procedure outlined above, we obtain the equation of motion for a massive test particle, considering the matter Lagrangian $L_m=p$, as
\be\label{1d_eq404}
\f{d^2x^\lambda}{ds^2}+\Gamma^\lambda_{~\mu\nu} U^\mu U^\nu=f^\lambda,
\ee
where the extra force acting on the test particles is given by
\bea\label{1d_eq405}
f^\lambda&=&\f{1}{\rho+p}\Big[-
\left(1+3f_T\right)\nabla_\nu p
-(\rho+p)f_{RT}R^{\sigma\rho}\left(\nabla_\nu
h_{\sigma\rho}-2\nabla_\rho h_{\sigma\nu}\right)
   \nonumber\\
&&+ \left(f_T+Rf_{RT}\right)\nabla_\nu\rho - f_{RT}
R_{\sigma\rho}h^{\sigma\rho}\nabla_\nu\left(\rho+p\right)\Big]
\f{h^{\lambda\nu}}{1+2f_T+Rf_{RT}}.
\eea
Contrary to the nonminimal coupling presented in \cite{Bertolami:2007gv}, and as can be seen from the above equations, the extra force does not vanish even with the Lagrangian $L_m=p$.

As in the previous cases, the extra-force is perpendicular to the four-velocity, satisfying the relation $f^{\mu} U_{\mu}=0$. In the absence of any coupling between matter and geometry, with $f_T=f_{RT}=0$, the extra-force takes the usual form of the standard general relativistic fluid motion, i.e., $f^{\lambda }=-h^{\lambda \nu}\nabla _{\nu }p/\left(\rho +p\right)$. Note, however, that in the case of  $f\left(R,T,R_{\mu\nu}T^{\mu\nu}\right)$ gravity, there is an explicit dependence of the extra-force on the Ricci tensor $R_{\sigma \rho }$, which makes the deviation from the geodesic motion more important for regions with strong curvatures.

%%%%%%%%%%%%%%%%%%%%%%%%%%%%%%%%%%%%%%%%%%%%%%%%%%%%%%%%%%%%%%%%%%%%%%%%%
\subsection{The Dolgov-Kawasaki instability in $f\left(R,T,R_{\mu\nu}T^{\mu\nu}\right)$ gravity}\label{sec5}
%%%%%%%%%%%%%%%%%%%%%%%%%%%%%%%%%%%%%%%%%%%%%%%%%%%%%%%%%%%%%%%%%%%%%%%%%%

Beside consistency with the Solar System tests, any gravitational theory should be stable against classical and quantum fluctuations. One of the important instabilities of modified theories of gravity is the Dolgov-Kawasaki instability \cite{Dolgov:2003px,Faraoni:2006sy,Sotiriou:2006sf,Faraoni:2007sn, Bertolami:2009cd, Wang:2012mws}, which we shall discuss in the present Section.

Let us assume that, in order to be consistent with the Solar System tests, the Lagrangian can be written as
\be
f(R,T,R_{\mu\nu}T^{\mu\nu})=R+\epsilon\Phi\left(R,T,R_{\mu\nu}T^{\mu\nu}\right),
 \ee
where $\epsilon$ is a small parameter. Following \cite{Dolgov:2003px}, we expand the space-time quantities around a constant curvature background with geometrical and physical parameters $\left(\eta _{\mu \nu},R_0, T_{\mu \nu }^0, T_0,L_0\right)$, so that
\be
R_{\mu\nu}=\f{1}{4}R_0\eta_{\mu\nu}+R^1_{\mu\nu},  \qquad  R=R_0+R_1,
\ee
 \be
T_{\mu\nu}=T^0_{\mu\nu}+T^1_{\mu\nu},  \qquad  T=T_0+T_1, \qquad  L_m=L_0+L_1,
 \ee
where we have locally expanded the metric tensor as
\be
g_{\mu\nu}=\eta_{\mu\nu}+h_{\mu\nu}.
\ee
In the above analysis, we have two types of approximations, as mentioned in \cite{Faraoni:2006sy,Sotiriou:2006sf,Faraoni:2007sn,Bertolami:2009cd,Wang:2012mws}. The first is an adiabatic expansion around a constant curvature space, which is justified on the time-scales much shorter than the Hubble time. The second approximation is a local expansion in the small regions of spacetime, which are locally flat. We emphasize that these approximations have been used extensively in $f(R)$ gravity \cite{Dolgov:2003px,Faraoni:2006sy,Sotiriou:2006sf,Faraoni:2007sn,Bertolami:2009cd,Wang:2012mws}.

The function $f\left(R,T,R_{\mu\nu}T^{\mu\nu}\right)$ can be expanded as
\begin{eqnarray}
f(R,T,R_{\mu\nu}T^{\mu\nu})
&=&R_0+R_1+\epsilon\bigg[\Phi(0)+\Phi_R(0)R_1
%   \nonumber\\
+\Phi_T(0)T_1+\Phi_{RT}(0)\left(\f{1}{4}R_0T^1+R^1_{\mu\nu}T_0^{\mu\nu}\right)\bigg]
	\nonumber\\
&=&R_0+\epsilon\Phi(0)+\big[1+\epsilon\Phi_R(0)\big]R_1+H^{(1)},
\end{eqnarray}
where the notation $(0)$ denotes the computation of the function at the background level, and for simplicity we have defined the first order quantity $H^{(1)}$ as
 \be
 H^{(1)}=\epsilon\left[\Phi_T(0)T_1+\Phi_{RT}(0)\left(\f{1}{4}R_0T^1+R^1_{\mu\nu}T_0^{\mu\nu}\right)\right].
 \ee
 We then obtain
\bea
f_R &=&1+\epsilon\Phi_R(0)+\epsilon\Phi_{R,R}(0)R_1+H_R^{(1)},  \\
%\ee
%\be
f_T &=&\epsilon\Phi_T(0)+\epsilon\Phi_{T,R}(0)R_1+H_T^{(1)},  \\
%\ee
%\be
f_{RT} &=&\epsilon\Phi_{RT}(0)+\epsilon\Phi_{RT,R}(0)R_1+H_{RT}^{(1)}.
\eea
Thus, expanding the trace equation \eqref{1d_trace} to first order, yields the following relation,
\begin{widetext}
\bea
\bigg(3\epsilon\Phi_{R,R}(0)+\f{1}{2}\epsilon T_0\Phi_{RT,R}(0)\bigg)\square R_1+\epsilon T_0^{\alpha\beta}\Phi_{RT,R}(0)\nabla_\alpha\nabla_\beta R_1+
\bigg[f_R(0)+\epsilon R_0\Phi_{R,R}(0)-\epsilon T_0\Phi_{T,R}(0)  -\f{1}{2}T_0f_{RT}(0)
   \nonumber\\
-\f{1}{2}\epsilon R_0T_0\Phi_{RT,R}(0)
+\f{1}{2}\epsilon R_0T_0\Phi_{RT,R}(0)+\epsilon R_0L_0\Phi_{RT,R}(0)+f_{RT}(0)L_0+4\epsilon L_0\Phi_{T,R}(0)-2-\epsilon\Phi_R(0)\bigg]R_1
    \nonumber\\
+3\square H_R^{(1)}+\f{1}{2}f_{RT}(0)\square T_1+\f{1}{2}T_0\square H_{RT}^{(1)}+f_{RT}(0)\nabla_\alpha\nabla_\beta T_1^{\alpha\beta}+T_0^{\alpha\beta}\nabla_\alpha\nabla_\beta H_{RT}^{(1)}+R_0 H_R^{(1)}-T_1 f_T(0)      \nonumber\\
-T_0 H_T^{(1)}+2R^1_{\mu\nu}T_0^{\mu\nu}f_{RT}(0)+R_0f_{RT}(0)L_1+R_0L_0 H_{RT}^{(1)}+4f_T(0)L_1+4L_0 H_T^{(1)}
   \nonumber\\
 -2 H^{(1)}+8\pi GT_1-2\eta^{\mu\nu}\eta^{\alpha\beta}\big[f_T(0)+\f{1}{4}R_0f_{RT}(0)\big]\f{\partial^2 L_1}{\partial g^{\mu\nu}\partial g^{\alpha\beta}}=0. \nonumber
\eea
%\end{widetext}

In the limit considered, one may write $\square=-\partial_t^2+\nabla^2$, thus obtaining
\be\label{}
T_0^{\alpha\beta}\nabla_\alpha\nabla_\beta R_1=T^{00}_0 \ddot{R}_1+T_0^{ij}\partial_i\partial_j R_1.
\ee
One can then rewrite the above equation as
\be
\ddot{R}_1+V_{\rm eff}^{ij}\nabla_i\nabla_j R_1+m_{\rm eff}^2 R_1 = H_{\rm eff},
\ee
where we have defined
\be
V_{\rm eff}^{ij}=\f{\big(3\epsilon\Phi_{R,R}(0)+\f{1}{2}\epsilon T_0\Phi_{RT,R}(0)\big)\delta^{ij}+\epsilon T_0^{ij}\Phi_{RT,R}(0)}{T_0^{00}-3\epsilon\Phi_{R,R}(0)-\f{1}{2}\epsilon T_0\Phi_{RT,R}(0)},
\ee
and
%\begin{widetext}
\bea
H_{\rm eff}&=&\big[3\epsilon\Phi_{R,R}(0)+\f{1}{2}\epsilon T_0\Phi_{RT,R}(0)-T_0^{00}\big]^{-1}\bigg\{ 3\square H_R^{(1)}+\f{1}{2}f_{RT}(0)\square T_1+\f{1}{2}T_0\square H_{RT}^{(1)}+f_{RT}(0)\nabla_\alpha\nabla_\beta T_1^{\alpha\beta}\nonumber\\
&&+T_0^{\alpha\beta}\nabla_\alpha\nabla_\beta H_{RT}^{(1)}+R_0 H_R^{(1)}-T_1 f_T(0)
-T_0 H_T^{(1)}+2R^1_{\mu\nu}T_0^{\mu\nu}f_{RT}(0)+R_0f_{RT}(0)L_1  + R_0L_0 H_{RT}^{(1)}
  \nonumber\\
&&+4f_T(0)L_1+4L_0 H_T^{(1)}-2 H^{(1)}+8\pi GT_1-2\eta^{\mu\nu}\eta^{\alpha\beta}\bigg[f_T(0)+\f{1}{4}R_0f_{RT}(0)\bigg]\f{\partial^2 L_1}{\partial g^{\mu\nu}\partial g^{\alpha\beta}}\bigg\},
\eea
%\end{widetext}
respectively, and the effective mass is given by
%\begin{widetext}
\bea
m_{\rm eff}^2&=&\left[\big(T_0^{00}-\f{1}{2} T_0\big) f_{RT,R}(0)-3f_{RR}(0)\right]^{-1}
\bigg[ \epsilon R_0\Phi_{R,R}(0)-\epsilon T_0\Phi_{T,R}(0)-\f{1}{2}\epsilon R_0T_0\Phi_{RT,R}(0)
\nonumber\\
&&- \f{1}{2}\epsilon T_0\Phi_{RT}(0)+
\f{1}{2}\epsilon R_0T_0\Phi_{RT,R}(0)+
\epsilon R_0L_0\Phi_{RT,R}(0)+\epsilon\Phi_{RT}(0)L_0+
4\epsilon L_0\Phi_{T,R}(0)-1-\epsilon\Phi_R(0)\bigg].
\eea
\end{widetext}

The dominant term in the above expression is $1/\big[3f_{RR}(0)+(\f{1}{2}T_0 -T_0^{00})f_{RT,R}(0)\big]$, and therefore the condition to avoid the Dolgov-Kawasaki instability is
\be
3f_{RR}(0)-\left(\rho_0-\f{1}{2}T_0 \right)f_{RT,R}(0)\geq0,
\ee
where $\rho _0$ is the background energy density of matter \cite{Dolgov:2003px}.

%%%%%%%%%%%%%%%%%%%%%%%%%%%%%%%%%%%%%%%%%%%%%%%%%%%%%%%%%%%%%%%%%%%%%%%%%%
\subsection{Specific cosmological application}\label{1d_sec6_3}
%%%%%%%%%%%%%%%%%%%%%%%%%%%%%%%%%%%%%%%%%%%%%%%%%%%%%%%%%%%%%%%%%%%%%%%%%%

We consider now cosmological applications of the theory, where  the functional form of the function $f\left(R,T,R_{\mu\nu}T^{\mu\nu}\right)$ is fixed in advance. In order to analyze the evolution and dynamics of the Universe, consider an isotropic and homogeneous spacetime, with the matter content described by a perfect fluid and where the matter Lagrangian chosen as $L_m=-\rho$. For the FLRW metric (\ref{metric}), we introduce the Hubble parameter $H=\dot{a}/a$, and the deceleration parameter $q$, defined as
\be
q=\frac{d}{dt}\frac{1}{H}-1.
\ee
Note that if $q<0$, the expansion of the Universe is accelerating, while positive values of $q> 0$, describe decelerating evolutions.

For simplicity, we consider the case in which the interaction between matter and geometry takes place only via the coupling between the energy-momentum and Ricci tensors, i.e.,
\begin{equation}\label{1d_113}
f=R+\alpha R_{\mu\nu}T^{\mu\nu} \,.
\end{equation}
In order to pass the Solar System and the other astrophysical tests, the correction parameter $\alpha$ in Eq.~(\ref{1d_113}) must be small. This simple case serves as an example to show the main differences of the present theory as compared to $f(R,T)$ gravity, considered before \cite{Harko:2011kv}.
The gravitational field equations for this form of $f$ are given by
\begin{eqnarray}\label{1d_eq203-1}
G_{\mu\nu}+\alpha\Big[2R_{\sigma(\mu}T^\sigma_{~\nu)}
-\f{1}{2}R_{
\rho\sigma}T^{\rho\sigma}g_{\mu\nu}
-\f{1}{2}RT_{\mu\nu}
-G_{\mu\nu}L_m
-2R^{\alpha\beta}\f{\partial^2 L_m}{\partial g^{\mu\nu}\partial
g^{\alpha\beta}}
    \nonumber\\
-\f{1}{2}\left(2 \nabla_\sigma\nabla_{(\nu}
T^\sigma_{~\mu)}-\square\,
T_{\mu\nu}-\nabla_\alpha\nabla_\beta T^{\alpha\beta}g_{\mu\nu}\right)
\Big]-
8\pi GT_{\mu\nu}=0.
\end{eqnarray}

For the case of the FLRW metric, the modified Friedmann equations are
\be\label{1d_c1}
3H^2=\frac{\kappa}{1-\alpha \rho }\rho +\frac{3}{2}\frac{\alpha }{1-\alpha \rho }H\left(\dot{\rho }-\dot{p}\right),
\ee
and
\be\label{1d_c2}
2\dot{H}+3H^2=\frac{2\alpha }{1+\alpha p}H\dot{\rho }-\frac{\kappa p}{1+\alpha p}+\frac{1}{2}\frac{\alpha }{1+\alpha p}\left(\ddot{\rho }-\ddot{p}\right),
\ee
respectively, where we have denoted $\kappa =8\pi G$ for simplicity. When $\alpha =0$ we recover the standard Friedmann equations. To remove the under-determinacy of the field equations, we impose an equation of state for the cosmological matter, $p=p(\rho)$. A standard form of the cosmological matter equation of state is $p=\omega \rho $, where $\omega ={\rm constant}$, and $0\leq \omega \leq 1$.

For simplicity, consider the case of low density cosmological  matter, with $p=0$. Moreover, we assume again that the condition $\alpha \rho  \ll 1$ holds. Then the gravitational field equations, corresponding to a FLRW Universe, take the approximate form
    \bea
    3H^2&=&\kappa \rho +\frac{3}{2}\alpha H\dot{\rho },  \label{c3} \\
    %\ee
   % \be
    2\dot{H}+3H^2&=&2\alpha H\dot{\rho }+\frac{1}{2}\alpha \ddot{\rho }. \label{c4}
    \eea

Consider a de Sitter-type solution of the field equations (\ref{c3})--(\ref{c4}) for the pressureless matter, by taking $H=H_0={\rm constant}$. Then it follows that, in order to have an accelerated expansion, the matter density must satisfy the equation
   \be
   \ddot{\rho }-H_0\dot{\rho }+\frac{2\kappa }{\alpha }\rho =0,
   \ee
with the general solution given by
   \bea\label{r1}
   \rho (t)&=&e^{\frac{1}{2} H_0 \left(t-t_0\right)}
   %\times \nonumber\\
   \left\{\frac{\sqrt{\alpha } \left(2 \rho
   _{01}-H_0 \rho _0\right) }{\sqrt{\alpha  H_0^2-8
   \kappa }}\sinh \left[\frac{ \sqrt{\alpha
  H_0^2-8 \kappa }}{2 \sqrt{\alpha }}\left(t-t_0\right)\right]
  %\nonumber\\   &&
  +\cosh \left[\frac{ \sqrt{\alpha  H_0^2-8 \kappa }}{2
   \sqrt{\alpha }}\left(t-t_0\right)\right]\right\},
   \eea
where we have used the initial conditions $\rho \left(t_0\right)=\rho _0$, and $\dot{\rho }\left(t_0\right)=\rho _{01}$, respectively.
  Therefore, in the presence of a non-trivial curvature-matter coupling, once the evolution of the matter density is given by Eq.~(\ref{r1}), the time evolution of the Universe is of the de Sitter type.
We refer the reader to \cite{Haghani:2013oma} for further cosmological applications.

%%%%%%%%%%%%%%%%%%%%%%%%%%%%%%%%%%%%%%%%%%%%%%%%%%%%%%%%%%%%%%%%%%%%%%%%%%
\section{Irreversible matter creation processes through a nonminimal curvature-matter coupling}\label{III:mattcreation}
%%%%%%%%%%%%%%%%%%%%%%%%%%%%%%%%%%%%%%%%%%%%%%%%%%%%%%%%%%%%%%%%%%%%%%%%%%

%%%%%%%%%%%%%%%%%%%%%%%%%%%%%%%%%%%%%%%%%%%%%%%%%%%%%%%%%%%%%%%%%%%%%%%%%%
\subsection{Thermodynamic open systems}
%%%%%%%%%%%%%%%%%%%%%%%%%%%%%%%%%%%%%%%%%%%%%%%%%%%%%%%%%%%%%%%%%%%%%%%%%%

An interesting application of curvature-matter coupling theories is to the thermodynamics of open systems \cite{Harko:2015pma,Lobo:2015awa}. In fact, Prigogine {\it et al.} \cite{Pri0,Pri} in the 1980s proposed an interesting type of cosmological history with a large-scale entropy production in the context of the cosmological thermodynamics of open systems. More specifically, they investigated the role of irreversible processes corresponding to matter creation in GR, by modifying the usual adiabatic energy conservation laws.
The cosmological models analyzed were based on a reinterpretation of the general relativistic matter energy-momentum tensor, where it was shown that the second law of thermodynamics allowed the creation of matter as an irreversible energy flow from the gravitational field to the created matter fluid, but forbidding the inverse process \cite{Calvao:1991wg}.
Prigogine {\it et al.} argued that the conventional initial big bang singularity appears to be structurally unstable with respect to irreversible matter creation, where the cosmological history in the framework of the thermodynamics of open systems starts from an instability of the quantum vacuum rather than from
a singularity. A remarkable fact is that the de Sitter stage appears to be an attractor independently of the initial fluctuation \cite{Pri0,Pri}.

Indeed, the final de Sitter phase is relevant due to the recent late-time cosmic acceleration \cite{Aghanim:2018eyx,Sotiriou:2006hs}, where the simplest explanation is to invoke a cosmological constant, which is usually associated to the quantum vacuum energy, which may decay into radiation and matter particles. This fact is particularly important in modified theories of gravity, with a nonminimal curvature-matter coupling \cite{Bertolami:2007gv}, which provide an alternative mechanism for the
gravitational particle production. As extensively mentioned above a general property of these theories is
the non-conservation of the energy-momentum tensor \cite{Bertolami:2007gv,Harko:2008qz,Harko:2010hw,Harko:2011kv}. Thus, this curvature-matter coupling, namely, the coupling between matter and the higher derivative curvature terms may be interpreted as an exchange of energy and momentum between both. it is in this context, that this coupling mechanism naturally induces a gravitational particle production \cite{Harko:2015pma,Lobo:2015awa}.

To this effect, consider the spatially-flat Friedman-Lemaitre-Robertson-Walker (FLRW), given by Eq. (\ref{metric}). Using the thermodynamics of open systems, we note that the number of particles is not conserved, in particular, the usual expression for the evolution of the number density, $\dot{n} \, + \, 3 H n = 0$, where $H = \dot{a}/a$, generalizes to $\dot{n}+3nH=\Gamma n$. The quantity $\Gamma$ is defined by the particle creation rate, using the second law of thermodynamics obeys $\Gamma \geq 0$. Therefore, taking into account the above considerations, the energy conservation equation describing an irreversible particle creation can be rewritten as an effective energy conservation equation, given by
\begin{equation}
\dot{\rho}+3\left( \rho +p+p_{c}\right) H=0,
  \label{comp}
\end{equation}%
where
\begin{equation}
 p_{c} = -\frac{\rho +p}{3}\frac{\Gamma }{H}.
  \label{comp2a}
\end{equation}%
where $p_{c}$ is a new thermodynamic quantity, $p_{c}$, denoted the creation pressure \cite{Pri}.

%%%%%%%%%%%%%%%%%%%%%%%%%%%%%%%%%%%%%%%%%%%%%%%%%%%%%%%%%%%%%%%%%%%%%%%%%%
\subsection{Thermodynamic interpretation in curvature-matter couplings}
%%%%%%%%%%%%%%%%%%%%%%%%%%%%%%%%%%%%%%%%%%%%%%%%%%%%%%%%%%%%%%%%%%%%%%%%%%

In the following, taking into account that with the second law of thermodynamics requires that space-time transforms into matter, we argue that the explicit curvature-matter coupling generates an irreversible energy flow from the gravitational field to newly created matter constituents.

Recall that the gravitational theory, with a nonminimal coupling between matter and curvature \cite{Bertolami:2007gv}, is given by the Lagrangian (\ref{eq:S_hybrid}), i.e., $L=\frac{1}{2}f_1(R)+\left[1+\lambda f_2(R)\right]{\ L}_{m}$. This is equivalent to an scalar-tensor representation with a two-potential scalar-tensor Brans-Dicke type theory \cite{Faraoni:2007sn}, given by the action (\ref{1b_300}), which for self-completeness on this section, we rewrite here:
%%%%%%%%%%%%%%%%%%%%%%%%%%%%%%%%%%%%%%%%%%%%%%%%%%%%%%%%%%%%%%%%%%%%%%%%%%%%%%%
\begin{equation}  \label{300}
S=\int  \left[ \frac{\psi R }{2} -V(\psi)\, +U(\psi) L_m \right]\,
\sqrt{-g} \, d^4x  \, ,
\end{equation}
%%%%%%%%%%%%%%%%%%%%%%%%%%%%%%%%%%%%%%%%%%%%%%%%%%%%%%%%%%%%%%%%%%%%%%%%%%%%%%
where the two potentials $V(\psi)$ and $U(\psi)$ are given as
%%%%%%%%%%%%%%%%%%%%%%%%%%%%%%%%%%%%%%%%%%%%%%%%%%%%%%%%%%%%%%%%%%%%%%%%%%%%%%%%%%%%%%%%%%%%%%%
\begin{equation}  \label{400}
V(\psi) = \frac{\phi(\psi) f_1^{\prime }\left[ \phi (\psi ) \right] -f_1%
\left[ \phi( \psi ) \right] }{2}, \qquad U( \psi) = 1+\lambda
f_2\left[ \phi( \psi ) \right]\,.
\end{equation}
%%%%%%%%%%%%%%%%%%%%%%%%%%%%%%%%%%%%%%%%%%%%%%%%%%%%%%%%%%%%%%%%%%%%%%%%%%%%%%%%%%%%%%%%%%%%%%%%%

Now, varying the action (\ref{300}) with respect to the metric $g_{\mu \nu}$ and to the scalar field $\psi$ provides the following field equations
%%%%%%%%%%%%%%%%%%%%%%%%%%%%%%%%%%%%%%%%%%%%%%%%%%%%%%%%%%%%%%%%%%%%%%%%%%%%%%%%%%%%%%%%%%%%%%%%%
\begin{equation}  \label{14}
\hspace{-0.1cm}\psi \left(R_{\mu \nu}-\frac{1}{2}g_{\mu \nu}R\right)+
\left(g_{\mu \nu }\nabla _{\alpha}\nabla^{\alpha} -\nabla _{\mu }\nabla _{\nu }\right)\psi =
U(\psi)T_{\mu \nu}+V(\psi)g_{\mu \nu},
\end{equation}
and
\begin{equation}  \label{15}
\frac{R}{2}-V^{\prime }(\psi)+U^{\prime }(\psi)L_m=0,
\end{equation}
respectively.

The covariant divergence of the energy-momentum tensor provides the following energy balance equation
%%%%%%%%%%%%%%%%%%%%%%%%%%%%%%%%%%%%%%%%%%%%%%%%%%%%%%%%%%%%%%%%%%%%%%%%%%%%%%%%%%%%%%%%%%%%%%%%%
%\begin{equation}
%\nabla _{\mu }T_{\nu }^{\mu }=-\left[ \nabla _{\mu }\ln U\left( \psi \right) %
%\right] T_{\nu }^{\mu }-\frac{2V^{\prime }\left( \psi \right) -U^{\prime
%}(\psi )L_{m}}{U(\psi )}\nabla _{\nu }\psi , \label{17}
%\end{equation}
%%%%%%%%%%%%%%%%%%%%%%%%%%%%%%%%%%%%%%%%%%%%%%%%%%%%%%%%%%%%%%%%%%%%%%%%%%%%%%%%%%%%%%%
%%%%%%%%%%%%%%%%%%%%%%%%%%%%%%%%%%%%%%%%%%%%%%%%%%%%%%%%%%%%%%%%%%%%%%%%%%%%%%%%%%%%%%%%%%%%%%%%%
\begin{equation}  \label{eneq}
\dot{\rho}+3H(\rho +p)+\rho \frac{d}{ds}\ln U(\psi )+\frac{2V^{\prime
}\left( \psi \right) -U^{\prime }(\psi )L_{m}}{U(\psi )}\dot{\psi}=0\,.
\end{equation}
%%%%%%%%%%%%%%%%%%%%%%%%%%%%%%%%%%%%%%%%%%%%%%%%%%%%%%%%%%%%%%%%%%%%%%%%%%%%%%%%%%%%%%%
which confronting with Eq. (\ref{comp}), yields the creation pressure, given by
\begin{equation}  \label{pc}
p_{c}=-\frac{1}{3H}\left\{\rho \frac{d}{dt}\ln U(\psi )+\frac{2V^{\prime
}\left( \psi \right) -U^{\prime }(\psi )L_{m}}{U(\psi )}\dot{\psi}\right\},
\end{equation}
where the particle creation rate $\Gamma $ is  a non-negative quantity defined as
\begin{equation}  \label{33}
\Gamma =-\frac{1}{\rho +p}\Bigg\{\rho \frac{d}{dt}\ln U(\psi )+\frac{%
2V^{\prime }\left( \psi \right) -U^{\prime }(\psi )L_{m}}{U(\psi )}\dot{\psi}%
\Bigg\}.
\end{equation}
%%%%%%%%%%%%%%%%%%%%%%%%%%%%%%%%%%%%%%%%%%%%%%%%%%%%%%%%%%%%%%%%%%%%%%%%%%%%%%%%%
In Eq. (\ref{eneq}), we note that the change in the particle number is due to the
transfer of energy-momentum from gravity to matter.

The entropy is defined through the positive particle creation rate $\Gamma $ by the relation $\dot{S}/S=\Gamma \geq 0$ \cite{Harko:2015pma,Lobo:2015awa}. This implies that in an ever expanding Universe with particle creation, the matter entropy will increase indefinitely. However, natural systems, left to themselves, tend to attain a state of thermodynamic equilibrium, which implies that the entropy of isolated systems never decreases, $\dot{S}\geq 0$. In addition to this, it is concave, at least in the last stage of  approaching thermodynamic equilibrium.

The validity of the second law of thermodynamics in cosmology was investigated in detail in \cite{Mimoso:2013zhp}, where it was shown that if the entropy, $S$, as measured by a comoving observer, is defined as the entropy of the apparent horizon plus that of matter and radiation within it, then the Universe approaches thermodynamic equilibrium as it nears the final de Sitter phase. Then it follows that $S$ increases, and is concave, thus validating the second law of thermodynamics, as one should expect given the strong connection between gravity and thermodynamics, for the case of the expanding Universe.

Taking into account a spatially-flat FLRW Universe filled with dust, one has
\begin{equation}
S = S_{\rm ah} + S_{\rm m} = \frac{\pi}{H^2} + \frac{4\pi}{3H^3}\,n(t)\;,
\label{MP}
\end{equation}
where the radius of the apparent horizon is $r_{\rm ah}= H^{-1}$ \cite{Bak:1999hd}.
Therefore, the thermodynamic requirements $S^{\prime}\geq 0 $ and $S^{\prime \prime }\leq 0$ impose specific constraints on the particle creation rate $\Gamma $, and its derivative with respect to the scale factor. A particularly important case is that of the de Sitter evolution of the Universe, with $H=H_{\star}=\mathrm{constant}$. In this case, we have
\begin{equation}
S^{\prime }= \frac{4\pi }{ 3H_{\star}^{4}}\frac{\left[ \Gamma (a)
-3H_{\star} \right]}{a}n(a) \geq 0  \,,
\end{equation}
and
\begin{eqnarray}
S^{\prime \prime}= \frac{4 \pi n(a)}{3 a^2 H_{\star}^5}
\Big\{\Gamma ^2(a)+H_{\star} \left[a \Gamma ^{\prime }(a)+12
H_{\star}\right] -7 H_{\star} \Gamma (a)\Big\} \leq 0 \,,
\end{eqnarray}
respectively.
%%%%%%%%%%%%%%%%%%%%%%%%%%%%%%%%%%%%%%%%%%%%%%%%%%%%%%%%%%%%%%%%%%%%%%%%%%%%%%%%%%%
Thus, the constraints $\Gamma \geq 3H_{\star}$ and $\Gamma ^{\prime
}(a)\leq \left[7\Gamma (a)-\Gamma
^2(a)/H_{\star}-12H_{\star}\right]/a$ are imposed on the particle creation
rate $\Gamma$. For the specific case of $\Gamma =3H_{\star}$, we obtain $S =
\mathrm{constant}$, showing that in this case the cosmological
evolution is isentropic. Here, $H_{\star}$ denotes the expansion
rate of the final de Sitter phase, where realistically it must be lower than the Hubble constant value, $H_{0}$. We refer the reader to \cite{Harko:2015pma,Lobo:2015awa} for more specific details.

%\subsubsection{Conclusion}

In conclusion, the thermodynamic interpretation suggests that the curvature-matter coupling may well be a leading feature of our Universe. In this hypothetical context, the current accelerated cosmic expansion may be seen as a hint for matter creation. In fact, the existence of some forms of the curvature-matter coupling leading to matter creation processes is not ruled out by the available
data. The functional forms of the potentials $V(\psi)$ and $U(\psi)$ that fully characterize the nonminimal curvature-matter coupling gravitational theory may be provided by fundamental quantum field theoretical models of the gravitational interaction, thus opening the possibility to an in-depth comparison of the predictions of the gravitational theory with cosmological and astrophysical data.

%%%%%%%%%%%%%%%%%%%%%%%%%%%%%%%%%%%%%%%%%%%%%%%%%%%%%%%%%%%%%%%%%%%%%%%%%%
\section{Quantum cosmology of $f(R,T)$ curvature-matter couplings}\label{1g_Part1:new2}
%%%%%%%%%%%%%%%%%%%%%%%%%%%%%%%%%%%%%%%%%%%%%%%%%%%%%%%%%%%%%%%%%%%%%%%%%%

The present Section presents a brief introduction to the study of the quantum cosmology of $f(R,T)$ gravity, and of some of its physical and theoretical implications. As a first step in our investigations, we derive the general form of the gravitational Hamiltonian in the $f(R,T)$ gravity theory, of the corresponding quantum potential, and of the canonical momenta associated to the Hamiltonian, respectively. Indeed, once these physical quantities are explicitly obtained, we introduce the full Wheeler-de Witt equation in $f(R,T)$ gravity, which describes the quantum properties of the very early Universe in the framework of the theory. The Wheeler-de Witt equation can provide a satisfactory description of the Universe during the period when quantum effects had a dominant influence on its dynamic evolution.

To present the subtleties of the theory, as well as the complexity of the physical applications, we consider only a specific form for the $f(R,T)$ theory. Once the gravitational action is fixed, we obtain the Hamiltonian form of the classical equations of motion, and the Wheeler-de Witt equation, which gives the unfolding of the wave function of the very early Universe, immediately after its birth. We then propose a time parameter for the $f(R,T)$ gravity quantum dynamical system, which allows us to introduce for the quantum-mechanical system under consideration a
 Schr\"{o}dinger--Wheeler--de Witt type equation for the description of its specific attributes. In order to solve the Schr\"{o}dinger--Wheeler--de Witt type equation of the quantum cosmology of $f(R,T)$ gravity, describing the properties of the early Universe, we use a perturbative approach for the study of this quantum cosmological equation. The energy levels of the Universe are obtained perturbatively by using a standard twofold degenerate perturbation method. Finally, we study the problem of the quantum time in $f(R,T)$ gravity by introducing a second quantization approach of time.

%%%%%%%%%%%%%%%%%%%%%%%%%%%%%%%%%%%%%%%%%%%%%%%%%%%%%%%%%%%%%%%%%%%%%%%%%%
\subsection{The Wheeler-de Witt equation in $f(R,T)$ gravity}\label{1g_sect2}
%%%%%%%%%%%%%%%%%%%%%%%%%%%%%%%%%%%%%%%%%%%%%%%%%%%%%%%%%%%%%%%%%%%%%%%%%%

In order to explore the quantum cosmology of $f(R,T)$ gravity, we need to define the classical parameters of the model. For this purpose, we adopt the FLRW metric as the classical background line element, which corresponds to a homogeneous and isotropic geometry. In addition to this, we assume that the matter content of the very early Universe can be modeled by a perfect cosmological fluid, described by two basic thermodynamic parameters, namely, the energy density and the thermodynamic pressure, respectively.
As a first step in our analysis, we develop the Hamiltonian formulation of $f(R,T)$ gravity, which will be useful to write down the Wheeler-de Witt equation, and gives the description of the evolution of the quantum Universe in its very early stages, in the presence of the curvature-matter coupling.

As we have already seen, the classical field equations of $f(R,T)$ gravity are given in an arbitrary geometry by Eq. (\ref{1d_field}), and for convenience, we reproduce here
 \bea\label{1g_f2h}
f_R(R,T) R_{\mu\nu} -\frac{1}{2} f(R,T) g_{\mu\nu} +(g_{\mu\nu}\square -\nabla_{\mu}\nabla_{\nu}) f_R(R,T)
 \nonumber\\
=-8\pi T_{\mu\nu} -f_T (R,T) T_{\mu\nu} -f_T (R,T) \Theta _{\mu\nu}.
\eea
 This yields the trace of the field equation Eq.~(\ref{1g_f2h}), given by
\be\label{1g_trh}
f_R(R,T) R -2f(R,T)+ 3\square f_R(R,T) =\frac{1}{2} T +f_T(R,T) T -4p f_T(R,T),
\ee
which will be useful below.

%%%%%%%%%%%%%%%%%%%%%%%%%%%%%%%%%%%%%%%%%%%%%%%%%%%%%%%%%%%%%%%%%%%%%%%%%%
\subsubsection{The effective cosmological Lagrangian and the potential}
%%%%%%%%%%%%%%%%%%%%%%%%%%%%%%%%%%%%%%%%%%%%%%%%%%%%%%%%%%%%%%%%%%%%%%%%%%

As mentioned above, the geometry of the classical spacetime is assumed to be homogeneous and isotropic, and described by the FLRW metric, given in spherical coordinates by
\begin{equation}\label{1g_FRW}
ds^2 = -N^2(t)dt^2 +a^2 (t) \left[\frac{dr^2}{1-kr^2}+r^2 (d\theta^2 +\sin ^2 \theta d\varphi ^2)\right],
\end{equation}
where $N(t)$ is the lapse function, $a(t)$ is the cosmological scale factor, and the constant $k$, taking the values $k=1,0,-1$, represents the closed, flat and open geometric models of the Universe, respectively.
The components of the Ricci tensor for the FLRW metric (\ref{1g_FRW}) are provided by
\begin{equation}
R_{00} = -3 \frac{\ddot{a}}{a} + 3 \frac{\dot{N}\dot{a}}{Na},
\end{equation}
\begin{equation}
R_{ii}=\frac{g_{ii}}{N^2}\left[\frac{\ddot{a}}{a}+2\left(\frac{\dot{a}}{a}\right)^2 +\frac{2kN^2}{a^2}-\frac{\dot{N}\dot{a}}{Na}\right],
\end{equation}
with $i=1,2,3$, and the Ricci scalar is given by
\begin{equation}
R =R_{\mu}^{\mu}= \frac{6}{N^2}\left[\frac{\ddot{a}}{a}+\left(\frac{\dot{a}}{a}\right)^2 +\frac{kN^2}{a^2}-\frac{\dot{N}\dot{a}}{Na}\right].
\end{equation}

 Since we assume that the matter content of the Universe consists of a perfect fluid, in the comoving frame, in which  the components of the four-velocity are $U_{\mu}=(N(t),0,0,0)$ and $U^{\mu}=\left(-1/N(t),0,0,0\right)$, respectively, the trace of the energy-momentum tensor is given by $T =-\rho +3p$.

Given this, we now derive the effective Lagrangian for the $f(R,T)$ theory. Thus, by varying the Lagrangian with respect to its dynamical variables gives the equations of motion for the system.
By taking into account the explicit expressions of the components of the Ricci tensor $R_{\mu \nu}$, and with the use of the identity (\ref{1g_trh}), the cosmological action for $f(R,T)$ gravity is given by
%\begin{widetext}
\bea\label{1g_L}
S_{\rm grav}&=&\int dt \Bigg\{Na^3 f(R,T)-\lambda \left[ R-\frac{6}{N}\left(\frac{\ddot{a}}{a}+\left(\frac{\dot{a}}{a}\right)^2 +\frac{kN^2}{a^2}-\frac{\dot{N\dot{a}}}{Na}\right)\right]
\nonumber\\
&&-\mu \left[\frac{1}{2} T +f_T(R,T)T-f_R(R,T)R -3\square f_R(R,T)+2f(R,T)-4pf_T(R,T)\right] \Bigg\}.
\eea
%\end{widetext}
Here, we introduced two parameters $\lambda$ and $\mu$ as Lagrange multipliers. The term containing the second Lagrange multiplier is chosen as the contracted field equation, since in this way it can be derived directly from the gravitational action, and no further assumptions are needed. Moreover, if we use other possible representations for the action, such as adopting for the term multiplying the second Lagrange multiplier $\mu $ the expression $T + \rho -3p$, important information about the geometry of the modified gravity will be lost.

By varying Eq.~(\ref{1g_L}) with respect to $R$ and $T$, we obtain the following expressions for the Lagrange multipliers $\lambda $ and $\mu$
\bea
\frac{\mu}{Na^3} = \frac{f_T}{1/2 +3f_T +f_{TT}T -f_{RT}R -3\square f_{RT}-4pf_{TT}} \equiv  \widetilde{\mu},
\eea
\bea
\frac{\lambda}{Na^3} = f_R -\frac{\mu (f_{RT} T -f_{RR} R + f_R -3\square f_{RR} -4pf_{RT})}{Na^3}
\equiv  \widetilde{\lambda}.
\eea

Therefore the gravitational part of the Lagrangian is obtained as
\begin{equation}
\mathcal{L}_{\rm grav} = -\frac{6}{N} a \dot{a}^2 \widetilde{\lambda} -\frac{6}{N}a^2 \dot{a}\dot{\widetilde{\lambda}} +6kNa\widetilde{\lambda}- Na^3 V,
\end{equation}
where the potential $V$ is given by the relation
\be
V = -f(R,T) +\widetilde{\lambda} R +\widetilde{\mu} \left[\frac{1}{2} T +f_T(R,T)T-f_R(R,T)R-3\square f_R(R,T)+2f(R,T)-4pf_T(R,T)\right].
\ee

In order to simplify the mathematical formalism of the problem, we consider the following definitions
\bea
f_R = A, \qquad f_T =B, \qquad f_{RR} = C, \qquad f_{RT} = D,
  \nonumber
  \eea
  \bea
f_{TT} = E, \qquad  \square f_{RR} = F, \qquad \square f_{RT} = G, \qquad T-4p=M.\nonumber
\eea
By using the newly introduced variables $\widetilde{\lambda}$ can be expressed  as
\begin{equation}
\widetilde{\lambda} = A - \frac{B(DM-CR+A-3F)}{1/2 +3B +EM-DR-3G}.
\end{equation}
 We introduce now two new quantities $\mathcal{A}$ and $\mathcal{Z}$ defined as
\begin{equation}
\mathcal{A}=\frac{1}{2} +3B +EM-DR-3G, \qquad \mathcal{Z}=DM-CR+A-3F.
\end{equation}
Thus, for $\widetilde{\lambda}$ we obtain the simple representation
\begin{equation}\label{1g_eqnx}
\widetilde{\lambda}= A - \frac{B\mathcal{Z}}{\mathcal{A}}.
\end{equation}
By taking the derivative with respect to the cosmological time of the above expression we find
\begin{equation}
\dot{\widetilde{\lambda}}=\dot{A} -\frac{B\dot{\mathcal{Z}}+\dot{B}\mathcal{Z}}{\mathcal{A}}+\frac{B\mathcal{Z}\dot{\mathcal{A}}}{\mathcal{A}^2}.
\end{equation}

%%%%%%%%%%%%%%%%%%%%%%%%%%%%%%%%%%%%%%%%%%%%%%%%%%%%%%%%%%%%%%%%%%%%%%%%%%
\subsubsection{The cosmological Hamiltonian}
%%%%%%%%%%%%%%%%%%%%%%%%%%%%%%%%%%%%%%%%%%%%%%%%%%%%%%%%%%%%%%%%%%%%%%%%%%

Note that by definition, the canonical momentum $P_{q}$ associated to the coordinate $q$ is obtained as $P_{q}=\partial\mathcal{L}/\partial \dot{q}$. Thus, by taking into account the definition of the momenta associated to the canonical variables, the cosmological Hamiltonian of $f(R,T)$ gravity is provided by
\bea
H_{\rm grav} = \dot{a}P_a +\dot{A} P_A + \dot{B}P_B +\dot{C}P_C+\dot{D}P_D+\dot{E}P_E		
	\nonumber\\
+\dot{F}P_F+\dot{G}P_G+\dot{R}P_R+\dot{M}P_M -\mathcal{L}_{\rm grav}\,.
\eea
The explicit forms of the canonical momenta associated to the cosmological action Eq.~(\ref{1g_L}) of $f(R,T)$ gravity can be found, after some simple calculations, as being given by
\begin{equation}
P_a = -2\frac{6}{N}a \dot{a} \widetilde{\lambda} -\frac{6}{N} a^2 \dot{\widetilde{\lambda}}, \qquad
 P_A =-\frac{6}{N} a^2 \dot{a} \left(1-\frac{B}{\mathcal{A}}\right),
 \qquad
 P_B =-\frac{6}{N} a^2 \dot{a} \left(-\frac{\mathcal{Z}}{\mathcal{A}} +\frac{3B\mathcal{Z}}{\mathcal{A}^2}\right),
\end{equation}
\begin{equation}
P_C =-\frac{6}{N} a^2 \dot{a} \left(\frac{BR}{\mathcal{A}}\right),
\qquad
P_D =-\frac{6}{N} a^2 \dot{a}\left[-\frac{B(T-4p)}{\mathcal{A}}-\frac{B\mathcal{Z}R}{\mathcal{A}^2}\right],
\qquad
P_E = -\frac{6}{N} a^2 \dot{a} \left[\frac{B\mathcal{Z}(T-4p)}{\mathcal{A}^2}\right],
\end{equation}
\begin{equation}
P_F = -\frac{6}{N} a^2 \dot{a} \left( \frac{3B}{\mathcal{A}}\right),
\qquad
P_G = -\frac{6}{N} a^2 \dot{a} \left(-\frac{3B\mathcal{Z}}{\mathcal{A}^2}\right),
\qquad
P_R = -\frac{6}{N} a^2 \dot{a} \left[\frac{BC}{\mathcal{A}}-\frac{B\mathcal{Z}D}{\mathcal{A}^2}\right],
\end{equation}
\begin{equation}
P_T =  -\frac{6}{N} a^2 \dot{a} \left(-\frac{BD}{\mathcal{A}}+\frac{B\mathcal{Z}E}{\mathcal{A}^2}\right),
\qquad
P_p =  -\frac{6}{N} a^2 \dot{a} \left(\frac{4DB}{\mathcal{A}}-\frac{4B\mathcal{Z}E}{\mathcal{A}^2}\right).
\end{equation}

With the use of the canonical momenta, we can construct, via the Legendre transformation, the cosmological Hamiltonian of $f(R,T)$ gravity as
%\begin{widetext}
\bea
H_{\rm grav} &=& \left(-\frac{6}{N} a \dot{a}^2  \widetilde{\lambda} -6kNa\widetilde{\lambda} + Na^3 V\right) -\frac{6}{N} a^2 \dot{a} \Bigg[\dot{A} -\frac{B\dot{A}}{\mathcal{A}}-\frac{\dot{B}\mathcal{Z}}{\mathcal{A}} +\frac{3B\dot{B}\mathcal{Z}}{\mathcal{A}^2}
+\frac{BR\dot{C}}{\mathcal{A}}
-\frac{B\dot{D}M}{\mathcal{A}}
	\nonumber\\
&&-\frac{B\mathcal{Z}R\dot{D}}{\mathcal{A}^2}+
\frac{B\mathcal{Z}\dot{E}M}{\mathcal{A}^2}+\frac{3B\dot{F}}{\mathcal{A}}-\frac{3B\mathcal{Z}\dot{G}}{\mathcal{A}^2}+\frac{BC\dot{R}}{\mathcal{A}}
	-\frac{B\mathcal{Z}D\dot{R}}{\mathcal{A}^2}-\frac{BD\dot{M}}{\mathcal{A}}+\frac{B\mathcal{Z}E\dot{M}}{\mathcal{A}^2} \Bigg]\,.
\eea
%\end{widetext}

In the following analysis, in order to simplify the notation, we represent the Hamiltonian in the form
\begin{equation}
H_{\rm grav}=(\cdots)-\frac{6}{N}a^2 \dot{a}[\cdots]\,.
\end{equation}
From the definition of the canonical momenta we can now easily derive the relation
\be
P_a P_A = \left(\frac{6}{N}\right)^2 a^3 \dot{a}^2 \widetilde{\lambda} \left(1-\frac{B}{\mathcal{A}}\right)+\left(\frac{6}{N}\right)^2 a^4 \dot{a} [\cdots] +
\left(\frac{6}{N}\right)^2 a^4 \dot{a} \left(-\frac{B}{\mathcal{A}}\right)\dot{\widetilde{\lambda}} \,.
\ee
Since
\begin{equation}
P_a P_F =\left(\frac{6}{N}\right)^2 2a^3 \dot{a}^2 \widetilde{\lambda} \left(3\frac{B}{\mathcal{A}}\right) +\left(\frac{6}{N}\right)^2 a^4 \dot{a} \dot{\widetilde{\lambda}} \left(\frac{3B}{\mathcal{A}}\right) \,,
\end{equation}
 the combination of the above two equations yields
\begin{equation}
-\frac{6}{N} a^2 \dot{a} [\cdots] = -\frac{N}{6a^2}\left(P_a P_A +\frac{1}{3}P_a P_F\right) + \frac{2\cdot 6}{N} a \dot{a}^2 \widetilde{\lambda}\,.
\end{equation}
Therefore, the gravitational Hamiltonian can be written as
\be
H_{\rm grav}= \frac{6}{N}a \dot{a}^2 \left(A-\frac{B\mathcal{Z}}{\mathcal{A}}\right) -\frac{N}{6a^2} \left(P_a P_A +\frac{1}{3}P_a P_F\right)+
 Na^3 V -6kNa \widetilde{\lambda} \,.
\ee
Since we have $P_A =-\frac{6}{N} a^2 \dot{a}\left(1-\frac{B}{\mathcal{A}}\right)$, after taking the square of it, we obtain
\bea
\frac{6}{N}a \dot{a}^2 A &=&\left[\frac{P_{A}^{2}}{\left(\frac{6}{N}\right)^2 a^4 \dot{a}^2}-\frac{B^2}{\mathcal{A}^2}+\frac{2B}{\mathcal{A}}\right]\, \frac{6}{N} a \dot{a}^2 A \nonumber\\
&=& \frac{NP_A^2}{6a^3} A +\frac{6}{N} a \dot{a}^2 A \left(\frac{-B^2}{\mathcal{A}^2} +\frac{2B}{\mathcal{A}}\right).
\eea

Thus, for the Hamiltonian we find the expression
\bea
H_{\rm grav}=\frac{N}{6a^3}P_A^2 A -\frac{N}{6a^2}(P_a P_A+\frac{1}{3}P_a P_F) -
\frac{6}{N}a\dot{a}^2 \frac{AB^2}{\mathcal{A}^2}
%\nonumber\\
-\frac{6}{N}a \dot{a}^2 \frac{B(\mathcal{Z}-2A)}{\mathcal{A}}+Na^3 V -6kNa\widetilde{\lambda} \,.
\eea
Considering now that
\begin{equation}
P_C P_R = \left(-\frac{6}{N} a^2 \dot{a}\right)^2 \left(\frac{B^2 R C}{\mathcal{A}^2}-\frac{B^2 \mathcal{Z}RD}{\mathcal{A}^3}\right),
\quad
P_D P_M =\left(-\frac{6}{N} a^2 \dot{a}\right)^2 \left(-\frac{BM}{\mathcal{A}}-\frac{B \mathcal{Z}R}{\mathcal{A}^2}\right) \left(-\frac{BD}{\mathcal{A}}+\frac{B\mathcal{Z}E}{\mathcal{A}^2}\right), \nonumber
\end{equation}
\begin{equation}
\frac{1}{3} P_E P_F =(-\frac{6}{N}a^2\dot{a})^2 \frac{B^2 \mathcal{Z}M}{\mathcal{A}^3},
\qquad
P_G^2 = \left(-\frac{6}{N}a^2 \dot{a}\right)^2 \frac{9B^2 \mathcal{Z}^2}{\mathcal{A}^4},
\qquad
P_C \left(P_A +\frac{P_F}{3}\right) =\left(-\frac{6}{N}a^2 \dot{a}\right)^2 \left(\frac{BR}{\mathcal{A}}\right), \nonumber
\end{equation}
we can also obtain the relation
\bea
-\frac{6}{N}a \dot{a}^2 \frac{B(\mathcal{Z}-2A)}{\mathcal{A}}&=&
-\frac{6}{N}a \dot{a}^2 \frac{B\left(DM-CR+A-3F-2A\right)}{1/2 +3B +EM -DR -3G}
\nonumber\\
&=&\frac{N}{6a^3} \left(P_A +\frac{P_F}{3}\right)\frac{P_F}{3} \left( A+3F\right) -
\frac{N}{6a^3} \frac{\mathcal{A}}{B} \left(P_C P_R + P_D P_M\right)
	\nonumber\\
&&+\frac{6a\dot{a}^2}{N} \frac{\mathcal{A}}{B} \frac{B\mathcal{Z}E}{\mathcal{A}^2} \left(-\frac{BM}{\mathcal{A}}-\frac{B\mathcal{Z}R}{\mathcal{A}^2}\right)-\frac{6a\dot{a}^2}{N}\left(-\frac{2BCR}{\mathcal{A}}\right).
\eea

Therefore, the gravitational part of the cosmological Hamiltonian of $f(R,T)$ gravity can be written as
\bea
H_{\rm grav} &=& \frac{N}{6a^3} \left(P_A +\frac{P_F}{3}\right)(A P_A +F P_F -a P_a)
+ \frac{N}{3a^3}P_C \left(P_A +\frac{1}{3}P_F\right)C
	\nonumber  \\
&&+
\frac{N}{6a^3} \frac{\mathcal{A}}{B} \left(P_C P_R + P_D P_M +\frac{P_E P_F}{3} E+\frac{P_G^2}{9}  RE\right)  +Na^3 V-
6kNa \left(A -\frac{B\mathcal{Z}}{\mathcal{A}}\right),
\eea
where we have introduced the quantum potential $V$ of the cosmological system defined as
\begin{equation}
V=-f +\widetilde{\lambda}R +\widetilde{\mu}\left(\frac{1}{2} T +f_T M -f_R R -3\square f_R +2f \right).
\end{equation}

For the matter part of the Hamiltonian we have \cite{Brown:1992kc}
\begin{equation}
H_{\rm matt}=-\mathcal{L}_{\rm matt}= - Na^3 p.
\end{equation}
Thus, the total Hamiltonian of the universe in $f(R,T)$ gravity is
\be\label{1g_H1}
H=H_{\rm grav}+H_{\rm matt}.
\ee

The gravitational Hamiltonian $H$ constructed  above is a very general one, since it contains all the canonical momenta associated to all variables of $f(R,T)$ gravity. This general representation can lead to the full description of the complex dynamics and evolution of all the gravitational field variables, as well as their associated canonical momenta.

%%%%%%%%%%%%%%%%%%%%%%%%%%%%%%%%%%%%%%%%%%%%%%%%%%%%%%%%%%%%%%%%%%%%%%%%%%
\subsubsection{The Wheeler-de Witt equation}
%%%%%%%%%%%%%%%%%%%%%%%%%%%%%%%%%%%%%%%%%%%%%%%%%%%%%%%%%%%%%%%%%%%%%%%%%%

The basic equation of quantum cosmology, namely, the Wheeler-de Witt equation, can be immediately obtained from the Hamiltonian (\ref{1g_H1}) of $f(R,T)$ gravity. Hence, by applying the standard quantization procedure we obtain the evolution equation of the quantum universe as
\begin{equation}
H \Psi = (H_{\rm grav} +H_{\rm matt})\Psi = N \mathcal{H} \Psi = 0.
\end{equation}

The Hamiltonian operator $\mathcal{H}$ for $f(R,T)$ gravity takes the form
\bea
\mathcal{H}&=& \frac{1}{6a^3}\left(P_A^2 A + P_A P_F F -P_A P_a a +\frac{A}{3} P_F P_A +
\frac{F}{3}P_F^2 -\frac{a}{3}P_F P_a\right)
+\frac{1}{3a^3}P_C \left(P_A +\frac{P_F}{3}\right)C
	\nonumber\\
&&-\frac{1}{6a^3} \frac{\mathcal{A}}{B} \left(P_C P_R +P_D P_M  +
\frac{P_E P_F}{3}E +\frac{P_G^2}{9}RE \right)
+a^3 V
	-6ka\left(A -\frac{B\mathcal{Z}}{\mathcal{A}}\right) - a^3 p.
\eea

To quantize the model, we need to carry out first parameter ordering. Several ways to perform it have been proposed in the literature \cite{Steigl:2005fk}.  In the following we adopt a method that maintains the Hamiltonian Hermitian \cite{Vakili:2010rf}. Hence we can obtain the following relationships (we perform the quantization via the substitution $P_q \rightarrow -i (\partial / \partial q)$),
\begin{equation}
q P_{q}^2  = \frac{1}{2}\left(q^u P_q q^v P_q q^w + q^w P_q q^v P_q q^u\right) =  -q\frac{\partial ^2}{\partial q ^2} +u w \frac{1}{q},
\end{equation}
where the parameters $u,v,w$, denoting the ambiguity in the ordering of the factors $q$ and $P_q$, satisfy the condition $u+v+w=1$, and the relation
\begin{equation}
q P_q  = \frac{1}{2} \left(q^r P_q q^s + q^s P_q q^r\right) = -i\left(q\frac{\partial}{\partial q}+1\right),
\end{equation}
where the parameters $r,s$ denote the ambiguity in the ordering of factors $q$ and $P_q$, and satisfy the condition $r+s =1$. Similarly one can show that
\begin{equation}
q^{-2} P_q = -i \left(-\frac{2}{q^3} +\frac{1}{q^2} \frac{\partial}{\partial q}\right).
\end{equation}

Therefore, the quantized cosmological Hamiltonian in $f(R,T)$ gravity can be obtained as
%\begin{widetext}
\bea
\mathcal{H} &=& -\frac{1}{6a^3}\left[\left(A\frac{\partial}{\partial A} +F\frac{\partial}{\partial F} +2C\frac{\partial}{\partial C} -a\frac{\partial}{\partial a}+5\right)\left(\frac{\partial}{\partial A}+\frac{1}{3}\frac{\partial}{\partial F}\right) -u_1 w_1 \frac{1}{A}
	 -u_2 w_2 \frac{1}{3F}\right]
\nonumber \\
&&	
	 +
	\frac{1}{6a^3}\Bigg[\frac{\mathcal{A}}{B}\left( \frac{\partial}{\partial C}\frac{\partial}{\partial R}  +\frac{\partial}{\partial D}\frac{\partial}{\partial M}+ \frac{E}{3}\frac{\partial}{\partial E} \frac{\partial}{\partial F} - \frac{RE}{9}\frac{\partial ^2}{\partial G ^2} \right)
	\nonumber\\
&&	+  \Bigg(- \frac{D}{B}\frac{\partial}{\partial C} -\frac{R}{B}\frac{\partial}{\partial M} +\frac{E}{B}\frac{\partial}{\partial D} +
	\frac{\mathcal{A}+EM}{3B}\frac{\partial}{\partial F}   +\frac{RE}{3BG}u_3w_3      \Bigg) \Bigg]
		+ a^3 V -6ka\left(A -\frac{B\mathcal{Z}}{\mathcal{A}}\right) -a^3 p.
\eea
%\end{widetext}
Here $u_1,w_1$, $u_2,w_2$ and $u_3,w_3$ denote the quantum mechanical ambiguity in the ordering of the factors $A , P_A$, $F, P_F$, and $G,P_G$, respectively.

In the next section, we apply the general formalism developed so far to investigate some particular quantum cosmological models in $f(R,T)$ gravity.

%%%%%%%%%%%%%%%%%%%%%%%%%%%%%%%%%%%%%%%%%%%%%%%%%%%%%%%%%%%%%%%%%%%%%%%%%%
\subsection{Specific cosmological application: $f(R,T)= F^0(R)+\theta RT$}\label{1g_sect3}
%%%%%%%%%%%%%%%%%%%%%%%%%%%%%%%%%%%%%%%%%%%%%%%%%%%%%%%%%%%%%%%%%%%%%%%%%%

In the previous section, we derived the general form of the Wheeler-de Witt equation in $f(R,T)$ gravity. As one can see from the form of this equation,  an analytic general solution of the Wheeler-de Witt equation for an arbitrary $f(R,T)$ would be extremely difficult to find. Instead of concentrating on the investigation of the general equation, in the present Section we consider a particular case of the $f(R,T)$ theory, in which the gravitational action takes the simple form
\be\label{1g_RTs}
f(R,T)= F^0(R) +\theta RT,
\ee
where $F^0(R)$ is an arbitrary function of the Ricci scalar only, and $\theta$ is an arbitrary function depending on the cosmological scale factor $a(t)$ of the classical universe. In this toy model, the term in the gravitational action involving the coupling of the curvature of spacetime with the trace of the matter energy-momentum tensor could give a hint of the implications of such a coupling on the quantum cosmological evolution of the Universe.

%%%%%%%%%%%%%%%%%%%%%%%%%%%%%%%%%%%%%%%%%%%%%%%%%%%%%%%%%%%%%%%%%%%%%%%%%%
\subsubsection{The Hamiltonian and the Wheeler-de Witt equation}
%%%%%%%%%%%%%%%%%%%%%%%%%%%%%%%%%%%%%%%%%%%%%%%%%%%%%%%%%%%%%%%%%%%%%%%%%%

For the model described by the Eq. (\ref{1g_RTs}) we can easily obtain
\be
 \mathcal{A}=\frac{1}{2} +2\theta R, \qquad B = \theta R, \qquad \frac{B}{\mathcal{A}}=\frac{\theta R}{1/2 +2\theta R}.
 \ee
Similarly, the other variables of physical and cosmological interest become
\be
A = F^0_R +\theta T, \qquad C = F^0_{RR}, \qquad D=\theta , \qquad E=0, \qquad G=\square \theta .
\ee

In the early quantum Universe after the initial Big Bang, the spacetime has a very high curvature, so that the condition $R\rightarrow \infty$ holds.  In the limit  $R\rightarrow \infty$, we have $B/\mathcal{A}=1/2$. By using the definitions of $P_A$ and $P_F$, we find $P_F = \left[3B/\left(\mathcal{A}-B\right)\right]P_A \approx 3P_A$. By also assuming that in the newly born quantum Universe the conditions $f_R \gg \square f_{RR} \rightarrow A \gg  F$ hold, we find for the gravitational Hamiltonian of the $f(R,T)=F^0(R)+\theta RT$ model the expression
\bea
H_{\rm grav} &=& -2 \frac{N}{6a^2} P_a P_A + 2\frac{N}{6a^3} P_A^2 A -2\frac{N}{6a^3} \left(P_C P_R +P_D P_M\right)
\nonumber  \\
&&
+\frac{4N}{6a^3}P_A P_C C +Na^3 V -6kNa\left(A - \frac{B\mathcal{Z}}{\mathcal{A}}\right),
\eea
where the quantum potential $V$ is defined as
\bea
V= \left(\frac{2R\theta}{1/2 +2R\theta}-1\right)\left(F^0-F^0_R R\right)+
\frac{R\theta}{1/2 +2R\theta}\left(\frac{1}{2} T +C R^2 +3F R -3\square f_R\right).
\eea
In the limit $R\rightarrow \infty$, and by assuming that $\theta \neq 0$, the quantum potential takes the form
\begin{equation}
V= \frac{1}{2}\left(\frac{1}{2} T + CR^2 +3F R - 3\square f_R\right).
\end{equation}

Therefore in the approximation of the large curvature $R$ the total Hamiltonian of the early Universe becomes
\bea\label{1g_48}
H=H_{\rm grav} +H_{\rm matt} =
\left(-\frac{2N}{6a^2}P_a P_A +\frac{2N}{6a^3}P_A^2 A -3kNaA\right)
%	\nonumber\\
+\left(-\frac{2N}{6a^3}P_C P_R +\frac{Na^3}{2}CR^2 -3kNaCR\right)
	\nonumber \\
+
\left(-\frac{2N}{6a^3}P_D P_M +\frac{Na^3}{4} M +3kNaDM  \right)
%	\nonumber \\
+
 \frac{4N}{6a^3}P_A P_C C + \frac{3Na^3}{2}\left(F R -\square f_R \right).
\eea

It is a general property of the Lagrangian/Hamiltonian systems that if we have some terms that can be ignored in the total action, they can also be ignored in  $L$ and $H$, without leading to any physical differences in the dynamical evolution  of the given system. Due to Gauss' theorem, in Eq.~(\ref{1g_48}) we have the relations
\bea
\int dt \sqrt{-g} \frac{3Na^3}{2} \square f_R &=& \int_{M} d^4x \sqrt{-g} \square f_R = \oint _{\partial M}f_R^{;\mu} \sqrt{-g} d\sigma_{\mu}^{3}.
\eea

Then we immediately see that the variational derivative of this term cancels out from the equation of motion,
\begin{equation}
\frac{2}{\sqrt{-g}}\frac{\delta}{\delta g^{\mu\nu}} \int \sqrt{-g}\square f_R d^4 x = 0,
\end{equation}
 and the term can be removed from the Hamiltonian function. Thus, we can now find the Wheeler-de Witt equation for the $f(R,T)=F^0(R)+\theta RT$ gravitational model, which in this specific case has the form
%\begin{widetext}
\bea
\mathcal{H}\Psi &=& \Bigg[\left(\frac{2}{6a^2}\frac{\partial ^2}{\partial a \partial A}- \frac{4}{6a^3}C\frac{\partial}{\partial A}\frac{\partial}{\partial C} -\frac{8}{6a^3}\frac{\partial}{\partial A} -\frac{2}{6a^3}A \frac{\partial ^2}{\partial A ^2}
	+\frac{2}{6}u_1 w_1 \frac{1}{Aa^3}-3kaA\right)
	\nonumber \\
&&
 + \left(\frac{2}{6a^3}\frac{\partial}{\partial C}\frac{\partial}{\partial R} +\frac{a^3}{2}CR^2 -3kaCR\right)
  + \left(\frac{2}{6a^3} \frac{\partial ^2}{\partial D \partial M} +\frac{a^3}{4} M + 3kaDM\right)
  +\frac{3a^3}{2}FR\Bigg]\Psi =0.
\eea

%%%%%%%%%%%%%%%%%%%%%%%%%%%%%%%%%%%%%%%%%%%%%%%%%%%%%%%%%%%%%%%%%%%%%%%%%%
\subsubsection{The Hamiltonian form of the field equations}
%%%%%%%%%%%%%%%%%%%%%%%%%%%%%%%%%%%%%%%%%%%%%%%%%%%%%%%%%%%%%%%%%%%%%%%%%%

It is a well known result of classical mechanics that the total time derivative of any function of the canonical variables can be calculated with the use of the Poisson bracket $\{,\}$ as
\begin{equation}
\frac{d}{dt}f =\frac{\partial f}{\partial t} + \{f,H \}.
\end{equation}
If the physical variables of the system do not depend explicitly on the time $t$, the Poisson brackets simplifies to
\begin{equation}
\frac{d}{dt}f = \{f,H \} .
\end{equation}
Therefore, since the Hamiltonian of $f(R,T)$ gravity has already been found, we can formulate the classical equations of motion of this modified gravity theory as
\begin{equation}
\dot{a} =\{a,H \} = -2\frac{N}{6a^3} P_A,
\qquad
\dot{A}= \{A,H \}=N\left[-\frac{1}{3a^2}P_a +\frac{2}{3a^3}AP_A  +\frac{4}{6a^3}P_C C \right],
\end{equation}
\bea
\dot{P_a}=\{P_a , H\} &=& N\Bigg\{\left(-\frac{2}{3a^3}P_A P_a +\frac{1}{a^4}P_A^2 A +3kA\right)
+\left( -\frac{1}{a^4}P_C P_R -\frac{3a^2}{2}CR^2 -3kCR\right)
\nonumber\\
&&
+\left[-\frac{1}{a^4}P_D P_M
-\frac{3a^2}{4}M +3kDM\right] + \frac{12}{6a^4}P_A P_C C  \Bigg\},
\eea
\begin{equation}
\dot{P_A}=\{P_A,H \}=N\left[-\frac{1}{3a^3}P_A^2 +3ka\right],
\qquad
\dot{C} = \{C,H \}=-\frac{N}{3a^3}\left[(P_R -2P_A C \right],
\end{equation}
\begin{equation}
\dot{P_C} =\{P_C , H \}= N\left[-\frac{a^3}{2}R^2 +3kaR -\frac{4}{6a^3}P_A P_C \right],
\qquad
\dot{R} = \{R,H \} =-\frac{N}{3a^3}P_C,
\end{equation}
\begin{equation}
\dot{P_R} =\{P_R , H\} = N\left[-a^3 CR + 3kaC -\frac{3a^3}{2}F\right],
\qquad
\dot{D} =\{D,H \}=-2 \frac{N}{6a^3} P_M,
\end{equation}
\begin{equation}
\dot{P_D} =-3kNaM,
\qquad
\dot{M} = -2\frac{N}{6a^3} P_D,
\qquad
\dot{P_M} = -N\left(\frac{a^3}{4} +3kaD\right).
\end{equation}

Now, we introduce a new time variable $\tau$, which is defined with the help of the original time variable $t$ by means of the relation
\begin{equation}
\tau = \int N(t) dt,
\end{equation}
or, in an equivalent formulation, as $d\tau/dt = N(t)$. On the other hand, by using the definition of $P_M$, we have
\begin{equation}
P_M = \frac{3a^2 \dot{a}}{N}D  = 3a^3 h D ,
\end{equation}
where $h \equiv  \dot{a}/(Na)$,
denotes the Hubble function associated to the cosmological model. Then in the new time variable $\tau$ the Hamilton equations of motion for the cosmological fluid, in the $f(R,T)$ gravity dominated universe, become
\begin{equation} \label{1g_Dprime}
D' = -h \theta =-\frac{a'}{a} D, \qquad
P_D' =-3kaM,
\qquad
M' = -2 \frac{1}{6a^3} P_D, \qquad
P_M' = \frac{a^3}{4} +3kaD ,
\end{equation}
where a prime denotes the derivative with respect to the modified time variable $\tau$. From the first equation of the above system we obtain the coupling constant $\theta =D$ of our toy gravitational model as
\begin{equation}
D= \frac{\delta}{a(\tau)} , \qquad \delta = {\rm constant}.
\end{equation}

This result shows us that the nonminimal coupling between the gravitational field and the matter field decreases as the scale factor of the Universe increases. This may be the physical reason why the nonminimal coupling between gravity and matter becomes so weak in the limit of large cosmological times, and that presently $f(R,T)$ gravity behaves as standard GR.

%%%%%%%%%%%%%%%%%%%%%%%%%%%%%%%%%%%%%%%%%%%%%%%%%%%%%%%%%%%%%%%%%%%%%%%%%%
\subsection{The problem of time}
%%%%%%%%%%%%%%%%%%%%%%%%%%%%%%%%%%%%%%%%%%%%%%%%%%%%%%%%%%%%%%%%%%%%%%%%%%

Our attempts of understanding quantum cosmology in a way similar to standard quantum mechanics, or quantum field theory, are seriously hampered by the lack of the time evolution of the wave function of the Universe in the Wheeler-de Witt equation. A possible way to transform the Wheeler-de Witt equation into a Schr\"{o}dinger-type equation can be formulated within the framework of $f(R,T)$ gravity, and such a transformation may be made possible by the presence of the curvature-matter coupling. To derive such a Schr\"{o}dinger-type equation we note first that the product $P_D P_M$ can be written as
 \begin{equation}
 -\frac{2}{6a^3} P_D P_M = -h D P_D.
 \end{equation}
 In the following, we introduce the fundamental hypothesis that this term can be expounded as  $P_{\tau}$, that is, the canonical momentum for time. This assumption can be justified in a convincing way, since after performing the standard quantization procedure $P_q = -i \frac{\partial}{\partial q}$, we obtain
\bea\label{1g_tq}
P_{\tau} = -i\frac{d}{d\tau} = \frac{d a(\tau)}{d\tau} \left(-i\frac{\partial}{\partial a}\right)=
 -\frac{a'}{a} \frac{\delta}{a}\left(-\frac{a^2}{\delta}\right)\left(-i \frac{\partial}{\partial a}\right) = -h D P_D.
\eea
The above result is obviously true if we accept that in the quantum cosmological model the scale factor $a(t)$ is the only time dependent variable. The relation introduced in Eq.~(\ref{1g_tq}) provides us with the possibility to introduce the further transformation
\begin{equation} \label{1g_timeT}
-\frac{2}{6a^3}P_D P_M \rightarrow P_{\tau}.
\end{equation}

The possibility of carrying out such a transformation indicates that the coupling between the gravitational field and the matter field may play an important role in the evolution of the very early Universe. We will accomplish now a further simplification of the gravitational action, by assuming it is of the form
\be\label{1g_frtc}
f(R,T) = R+\theta R T,
\ee
and whose main properties we have already discussed. Then we have $C=F= P_R=0$, while the cosmological Hamiltonian of the gravitational system is given by
 \bea
H&=&H_{\rm grav} +H_{\rm matt}
	\nonumber  \\
&=& -\frac{2N}{6a^2}P_a P_A +\frac{2N}{6a^3}P_A^2 A -
3kNaA 	-\frac{N}{6a^3}P_D P_M +\frac{Na^3}{4} M +3kNaM.
\eea

The above Hamiltonian is very similar in its structure to the Hamiltonian obtained in the framework of $f(R)$ gravity, and discussed in \cite{Vakili:2009he}, except for the presence of a new term
\begin{equation*}
-2\frac{N}{6a^3}P_D P_T +N(\frac{a^3}{4} +3ka)M .
\end{equation*}

This term shows us the effect of the coupling between curvature and matter. Then the Wheeler-de Witt equation $\mathcal{H}\Psi =0$ for this $f(R,T)$ gravity model can be written down as
\be
\mathcal{H}\Psi = \left[ -\frac{2}{6a^2}P_a P_A +\frac{2}{6a^3}P_A^2 A -3kaA
-\frac{2}{6a^3}P_D P_M +
\frac{a^3}{4}M +3kaDM \right]\Psi =0.
\ee
With the help of the transformation defined in Eq.~(\ref{1g_timeT}), we immediately find
\be
\mathcal{H}_{\rm eff}\Psi = \left[ -\frac{2}{6a^2}P_a P_A +\frac{2}{6a^3}P_A^2 A -3kaA
 +
 \frac{a^3}{4} M + 3kaDM \right]\Psi = -P_{\tau}\Psi.
\ee

To quantize the model, we substitute $P_q = -i \partial/\partial q$  in the above Hamiltonian. Thus, we obtain the  Schr\"{o}dinger-Wheeler-de Witt (SWDW) equation, describing the quantum evolution of the very early Universe in the specific $f(R,T)$ gravity model (\ref{1g_frtc}) as
\bea \label{1g_H}
\mathcal{H}_{\rm eff}\Psi = \Bigg[\Bigg(\frac{2}{6a^2}\frac{\partial ^2}{\partial a \partial A} -\frac{4}{6a^3}\frac{\partial}{\partial A} -\frac{2}{6a^3}A \frac{\partial ^2}{\partial A ^2} +
\frac{2}{6}u_1 w_1 \frac{1}{Aa^3} -3kaA\Bigg)
	+\frac{a^3}{4} M +3kaDM\Bigg]\Psi = i\frac{\partial \Psi}{\partial \tau}.
\eea
This quantum evolution equation has the form of the standard quantum-mechanical Schr\"{o}dinger equation,
\begin{equation}
\mathcal{H}_{\rm eff}\Psi = i\frac{\partial \Psi}{\partial \tau}.
\end{equation}

Therefore, we have obtained the important result that in the framework of $f(R,T)$ gravity, we can generate a Schr\"{o}dinger-type quantum mechanical evolution equation from the Wheeler-de Witt equation, which can find an answer to the challenge of time in quantum gravity. In the limiting case $d\tau/dt= N(t) = 1$, the  SWDW equation takes the standard form of the Schr\"{o}dinger equation with which we are acquainted with, namely,
\begin{equation}
\mathcal{H}\Psi = i\frac{\partial\Psi}{\partial t}.
\end{equation}

%%%%%%%%%%%%%%%%%%%%%%%%%%%%%%%%%%%%%%%%%%%%%%%%%%%%%%%%%%%%%%%%%%%%%%%%%%
\subsection{The physical interpretation of the effective Hamiltonian}
%%%%%%%%%%%%%%%%%%%%%%%%%%%%%%%%%%%%%%%%%%%%%%%%%%%%%%%%%%%%%%%%%%%%%%%%%%

Let us analyze now in more detail the physical meaning of the time $\tau$, and of the effective Hamiltonian $H_{\rm eff}$ we have just introduced. This would also allow us to take a more profound sight into the fundamental quantum gravitational time problem. In the standard approach to quantum cosmology, as described by the Wheeler-de Witt equation, which is based on the condition $\mathcal{H}\psi =0$, there seems to be no time evolution, or dynamics, of the quantum dominated early Universe. Therefore it follows that the wave function of the very early Universe (or more specifically, the corresponding physical states) does not characterize quantum gravitational states of the Universe at a
particular time instance, as it happens in standard quantum mechanics.
Rather, the solution of the WDW equation describes states for all
times or, in a more precise formulation, the wave function provides
only information about the state of the very early quantum Universe
that is invariant with respect to all spacetime diffeomorphism
transformations \cite{Baez:1995sj}

However, in the modified gravity model $f(R,T) =R +\theta RT$, the concept of time can  be introduced locally by taking into account the coupling of the geometry and of the matter fields. This is possible since the interaction between gravity (geometry) and  matter is also local.  The deep relation between the two fundamental fields of physics, thermodynamics and gravity, tells us that the unidirectional arrow of time may be generated via the second law of thermodynamics, since both processes are closely related to some forms of irreversible evolution. If we restrict ourselves to the consideration of a small volume of the Universe, and interpret it as an adiabatic system, with the WDW equation $\mathcal{H} \Psi =0$ still holding in it, the coupling between curvature and matter will generate an arrow of time that would allow us to measure the increase of the entropy of this local volume of the Universe, as determined  by its matter content. The other components of the Hamiltonian function, included in the effective Hamiltonian $\mathcal{H_{\rm eff}}$, determine the dynamics of the gravitational field, and of its normal matter components. Consequently, the effective Hamiltonian can be written in the general form
\begin{equation}
\mathcal{H}_{\rm eff} \sim b_1 p_i p_j + b_2 x_i x_j \,,
\end{equation}
%$\mathcal{H} _{eff} \sim b_1 p_i p_j + b_2 x_i x_j$,
which is similar to the Hamiltonian form we usually deal with in ordinary quantum mechanics. Therefore, we can propose that the standard quantum mechanical Schr\"{o}dinger equation just represents an effective, locally valid version, of the Schr\"{o}dinger-Wheeler-de Witt equation. In other words, we may propose the conjecture according to which the Wheeler-de Witt equation in the presence of a curvature-matter coupling supplies the global quantum description for the Universe, while the standard
Schr\"{o}dinger equation, obtained in the limit of the vanishing curvature-matter intermixing, provides just the local quantum description of the microscopic regions of the present Universe.

\newpage

%%%%%%%%%%%%%%%%%%%%%%%%%%%%%%%%%%%%%%%%%%%%%%%%%%%%%%%%%%%%%%%%%%%%%%%%%%
\part{Conclusions}\label{Conclusions}
%%%%%%%%%%%%%%%%%%%%%%%%%%%%%%%%%%%%%%%%%%%%%%%%%%%%%%%%%%%%%%%%%%%%%%%%%%

In this review article, we have extensively analyzed the cosmological and astrophysical applications of two recently proposed classes of modified theories of gravity, namely, hybrid metric-Palatini gravity, building a bridge between the metric and Palatini formalism, and generalized gravity theories with curvature-matter couplings, in which geometry is coupled with the matter Lagrangian, or with the trace of the energy-momentum tensor, respectively. The starting point of our investigations was the ``simple'' $f(R)$ gravity theory, which, despite its promising features also possesses several drawbacks, and seems to be unfit to provide a full, accurate and consistent description of gravitational phenomena. Hence going beyond the $f(R)$ gravity theory may be necessary in order to account for the high complexity of the recent observational phenomena. From a theoretical point of view we are also facing the problem of the maximal extension of the Hilbert-Einstein Lagrangian, which has a simple additive structure in its combination of geometry and matter.

In the present day approach to gravitational physics one can distinguish between two possible avenues for the interpretation of the complex and intriguing structure of the Universe. The first approach may be called the dark components model, and generalizes the Einstein gravitational field equations by adding two more terms to the total energy momentum tensor, corresponding to dark energy and dark matter, respectively. Hence in this approach gravity is described by the equation $G_{\mu \nu}=\kappa ^2 T^{\rm bar}_{\mu \nu}+\kappa ^2T^{\rm DM}_{\mu \nu}(\phi, \psi _{\mu},...)+\kappa ^2T^{\rm DE}_{\mu \nu}(\phi, \psi _{\mu},...)$, where $T^{\rm bar}_{\mu \nu}$, $T^{\rm DM}_{\mu \nu}(\phi, \psi_{\mu},...)$, and $T^{\rm DE}_{\mu \nu}(\phi, \psi_{\mu},...)$ are the energy-momentum tensors of baryonic matter, dark matter and dark energy (also interpreted as a form of matter, since according to the theory of relativity mass and energy are equivalent), which are both functions of some (scalar or vector) fields.  Despite its remarkable success, the dark component approach is still facing its own problems, the most serious one being the lack of any direct detection of dark matter, despite extensive experimental efforts. Moreover, the plethora of the field type models for dark energy, each with its own problems, raises the question of the possibility of a unique and consistent field theoretical description of the major constituent of the Universe.

The second approach, which may be called the dark gravity approach, adopts a purely geometrical point of view on the gravitational phenomena, and explains their dynamics by modifying the geometric structure of the Einstein field equations. In dark gravity, the Einstein equations may be written as $G_{\mu \nu}=\kappa ^2T_{\mu \nu}+\kappa ^2 T_{\mu \nu}^{(\rm geom)}\left(g_{\mu \nu}, R, \square R,...\right)$,  where $T_{\mu \nu}^{(\rm geom)}\left(g_{\mu \nu}, R, \square R,...\right)$ is a purely geometric quantity, constructed from the metric, playing the role of an effective geometric energy-momentum tensor, and which may act as dark energy, dark matter, or both. $f(R)$ gravity belongs to this class of theories, and it introduces a purely geometric perspective on the gravitational phenomena going beyond the Einsteinian description. The hybrid metric-Palatini gravity theory also resides in the dark gravity class of theories, since it is constructed on purely geometric grounds, and it generalizes and combines two geometric approaches, the metric and the Palatini ones, respectively.

However, a third way to gravitation is also possible, and it is represented by the second class of models considered in the present review. From a theoretical point of view the basic idea is the replacement of the simple additive structure of the Hilbert-Einstein Lagrangian density with a more general algebraic structure. The maximal extension of the gravitational Lagrangian can be constructed by assuming that the gravitational action is an arbitrary function of the curvature scalar and of the matter Lagrangian. Such a construction implies automatically the existence of a nonminimal geometry-matter coupling. Other types of such a coupling are also possible, and in particular the coupling between the Ricci scalar and the trace of the matter energy-momentum tensor has opened novel and interesting avenues for the study of gravity.

Hence, one can assume the existence of a third direction for the understanding of the gravitational interaction, which we may call the dark coupling approach, and in which the Einstein gravitational equations can be generalized to the form  $G_{\mu \nu}=\kappa ^2T_{\mu \nu}+\kappa ^2 T_{\mu \nu}^{(\rm coup)}\left(g_{\mu \nu}, R, L_m, T, \square R, \square T,... \right)$,  where the effective energy-momentum tensor of the theory $T_{\mu \nu}^{(\rm coup)}\left(g_{\mu \nu}, R, L_m, T, \square R, \square T,... \right)$ is constructed from the maximal extension of the Hilbert-Einstein Lagrangian by taking into account its nonadditive structure, which automatically implies the presence of geometry-matter couplings. Note that in both dark gravity and dark coupling theories the Einstein gravitational constant $\kappa ^2$ becomes an effective quantity, which may be a function of the field parameters, and of the couplings.
In\emph{} this review we have presented and summarized the basic ideas, formalisms, and applications of a dark gravity theory, and of several dark coupling type theories. The gravitational field equations were presented, as well as the equations of motion for test particles. We have also explored and presented in detail a large number of astrophysical and cosmological applications of the considered theories, ranging from the scale of the Solar System to the cosmological dimensions of the present day Universe.

The first considered theory, the HMPG theory, is a purely geometric (dark gravity type) theory, interpolating between the metric and Palatini formulations of general relativity. The gravitational action is constructed by adding to the standard Ricci scalar a new term obtained from the Palatini curvature. The theory admits a scalar-tensor representation that leads to some important physical and cosmological results. In particular, the most attractive feature of the theory is its ability to cover in a physical way length scales ranging from the Solar System to galaxies and to the cosmological level, and to pass all observational tests on these scales. This is possible since in HMPG there is no definite requirement on the mass of the scalar field, which can also be very small. This result is in strong opposition with the requirement of a large mass for the scalar field that imposes severe limitations on the $f(R)$ gravity theory. From a cosmological point of view the cosmological models constructed in HMPG have the potential of explaining the recent acceleration of the Universe without resorting to the cosmological constant, or to the equally mysterious dark energy. Moreover, the problem of the possibility of the description of dark matter by the HMPG has been also fully addressed, and we have found some strong indications that such an interpretation of dark matter as a geometric effect is possible, on both galactic and clusters of galaxies levels. This would open the possibility of the explanation of all observed features of the Universe in a framework that unifies the standard metric and the intriguing Palatini formalisms. The HMPG theory has also profound implications on the structure of compact stellar type objects, and on black holes, including their thermodynamic properties. In particular, the masses and the maximum masses of the neutron stars are significantly increased due to the presence of the scalar field. This result may prove to be important especially in the context of the recently reported  LIGO/Virgo  discovery of a merging binary system whose components consist of a $23M_{\odot}$ mass black hole and a $2.6M_{\odot}$ mass compact companion star (the gravitational wave event GW190814) \cite{Abbott:2020khf}. This extremely intriguing observation has lead to a strong debate about the nature of the secondary star, whose mass is located in the so-called mass gap of the neutron stars \cite{Tsokaros:2020hli}. On the other hand this observation can be easily explained by assuming that the companion star of the massive black hole  is a HMPG star. However, more investigations of the structure of stars in HMPG gravity is necessary, and such investigations may lead to the possibility of testing modified theories of gravity via gravitational wave observations.

We have also discussed in detail several classes of dark coupling theories, the $f\left(R,L_m\right)$, the $f(R,T)$ and the $f(R,T,R_{\mu \nu}T^{\mu \nu})$ theories. Each of these theories involve the coupling between matter and geometry, but the forms of the couplings are different. Generally, in the context of the dark coupling theories, in the presence of curvature-matter couplings, the motion of test particles is non-geodesic, with an extra force orthogonal to the four-velocity of the particles being induced by the interaction between matter and geometry.
The cosmological implications of these theories have been investigated in detail, and they indicate some very promising results in the explanation of the late acceleration of the Universe. Moreover, it is interesting to note that the $f(R,T,R_{\mu \nu}T^{\mu \nu})$ gravity theory is not plagued by the Dolgov-Kawasaki instability, and hence it successfully passes this important theoretical test.
An interesting characteristic of the presence of the coupling in the gravitational action is the non-conservation of the matter energy-momentum tensor, which points towards a deep modification of some fundamental physical processes in these theories. The nonconservation of the matter energy-momentum tensor can be interpreted as describing matter and entropy creation, with the gravitational field acting as a source of both matter and entropy. From a thermodynamic point of view, irreversible matter creation induces a thermodynamic arrow of time, and hence geometry-matter coupling and the corresponding particle creation may serve as an indicator of distinguishing the past from the future, and of the direction of the time evolution of physical systems and processes.

Particle creation is also the distinguishing characteristics of quantum field theory in curved space-times \cite{Parker:2009uva}.  Even in ordinary quantum field theory, formulated in flat Minkowski spacetime, locally the number of particles is not well-defined. For curved geometries induced by the presence of non-zero cosmological constants, quantum fields lose their standard physical interpretation as asymptotic particles. The presence of particle creation in quantum theoretical approaches to describe physics and in curved spacetime and quantum gravity  leads to the intriguing possibility that one may interpret non-conservative gravitational models as a first order approximation to the physical phenomenology of quantum gravity \cite{Haghani:2017vqx}.  Thus, we may conjecture that gravitational field theories that intrinsically contain effective particle creation processes can provide at least a phenomenological description and perspective  of the quantum processes associated to the gravitational field. Similar lines of thought were investigated in \cite{Lobato:2018vpq}, where it was pointed out that energy nonconservation is also a prediction of quantum theory with time-space noncommutativity. If the time variable is considered as an operator, and if compact spatial coordinates that do not commute with the time do exist, then the time evolution is quantized, also leading to a violation of energy conservation. A model in a 5-dimensional flat spacetime, consisting of 3 commutative spatial dimensions, taken together with 1 compact spatial dimension was considered, with the spatial coordinates not commuting with time. In this case the energy flows from the 3-dimensional commutative slice into the compact extra dimension (and vice versa), thus leading to a restoration of energy conservation. Since the energy flux is proportional to the energy density of the matter content, one can derive  a differential equation for $f( R, T)$, and thus a physical criterion for constraining the functional form of f( R, T) can be obtained.

Dark coupling gravitational theories essentially depend on the choice of the form of the coupling between matter and geometry, that is, of the function $f$ appearing in the gravitational action. Different (generally simple) functional forms of $f$ have been proposed, and they can be tested observationally, also obtaining in this way constraints on the parameters of the models. It would be important to obtain some better theoretical insights into the structure of dark coupling theories that would allow to fix the forms of the coupling between matter and geometry.
The predictions of the gravity theories considered in the present review lead to some major differences, as compared to the predictions of standard General Relativity, or to its generalizations that neglect the role of matter in structuring spacetime geometry.  These lead to the possibility of confronting modified gravity theory through the confrontation of the predictions of different competing models in the fundamental fields of gravitational collapse, gravitational waves, dark matter, and dark energy, respectively.

The greatest theoretical challenge present day physics faces is the solution of the problem of quantum gravity, a still unsolved question, with very little progress being made after almost 80 years of intensive research. Since an exact solution is missing, due to the complexity of the problem, employing some approximate methods for the study of the quantum effects in gravity seems to be the best approach to follow in the search for a quantum version of General Relativity. Among the many proposed approaches a promising path for understanding the role that quantum effects may play in gravity consists in the inclusion of some fluctuating tensor terms in the gravitational metric, which are quantum in their nature, and which are essentially stochastic in their physical and mathematical structure.  In such an approach  one can  represent  the quantum metric operator $\hat{g}_{\mu\nu}$ as a sum of two terms, with the first given by the mean value of the classical metric $g_{\mu \nu}$, while the second represents a fluctuating part $\delta\hat{g}_{\mu\nu}$. Consequently, we can decompose the quantum metric according to \cite{Dzhunushaliev:2013nea}
\begin{eqnarray}\label{c1}
\hat{g}_{\mu\nu}=g_{\mu\nu}+\delta\hat{g}_{\mu\nu}.
\end{eqnarray}

Moreover, in order to obtain some workable models one can assume that the average of the stochastically fluctuating part of the metric, which has its origin in the quantum structure of the spacetime,  can be described in terms of a classical tensor quantity  $K_{\mu\nu}$, as
\be\label{c2}
\left<\delta \hat{g}_{\mu\nu}\right>=K_{\mu\nu}\ne 0.
\ee

The gravitational Lagrangian corresponding this theoretical approach to the quantum gravity problem is given by
\bea
L&=&=\sqrt{-g}\Bigg[\frac{1}{2\kappa ^2} \left(R+G_{\mu\nu}\delta\hat{g}^{\mu\nu}\right)+
\mathcal{L}_m-\frac{1}{2}T_{\mu\nu}\delta\hat{g}^{\mu\nu}\Bigg],
\eea
and in the semiclassical approximation it leads to a gravitational theory with geometry-matter coupling, as pointed out in \cite{Yang:2015jla}. For a detailed analysis of the cosmological implications of semiclassical quantum gravitational models in the presence of metric fluctuations see \cite{Liu:2016qfx}.

It is interesting that a conceptualization of the problem of quantum gravity in terms of a fluctuating metric tensor leads us towards classical gravity theories with a compelling curvature-matter coupling, as well as the non-conservation of the energy-momentum tensors of the matter. Consequently, all such semiclassical gravity theories involve irreversible matter production processes. Hence a better comprehension of the physical and theoretical basis of the modified gravity theories with a curvature-matter coupling can be obtained even from the study of the quantum gravitational theories with first order quantum corrections induced by a stochastically fluctuating metric.  The examination of the physical, astrophysical and cosmological implications of these theories originating in the quantum structure of the spacetime can shed some light on the structure and implications of the modified gravity theories with geometry-matter coupling, and these classical theories can be put to work for the further phenomenological investigation of the quantum effects in gravity.

As future avenues of research, one should aim to characterize as much as possible the phenomenology predicted by these theories in order to find constraints arising from observations. The study of these phenomena may also provide some specific signatures and effects, which could distinguish and discriminate between the various theories of modified gravity. We also propose to use the data of SNIa, BAO, CMB shift parameter to obtain restrictions for the respective models, and explore in detail the implications of HMPG theory and the theories with geometry-matter coupling on the analysis of structure formation in the Universe.

%%%%%%%%%%%%%%%%%%%%%%%%%%%%%%%%%%%%%%%%%%%%%%%%%%%%%%%%%%%%%%%%%%%%%%%%%%
\section*{Acknowledgements}

We would like to thank the anonymous reviewer for comments and suggestions that helped us to significantly improve our manuscript. FSNL acknowledges support from the Funda\c{c}\~{a}o para a Ci\^{e}ncia e a Tecnologia (FCT) Scientific Employment Stimulus contract with reference 
CEECIND/04057/2017. FSNL also thanks funding from the research grants No. UID/FIS/04434/2020 and No. PTDC/FIS-OUT/29048/2017.
%%%%%%%%%%%%%%%%%%%%%%%%%%%%%%%%%%%%%%%%%%%%%%%%%%%%%%%%%%%%%%%%%%%%%%%%%%

\newpage

%%%%%%%%%%%%%%%%%%%%%%%%%%%%%%%%%%%%%%%%%%%%%%%%%%%%%%%%%%%%%%%%%%%%%%%%%%


\begin{thebibliography}{99}
%%%%%%%%%%%%%%%%%%%%%%%%%%%%%%%%%%%%%%%%%%%%%%%%%%%%%%%%%%%%%%%%%%%%%%%%%%


%\cite{Perlmutter:1998np}
\bibitem{Perlmutter:1998np}
S.~Perlmutter \textit{et al.} [Supernova Cosmology Project],
``Measurements of $\Omega$ and $\Lambda$ from 42 high redshift supernovae,''
Astrophys. J. \textbf{517} (1999), 565-586
%doi:10.1086/307221
[arXiv:astro-ph/9812133 [astro-ph]].
%12625 citations counted in INSPIRE as of 24 Jul 2020

%\cite{Riess:1998cb}
\bibitem{Riess:1998cb}
A.~G.~Riess \textit{et al.} [Supernova Search Team],
``Observational evidence from supernovae for an accelerating universe and a cosmological constant,''
Astron. J. \textbf{116} (1998), 1009-1038
%doi:10.1086/300499
[arXiv:astro-ph/9805201 [astro-ph]].
%12540 citations counted in INSPIRE as of 24 Jul 2020

%\cite{Avelino:2016lpj}
\bibitem{Avelino:2016lpj}
P.~Avelino, T.~Barreiro, C.~S.~Carvalho, A.~da Silva, F.~S.~N.~Lobo, P.~Martin-Moruno, J.~P.~Mimoso, N.~J.~Nunes, D.~Rubiera-Garcia, D.~Saez-Gomez, L.~Sousa, I.~Tereno and A.~Trindade,
``Unveiling the Dynamics of the Universe,''
Symmetry \textbf{8} (2016) no.8, 70
%doi:10.3390/sym8080070
[arXiv:1607.02979 [astro-ph.CO]].
%21 citations counted in INSPIRE as of 08 Jul 2020

%\cite{Nojiri:2010wj}
\bibitem{Nojiri:2010wj}
S.~Nojiri and S.~D.~Odintsov,
``Unified cosmic history in modified gravity: from $F(R)$ theory to Lorentz non-invariant models,''
Phys. Rept. \textbf{505} (2011), 59-144
%doi:10.1016/j.physrep.2011.04.001
[arXiv:1011.0544 [gr-qc]].
%2268 citations counted in INSPIRE as of 24 Jul 2020

%\cite{Nojiri:2006ri}
\bibitem{Nojiri:2006ri}
S.~Nojiri and S.~D.~Odintsov,
``Introduction to modified gravity and gravitational alternative for dark energy,''
eConf \textbf{C0602061} (2006), 06
%doi:10.1142/S0219887807001928
[arXiv:hep-th/0601213 [hep-th]].
%2151 citations counted in INSPIRE as of 24 Jul 2020

%\cite{Nojiri:2003ft}
\bibitem{Nojiri:2003ft}
S.~Nojiri and S.~D.~Odintsov,
``Modified gravity with negative and positive powers of the curvature: Unification of the inflation and of the cosmic acceleration,''
Phys. Rev. D \textbf{68} (2003), 123512
%doi:10.1103/PhysRevD.68.123512
[arXiv:hep-th/0307288 [hep-th]].
%1552 citations counted in INSPIRE as of 23 Jul 2020

%\cite{Sotiriou:2008rp}
\bibitem{Sotiriou:2008rp}
T.~P.~Sotiriou and V.~Faraoni,
``$f(R)$ Theories Of Gravity,''
Rev. Mod. Phys. \textbf{82} (2010), 451-497
%doi:10.1103/RevModPhys.82.451
[arXiv:0805.1726 [gr-qc]].
%2598 citations counted in INSPIRE as of 23 Jul 2020

%\cite{DeFelice:2010aj}
\bibitem{DeFelice:2010aj}
A.~De Felice and S.~Tsujikawa,
``$f(R)$ theories,''
Living Rev. Rel. \textbf{13} (2010), 3
%doi:10.12942/lrr-2010-3
[arXiv:1002.4928 [gr-qc]].
%2103 citations counted in INSPIRE as of 23 Jul 2020

%\cite{Capozziello:2011et}
\bibitem{Capozziello:2011et}
S.~Capozziello and M.~De Laurentis,
``Extended Theories of Gravity,''
Phys. Rept. \textbf{509} (2011), 167-321
%doi:10.1016/j.physrep.2011.09.003
[arXiv:1108.6266 [gr-qc]].
%1550 citations counted in INSPIRE as of 24 Jul 2020

%\cite{Capozziello:2002rd}
\bibitem{Capozziello:2002rd}
S.~Capozziello,
``Curvature quintessence,''
Int. J. Mod. Phys. D \textbf{11} (2002), 483-492
%doi:10.1142/S0218271802002025
[arXiv:gr-qc/0201033 [gr-qc]].
%855 citations counted in INSPIRE as of 21 Jul 2020

%\cite{Capozziello:2003tk}
\bibitem{Capozziello:2003tk}
S.~Capozziello, S.~Carloni and A.~Troisi,
``Quintessence without scalar fields,''
Recent Res. Dev. Astron. Astrophys. \textbf{1} (2003), 625
[arXiv:astro-ph/0303041 [astro-ph]].
%750 citations counted in INSPIRE as of 21 Jul 2020

%\cite{Capozziello:2007ec}
\bibitem{Capozziello:2007ec}
S.~Capozziello and M.~Francaviglia,
``Extended Theories of Gravity and their Cosmological and Astrophysical Applications,''
Gen. Rel. Grav. \textbf{40} (2008), 357-420
%doi:10.1007/s10714-007-0551-y
[arXiv:0706.1146 [astro-ph]].
%635 citations counted in INSPIRE as of 23 Jul 2020

%\cite{Aghanim:2018eyx}
\bibitem{Aghanim:2018eyx}
N.~Aghanim \textit{et al.} [Planck],
``Planck 2018 results. VI. Cosmological parameters,''
[arXiv:1807.06209 [astro-ph.CO]].
%3147 citations counted in INSPIRE as of 24 Jul 2020

%\cite{Sotiriou:2006hs}
\bibitem{Sotiriou:2006hs}
T.~P.~Sotiriou,
``f(R) gravity and scalar-tensor theory,''
Class. Quant. Grav. \textbf{23} (2006), 5117-5128
%doi:10.1088/0264-9381/23/17/003
[arXiv:gr-qc/0604028 [gr-qc]].
%276 citations counted in INSPIRE as of 22 Jul 2020

%\cite{Cognola:2007zu}
\bibitem{Cognola:2007zu}
G.~Cognola, E.~Elizalde, S.~Nojiri, S.~D.~Odintsov, L.~Sebastiani and S.~Zerbini,
``A Class of viable modified f(R) gravities describing inflation and the onset of accelerated expansion,''
Phys. Rev. D \textbf{77} (2008), 046009
%doi:10.1103/PhysRevD.77.046009
[arXiv:0712.4017 [hep-th]].
%525 citations counted in INSPIRE as of 22 Jul 2020

%\cite{Capozziello:2008rq}
\bibitem{Capozziello:2008rq}
S.~Capozziello, C.~Corda and M.~F.~De Laurentis,
``Massive gravitational waves from f(R) theories of gravity: Potential detection with LISA,''
Phys. Lett. B \textbf{669} (2008), 255-259
%doi:10.1016/j.physletb.2008.10.001
[arXiv:0812.2272 [astro-ph]].
%93 citations counted in INSPIRE as of 08 Jul 2020

%\cite{Motohashi:2017vdc}
\bibitem{Motohashi:2017vdc}
H.~Motohashi and A.~A.~Starobinsky,
``$f(R)$ constant-roll inflation,''
Eur. Phys. J. C \textbf{77} (2017) no.8, 538
%doi:10.1140/epjc/s10052-017-5109-x
[arXiv:1704.08188 [astro-ph.CO]].
%51 citations counted in INSPIRE as of 08 Jul 2020

%\cite{Motohashi:2019tyj}
\bibitem{Motohashi:2019tyj}
H.~Motohashi and A.~A.~Starobinsky,
``Constant-roll inflation in scalar-tensor gravity,''
JCAP \textbf{11} (2019), 025
%doi:10.1088/1475-7516/2019/11/025
[arXiv:1909.10883 [gr-qc]].
%6 citations counted in INSPIRE as of 08 Jul 2020

%\cite{Lobo:2008sg}
\bibitem{Lobo:2008sg}
F.~S.~N.~Lobo,
``The Dark side of gravity: Modified theories of gravity,''
[arXiv:0807.1640 [gr-qc]].
%222 citations counted in INSPIRE as of 15 Jul 2020

%\cite{Li:2006ag}
\bibitem{Li:2006ag}
B.~Li, K.~C.~Chan and M.~C.~Chu,
``Constraints on f(R) Cosmology in the Palatini Formalism,''
Phys. Rev. D \textbf{76} (2007), 024002
%doi:10.1103/PhysRevD.76.024002
[arXiv:astro-ph/0610794 [astro-ph]].
%90 citations counted in INSPIRE as of 08 Jul 2020

%\cite{Borowiec:2011wd}
\bibitem{Borowiec:2011wd}
A.~Borowiec, M.~Kamionka, A.~Kurek and M.~Szydlowski,
``Cosmic acceleration from modified gravity with Palatini formalism,''
JCAP \textbf{02} (2012), 027
%doi:10.1088/1475-7516/2012/02/027
[arXiv:1109.3420 [gr-qc]].
%28 citations counted in INSPIRE as of 21 Jul 2020

%\cite{Stachowski:2016zio}
\bibitem{Stachowski:2016zio}
A.~Stachowski, M.~Szydłowski and A.~Borowiec,
``Starobinsky cosmological model in Palatini formalism,''
Eur. Phys. J. C \textbf{77} (2017) no.6, 406
%doi:10.1140/epjc/s10052-017-4981-8
[arXiv:1608.03196 [gr-qc]].
%24 citations counted in INSPIRE as of 08 Jul 2020

%\cite{Olmo:2011uz}
\bibitem{Olmo:2011uz}
G.~J.~Olmo,
``Palatini Approach to Modified Gravity: f(R) Theories and Beyond,''
Int. J. Mod. Phys. D \textbf{20} (2011), 413-462
%doi:10.1142/S0218271811018925
[arXiv:1101.3864 [gr-qc]].
%353 citations counted in INSPIRE as of 24 Jul 2020

%\cite{Capozziello:2007eu}
\bibitem{Capozziello:2007eu}
S.~Capozziello and S.~Tsujikawa,
``Solar system and equivalence principle constraints on f(R) gravity by chameleon approach,''
Phys. Rev. D \textbf{77} (2008), 107501
%doi:10.1103/PhysRevD.77.107501
[arXiv:0712.2268 [gr-qc]].
%215 citations counted in INSPIRE as of 08 Jul 2020

%\cite{Khoury:2003rn}
\bibitem{Khoury:2003rn}
J.~Khoury and A.~Weltman,
``Chameleon cosmology,''
Phys. Rev. D \textbf{69} (2004), 044026
%doi:10.1103/PhysRevD.69.044026
[arXiv:astro-ph/0309411 [astro-ph]].
%1200 citations counted in INSPIRE as of 21 Jul 2020

%\cite{Harko:2011nh}
\bibitem{Harko:2011nh}
T.~Harko, T.~S.~Koivisto, F.~S.~N.~Lobo and G.~J.~Olmo,
``Metric-Palatini gravity unifying local constraints and late-time cosmic acceleration,''
Phys. Rev. D \textbf{85} (2012), 084016
%doi:10.1103/PhysRevD.85.084016
[arXiv:1110.1049 [gr-qc]].
%111 citations counted in INSPIRE as of 15 Jul 2020

%\cite{Capozziello:2013uya}
\bibitem{Capozziello:2013uya}
S.~Capozziello, T.~Harko, F.~S.~N.~Lobo and G.~J.~Olmo,
``Hybrid modified gravity unifying local tests, galactic dynamics and late-time cosmic acceleration,''
Int. J. Mod. Phys. D \textbf{22} (2013), 1342006
%doi:10.1142/S0218271813420066
[arXiv:1305.3756 [gr-qc]].
%38 citations counted in INSPIRE as of 08 Jul 2020

%\cite{Capozziello:2012ny}
\bibitem{Capozziello:2012ny}
S.~Capozziello, T.~Harko, T.~S.~Koivisto, F.~S.~N.~Lobo and G.~J.~Olmo,
``Cosmology of hybrid metric-Palatini f(X)-gravity,''
JCAP \textbf{04} (2013), 011
%doi:10.1088/1475-7516/2013/04/011
[arXiv:1209.2895 [gr-qc]].
%66 citations counted in INSPIRE as of 14 Jul 2020

%\cite{Capozziello:2015lza}
\bibitem{Capozziello:2015lza}
S.~Capozziello, T.~Harko, T.~S.~Koivisto, F.~S.~N.~Lobo and G.~J.~Olmo,
``Hybrid metric-Palatini gravity,''
Universe \textbf{1} (2015) no.2, 199-238
%doi:10.3390/universe1020199
[arXiv:1508.04641 [gr-qc]].
%66 citations counted in INSPIRE as of 14 Jul 2020

%\cite{Harko:2018ayt}
\bibitem{Harko:2018ayt}
T.~Harko and F.~S.~N.~Lobo,
%``Extensions of f(R) Gravity,''
Extensions of $f(R)$ Gravity: Curvature-Matter Couplings and Hybrid Metric-Palatini Theory, Cambridge Monographs on Mathematical Physics, Cambridge, Cambridge University Press (2018).


%\cite{Capozziello:2013wq}
\bibitem{Capozziello:2013wq}
S.~Capozziello, T.~Harko, T.~S.~Koivisto, F.~S.~N.~Lobo and G.~J.~Olmo,
``Hybrid $f(R)$ theories, local constraints, and cosmic speedup,''
%doi:10.1142/9789814623995_0106
[arXiv:1301.2209 [gr-qc]].
%4 citations counted in INSPIRE as of 08 Jul 2020

%\cite{Bertolami:2007gv}
\bibitem{Bertolami:2007gv}
O.~Bertolami, C.~G.~Boehmer, T.~Harko and F.~S.~N.~Lobo,
``Extra force in f(R) modified theories of gravity,''
Phys. Rev. D \textbf{75} (2007), 104016
%doi:10.1103/PhysRevD.75.104016
[arXiv:0704.1733 [gr-qc]].
%497 citations counted in INSPIRE as of 21 Jul 2020

%\cite{Harko:2014gwa}
\bibitem{Harko:2014gwa}
T.~Harko and F.~S.~N.~Lobo,
``Generalized curvature-matter couplings in modified gravity,''
Galaxies \textbf{2} (2014) no.3, 410-465
%doi:10.3390/galaxies2030410
[arXiv:1407.2013 [gr-qc]].
%92 citations counted in INSPIRE as of 08 Jul 2020

%\cite{Gonner:1984zx}
\bibitem{Gonner:1984zx}
H.~F.~M.~Gonner,
``THEORIES OF GRAVITATION WITH NONMINIMAL COUPLING OF MATTER AND THE GRAVITATIONAL FIELD,''
Found. Phys. \textbf{14} (1984), 865-881
%doi:10.1007/BF00737554
%37 citations counted in INSPIRE as of 08 Jul 2020

%\cite{Nojiri:2004fw}
\bibitem{Nojiri:2004fw}
S.~Nojiri and S.~D.~Odintsov,
``Dark energy and cosmic speed-up from consistent modified gravity,''
PoS \textbf{WC2004} (2004), 024
%doi:10.22323/1.013.0024
[arXiv:hep-th/0412030 [hep-th]].
%57 citations counted in INSPIRE as of 08 Jul 2020

%\cite{Bamba:2008ja}
\bibitem{Bamba:2008ja}
K.~Bamba and S.~D.~Odintsov,
``Inflation and late-time cosmic acceleration in non-minimal Maxwell-$F(R)$ gravity and the generation of large-scale magnetic fields,''
JCAP \textbf{04} (2008), 024
%doi:10.1088/1475-7516/2008/04/024
[arXiv:0801.0954 [astro-ph]].
%157 citations counted in INSPIRE as of 10 Jul 2020

%\cite{Bamba:2008xa}
\bibitem{Bamba:2008xa}
K.~Bamba, S.~Nojiri and S.~D.~Odintsov,
``Inflationary cosmology and the late-time accelerated expansion of the universe in non-minimal Yang-Mills-F(R) gravity and non-minimal vector-F(R) gravity,''
Phys. Rev. D \textbf{77} (2008), 123532
%doi:10.1103/PhysRevD.77.123532
[arXiv:0803.3384 [hep-th]].
%117 citations counted in INSPIRE as of 20 Jul 2020

%\cite{Harko:2010mv}
\bibitem{Harko:2010mv}
T.~Harko and F.~S.~N.~Lobo,
``$f(R,L_{m})$ gravity,''
Eur. Phys. J. C \textbf{70} (2010), 373-379
%doi:10.1140/epjc/s10052-010-1467-3
[arXiv:1008.4193 [gr-qc]].
%164 citations counted in INSPIRE as of 08 Jul 2020

%\cite{Harko:2011kv}
\bibitem{Harko:2011kv}
T.~Harko, F.~S.~N.~Lobo, S.~Nojiri and S.~D.~Odintsov,
``$f(R,T)$ gravity,''
Phys. Rev. D \textbf{84} (2011), 024020
%doi:10.1103/PhysRevD.84.024020
[arXiv:1104.2669 [gr-qc]].
%859 citations counted in INSPIRE as of 22 Jul 2020

%\cite{Faraoni:2004pi}
\bibitem{Faraoni:2004pi}
V.~Faraoni,
``Cosmology in scalar tensor gravity,''
Fundam. Theor. Phys. \textbf{139} (2004)
%doi:10.1007/978-1-4020-1989-0
%171 citations counted in INSPIRE as of 16 Jul 2020

%\cite{Will:2014kxa}
\bibitem{Will:2014kxa}
C.~M.~Will,
``The Confrontation between General Relativity and Experiment,''
Living Rev. Rel. \textbf{17} (2014), 4
%doi:10.12942/lrr-2014-4
[arXiv:1403.7377 [gr-qc]].
%1184 citations counted in INSPIRE as of 21 Jul 2020

%\cite{Bertolami:2007zm}
\bibitem{Bertolami:2007zm}
O.~Bertolami, F.~Gil Pedro and M.~Le Delliou,
``Dark Energy-Dark Matter Interaction and the Violation of the Equivalence Principle from the Abell Cluster A586,''
Phys. Lett. B \textbf{654} (2007), 165-169
%doi:10.1016/j.physletb.2007.08.046
[arXiv:astro-ph/0703462 [astro-ph]].
%287 citations counted in INSPIRE as of 08 Jul 2020

%\cite{Damour:2001fn}
\bibitem{Damour:2001fn}
T.~Damour,
``Questioning the equivalence principle,''
Compt. Rend. Acad. Sci. Ser. IV Phys. Astrophys. \textbf{2} (2001) no.9, 1249-1256
%doi:10.1016/S1296-2147(01)01272-0
[arXiv:gr-qc/0109063 [gr-qc]].
%45 citations counted in INSPIRE as of 08 Jul 2020

%\cite{Damour:1996xt}
\bibitem{Damour:1996xt}
T.~Damour,
``Testing the equivalence principle: Why and how?,''
%doi:10.1088/0264-9381/13/11A/005
[arXiv:gr-qc/9606080 [gr-qc]].
%65 citations counted in INSPIRE as of 08 Jul 2020

%\cite{Damour:2010rp}
\bibitem{Damour:2010rp}
T.~Damour and J.~F.~Donoghue,
``Equivalence Principle Violations and Couplings of a Light Dilaton,''
Phys. Rev. D \textbf{82} (2010), 084033
%doi:10.1103/PhysRevD.82.084033
[arXiv:1007.2792 [gr-qc]].
%98 citations counted in INSPIRE as of 10 Jul 2020

%\cite{Bojowald:2011zzb}
\bibitem{Bojowald:2011zzb}
M.~Bojowald,
``Quantum cosmology,''
Lect. Notes Phys. \textbf{835} (2011), pp.1-308
%doi:10.1007/978-1-4419-8276-6
%18 citations counted in INSPIRE as of 08 Jul 2020

%\cite{Bojowald:2015iga}
\bibitem{Bojowald:2015iga}
M.~Bojowald,
``Quantum cosmology: a review,''
Rept. Prog. Phys. \textbf{78} (2015), 023901
%doi:10.1088/0034-4885/78/2/023901
[arXiv:1501.04899 [gr-qc]].
%62 citations counted in INSPIRE as of 09 Jul 2020

%\cite{Mukhanov:2007zz}
\bibitem{Mukhanov:2007zz}
V.~Mukhanov and S.~Winitzki,
``Introduction to quantum effects in gravity,''
%111 citations counted in INSPIRE as of 13 Jul 2020

%\cite{DeWitt:1967yk}
\bibitem{DeWitt:1967yk}
B.~S.~DeWitt,
``Quantum Theory of Gravity. 1. The Canonical Theory,''
Phys. Rev. \textbf{160} (1967), 1113-1148
%doi:10.1103/PhysRev.160.1113
%2549 citations counted in INSPIRE as of 23 Jul 2020

%\cite{DeWitt:1967ub}
\bibitem{DeWitt:1967ub}
B.~S.~DeWitt,
``Quantum Theory of Gravity. 2. The Manifestly Covariant Theory,''
Phys. Rev. \textbf{162} (1967), 1195-1239
doi:10.1103/PhysRev.162.1195
%1475 citations counted in INSPIRE as of 23 Jul 2020

%\cite{DeWitt:1967uc}
\bibitem{DeWitt:1967uc}
B.~S.~DeWitt,
%``Quantum Theory of Gravity. 3. Applications of the Covariant Theory,''
Phys. Rev. \textbf{162} (1967), 1239-1256
%doi:10.1103/PhysRev.162.1239
%845 citations counted in INSPIRE as of 23 Jul 2020

%\cite{Banerjee:2011qu}
\bibitem{Banerjee:2011qu}
K.~Banerjee, G.~Calcagni and M.~Martin-Benito,
``Introduction to loop quantum cosmology,''
SIGMA \textbf{8} (2012), 016
%doi:10.3842/SIGMA.2012.016
[arXiv:1109.6801 [gr-qc]].
%150 citations counted in INSPIRE as of 08 Jul 2020

%\cite{MartinBenito:2008ej}
\bibitem{MartinBenito:2008ej}
M.~Martin-Benito, L.~J.~Garay and G.~A.~Mena Marugan,
``Hybrid Quantum Gowdy Cosmology: Combining Loop and Fock Quantizations,''
Phys. Rev. D \textbf{78} (2008), 083516
%doi:10.1103/PhysRevD.78.083516
[arXiv:0804.1098 [gr-qc]].
%113 citations counted in INSPIRE as of 15 Jul 2020

%\cite{Vakili:2010rf}
\bibitem{Vakili:2010rf}
B.~Vakili,
``Classical and quantum dynamics of a perfect fluid scalar-metric cosmology,''
Phys. Lett. B \textbf{688} (2010), 129-136
%doi:10.1016/j.physletb.2010.04.007
[arXiv:1004.0306 [gr-qc]].
%22 citations counted in INSPIRE as of 08 Jul 2020

%\cite{Vakili:2009he}
\bibitem{Vakili:2009he}
B.~Vakili,
``Quadratic quantum cosmology with Schutz' perfect fluid,''
Class. Quant. Grav. \textbf{27} (2010), 025008
%doi:10.1088/0264-9381/27/2/025008
[arXiv:0908.0998 [gr-qc]].
%16 citations counted in INSPIRE as of 08 Jul 2020

%\cite{Xu:2016rdf}
\bibitem{Xu:2016rdf}
M.~X.~Xu, T.~Harko and S.~D.~Liang,
``Quantum Cosmology of $f(R,T)$ gravity,''
Eur. Phys. J. C \textbf{76} (2016) no.8, 449
%doi:10.1140/epjc/s10052-016-4303-6
[arXiv:1608.00113 [gr-qc]].
%11 citations counted in INSPIRE as of 08 Jul 2020

%\cite{Capozziello:2013gza}
\bibitem{Capozziello:2013gza}
S.~Capozziello, T.~Harko, F.~S.~N.~Lobo, G.~J.~Olmo and S.~Vignolo,
``The Cauchy problem in hybrid metric-Palatini $f(X)$-gravity,''
Int. J. Geom. Meth. Mod. Phys. \textbf{11} (2014) no.5, 1450042
%doi:10.1142/S021988781450042X
[arXiv:1312.1320 [gr-qc]].
%12 citations counted in INSPIRE as of 08 Jul 2020

%\cite{Olmo:2005zr}
\bibitem{Olmo:2005zr}
G.~J.~Olmo,
``The Gravity Lagrangian according to solar system experiments,''
Phys. Rev. Lett. \textbf{95} (2005), 261102
%doi:10.1103/PhysRevLett.95.261102
[arXiv:gr-qc/0505101 [gr-qc]].
%241 citations counted in INSPIRE as of 16 Jul 2020

%\cite{Olmo:2005hc}
\bibitem{Olmo:2005hc}
G.~J.~Olmo,
``Post-Newtonian constraints on f(R) cosmologies in metric and Palatini formalism,''
Phys. Rev. D \textbf{72} (2005), 083505
%doi:10.1103/PhysRevD.72.083505
[arXiv:gr-qc/0505135 [gr-qc]].
%291 citations counted in INSPIRE as of 13 Jul 2020

%\cite{Dyadina:2019yon}
\bibitem{Dyadina:2019yon}
P.~I.~Dyadina, S.~P.~Labazova and S.~O.~Alexeyev,
``Post-Newtonian limit of hybrid metric-Palatini f(R)-gravity,''
%doi:10.1134/S0044451019110087
[arXiv:1907.06919 [gr-qc]].
%0 citations counted in INSPIRE as of 08 Jul 2020

%\cite{Will:2018mcj}
\bibitem{Will:2018mcj}
C.~M.~Will,
``New General Relativistic Contribution to Mercury’s Perihelion Advance,''
Phys. Rev. Lett. \textbf{120} (2018) no.19, 191101
%doi:10.1103/PhysRevLett.120.191101
[arXiv:1802.05304 [gr-qc]].
%20 citations counted in INSPIRE as of 08 Jul 2020

%\cite{Dyadina:2018ryl}
\bibitem{Dyadina:2018ryl}
P.~I.~Dyadina, N.~A.~Avdeev and S.~O.~Alexeyev,
``Horndeski gravity without screening in binary pulsars,''
Mon. Not. Roy. Astron. Soc. \textbf{483} (2019) no.1, 947-963
%doi:10.1093/mnras/sty3094
[arXiv:1811.05393 [astro-ph.HE]].
%7 citations counted in INSPIRE as of 08 Jul 2020

%\cite{Kausar:2019iwu}
\bibitem{Kausar:2019iwu}
H.~R.~Kausar, R.~Saleem and A.~Ilyas,
``Cosmological inflation in f(X) gravity theory,''
Phys. Dark Univ. \textbf{26} (2019), 100401
%doi:10.1016/j.dark.2019.100401
%1 citations counted in INSPIRE as of 08 Jul 2020

%\cite{Boehmer:2013oxa}
\bibitem{Boehmer:2013oxa}
C.~G.~Böhmer, F.~S.~N.~Lobo and N.~Tamanini,
``Einstein static Universe in hybrid metric-Palatini gravity,''
Phys. Rev. D \textbf{88} (2013) no.10, 104019
%doi:10.1103/PhysRevD.88.104019
[arXiv:1305.0025 [gr-qc]].
%42 citations counted in INSPIRE as of 14 Jul 2020

%\cite{Santos:2016tds}
\bibitem{Santos:2016tds}
J.~Santos, M.~J.~Rebouças and A.~F.~F.~Teixeira,
``Homogeneous Gödel-type solutions in hybrid metric-Palatini gravity,''
Eur. Phys. J. C \textbf{78} (2018) no.7, 567
%doi:10.1140/epjc/s10052-018-6025-4
[arXiv:1611.03985 [gr-qc]].
%9 citations counted in INSPIRE as of 08 Jul 2020

%\cite{Coley:2003mj}
\bibitem{Coley:2003mj}
A.~A.~Coley,
``Dynamical systems and cosmology,''
Astrophys. Space Sci. Libr. \textbf{291} (2003)
%doi:10.1007/978-94-017-0327-7
%65 citations counted in INSPIRE as of 24 Jul 2020

%\cite{Carloni:2015bua}
\bibitem{Carloni:2015bua}
S.~Carloni, T.~Koivisto and F.~S.~N.~Lobo,
``Dynamical system analysis of hybrid metric-Palatini cosmologies,''
Phys. Rev. D \textbf{92} (2015) no.6, 064035
%doi:10.1103/PhysRevD.92.064035
[arXiv:1507.04306 [gr-qc]].
%26 citations counted in INSPIRE as of 08 Jul 2020

%\cite{Odintsov:2017tbc}
\bibitem{Odintsov:2017tbc}
S.~D.~Odintsov and V.~K.~Oikonomou,
``Autonomous dynamical system approach for $f(R)$ gravity,''
Phys. Rev. D \textbf{96} (2017) no.10, 104049
%doi:10.1103/PhysRevD.96.104049
[arXiv:1711.02230 [gr-qc]].
%39 citations counted in INSPIRE as of 22 Jul 2020

%\cite{Odintsov:2018uaw}
\bibitem{Odintsov:2018uaw}
S.~D.~Odintsov and V.~K.~Oikonomou,
``Dynamical Systems Perspective of Cosmological Finite-time Singularities in $f(R)$ Gravity and Interacting Multifluid Cosmology,''
Phys. Rev. D \textbf{98} (2018) no.2, 024013
%doi:10.1103/PhysRevD.98.024013
[arXiv:1806.07295 [gr-qc]].
%29 citations counted in INSPIRE as of 22 Jul 2020

%\cite{Lima:2014aza}
\bibitem{Lima:2014aza}
N.~A.~Lima,
``Dynamics of Linear Perturbations in the hybrid metric-Palatini gravity,''
Phys. Rev. D \textbf{89} (2014) no.8, 083527
%doi:10.1103/PhysRevD.89.083527
[arXiv:1402.4458 [astro-ph.CO]].
%16 citations counted in INSPIRE as of 08 Jul 2020

%\cite{Lima:2015nma}
\bibitem{Lima:2015nma}
N.~A.~Lima and V.~S.-Barreto,
``Constraints on Hybrid Metric-palatini Gravity from Background Evolution,''
Astrophys. J. \textbf{818} (2016) no.2, 186
%doi:10.3847/0004-637X/818/2/186
[arXiv:1501.05786 [astro-ph.CO]].
%13 citations counted in INSPIRE as of 08 Jul 2020

%\cite{Leanizbarrutia:2017xyd}
\bibitem{Leanizbarrutia:2017xyd}
I.~Leanizbarrutia, F.~S.~N.~Lobo and D.~Saez-Gomez,
``Crossing SNe Ia and BAO observational constraints with local ones in hybrid metric-Palatini gravity,''
Phys. Rev. D \textbf{95} (2017) no.8, 084046
%doi:10.1103/PhysRevD.95.084046
[arXiv:1701.08980 [gr-qc]].
%9 citations counted in INSPIRE as of 08 Jul 2020

%\cite{Boehmer:2007kx}
\bibitem{Boehmer:2007kx}
C.~G.~Boehmer, T.~Harko and F.~S.~N.~Lobo,
``Dark matter as a geometric effect in $f(R)$ gravity,''
Astropart. Phys. \textbf{29} (2008), 386-392
%doi:10.1016/j.astropartphys.2008.04.003
[arXiv:0709.0046 [gr-qc]].
%169 citations counted in INSPIRE as of 16 Jul 2020

%\cite{Bohmer:2007fh}
\bibitem{Bohmer:2007fh}
C.~G.~Boehmer, T.~Harko and F.~S.~N.~Lobo,
``Generalized virial theorem in $f(R)$ gravity,''
JCAP \textbf{03} (2008), 024
%doi:10.1088/1475-7516/2008/03/024
[arXiv:0710.0966 [gr-qc]].
%97 citations counted in INSPIRE as of 08 Jul 2020

%\cite{Borka:2015vqa}
\bibitem{Borka:2015vqa}
D.~Borka, S.~Capozziello, P.~Jovanović and V.~Borka Jovanović,
``Probing hybrid modified gravity by stellar motion around Galactic Center,''
Astropart. Phys. \textbf{79} (2016), 41-48
%doi:10.1016/j.astropartphys.2016.03.002
[arXiv:1504.07832 [gr-qc]].
%12 citations counted in INSPIRE as of 08 Jul 2020

%\cite{Capozziello:2013yha}
\bibitem{Capozziello:2013yha}
S.~Capozziello, T.~Harko, T.~S.~Koivisto, F.~S.~N.~Lobo and G.~J.~Olmo,
``Galactic rotation curves in hybrid metric-Palatini gravity,''
Astropart. Phys. \textbf{50-52} (2013), 65-75
%doi:10.1016/j.astropartphys.2013.09.005
[arXiv:1307.0752 [gr-qc]].
%20 citations counted in INSPIRE as of 08 Jul 2020

%\cite{Capozziello:2012qt}
\bibitem{Capozziello:2012qt}
S.~Capozziello, T.~Harko, T.~S.~Koivisto, F.~S.~N.~Lobo and G.~J.~Olmo,
``The virial theorem and the dark matter problem in hybrid metric-Palatini gravity,''
JCAP \textbf{07} (2013), 024
%doi:10.1088/1475-7516/2013/07/024
[arXiv:1212.5817 [physics.gen-ph]].
%55 citations counted in INSPIRE as of 08 Jul 2020

%\cite{Danila:2016lqx}
\bibitem{Danila:2016lqx}
B.~Danila, T.~Harko, F.~S.~N.~Lobo and M.~K.~Mak,
``Hybrid metric-Palatini stars,''
Phys. Rev. D \textbf{95} (2017) no.4, 044031
%doi:10.1103/PhysRevD.95.044031
[arXiv:1608.02783 [gr-qc]].
%12 citations counted in INSPIRE as of 08 Jul 2020

%\cite{Antoniadis:2013pzd}
\bibitem{Antoniadis:2013pzd}
J.~Antoniadis, P.~C.~C.~Freire, N.~Wex, T.~M.~Tauris, R.~S.~Lynch, M.~H.~van Kerkwijk, M.~Kramer, C.~Bassa, V.~S.~Dhillon, T.~Driebe, J.~W.~T.~Hessels, V.~M.~Kaspi, V.~I.~Kondratiev, N.~Langer, T.~R.~Marsh, M.~A.~McLaughlin, T.~T.~Pennucci, S.~M.~Ransom, I.~H.~Stairs, J.~van Leeuwen, J.~P.~W.~Verbiest and D.~G.~Whelan,
``A Massive Pulsar in a Compact Relativistic Binary,''
Science \textbf{340} (2013), 6131
%doi:10.1126/science.1233232
[arXiv:1304.6875 [astro-ph.HE]].
%1356 citations counted in INSPIRE as of 23 Jul 2020

%\cite{Linares:2018ppq}
\bibitem{Linares:2018ppq}
M.~Linares, T.~Shahbaz and J.~Casares,
``Peering into the dark side: Magnesium lines establish a massive neutron star in PSR J2215+5135,''
Astrophys. J. \textbf{859} (2018) no.1, 54
%doi:10.3847/1538-4357/aabde6
[arXiv:1805.08799 [astro-ph.HE]].
%73 citations counted in INSPIRE as of 23 Jul 2020

%\cite{Cromartie:2019kug}
\bibitem{Cromartie:2019kug}
H.~T.~Cromartie, E.~Fonseca, S.~M.~Ransom, P.~B.~Demorest, Z.~Arzoumanian, H.~Blumer, P.~R.~Brook, M.~E.~DeCesar, T.~Dolch, J.~A.~Ellis, R.~D.~Ferdman, E.~C.~Ferrara, N.~Garver-Daniels, P.~A.~Gentile, M.~L.~Jones, M.~T.~Lam, D.~R.~Lorimer, R.~S.~Lynch, M.~A.~McLaughlin, C.~Ng, D.~J.~Nice, T.~T.~Pennucci, R.~Spiewak, I.~H.~Stairs, K.~Stovall, J.~K.~Swiggum and W.~Zhu,
``Relativistic Shapiro delay measurements of an extremely massive millisecond pulsar,''
Nature Astron. \textbf{4} (2019) no.1, 72-76
%doi:10.1038/s41550-019-0880-2
[arXiv:1904.06759 [astro-ph.HE]].
%259 citations counted in INSPIRE as of 23 Jul 2020

%\cite{Capozziello:2015yza}
\bibitem{Capozziello:2015yza}
S.~Capozziello, M.~De Laurentis, R.~Farinelli and S.~D.~Odintsov,
``Mass-radius relation for neutron stars in f(R) gravity,''
Phys. Rev. D \textbf{93} (2016) no.2, 023501
%doi:10.1103/PhysRevD.93.023501
[arXiv:1509.04163 [gr-qc]].
%102 citations counted in INSPIRE as of 22 Jul 2020

%\cite{Danila:2018xya}
\bibitem{Danila:2018xya}
B.~Dǎnilǎ, T.~Harko, F.~S.~N.~Lobo and M.~K.~Mak,
``Spherically symmetric static vacuum solutions in hybrid metric-Palatini gravity,''
Phys. Rev. D \textbf{99} (2019) no.6, 064028
%doi:10.1103/PhysRevD.99.064028
[arXiv:1811.02742 [gr-qc]].
%10 citations counted in INSPIRE as of 14 Jul 2020

%\cite{Bronnikov:2019ugl}
\bibitem{Bronnikov:2019ugl}
K.~A.~Bronnikov,
``Spherically symmetric black holes and wormholes in hybrid metric-Palatini gravity,''
%doi:10.1134/S0202289319040030
[arXiv:1908.02012 [gr-qc]].
%4 citations counted in INSPIRE as of 08 Jul 2020

%\cite{Bronnikov:2020vgg}
\bibitem{Bronnikov:2020vgg}
K.~A.~Bronnikov, S.~V.~Bolokhov and M.~V.~Skvortsova,
``Hybrid metric-Palatini gravity: black holes, wormholes, singularities and instabilities,''
[arXiv:2006.00559 [gr-qc]].
%1 citations counted in INSPIRE as of 08 Jul 2020

%\cite{Capozziello:2013vna}
\bibitem{Capozziello:2013vna}
S.~Capozziello, F.~S.~N.~Lobo and J.~P.~Mimoso,
``Energy conditions in modified gravity,''
Phys. Lett. B \textbf{730} (2014), 280-283
%doi:10.1016/j.physletb.2014.01.066
[arXiv:1312.0784 [gr-qc]].
%54 citations counted in INSPIRE as of 08 Jul 2020

%\cite{Capozziello:2014bqa}
\bibitem{Capozziello:2014bqa}
S.~Capozziello, F.~S.~N.~Lobo and J.~P.~Mimoso,
``Generalized energy conditions in Extended Theories of Gravity,''
Phys. Rev. D \textbf{91} (2015) no.12, 124019
%doi:10.1103/PhysRevD.91.124019
[arXiv:1407.7293 [gr-qc]].
%58 citations counted in INSPIRE as of 08 Jul 2020

%\cite{Mimoso:2014ofa}
\bibitem{Mimoso:2014ofa}
J.~P.~Mimoso, F.~S.~N.~Lobo and S.~Capozziello,
``Extended Theories of Gravity with Generalized Energy Conditions,''
J. Phys. Conf. Ser. \textbf{600} (2015), 012047
%doi:10.1088/1742-6596/600/1/012047
[arXiv:1412.6670 [gr-qc]].
%8 citations counted in INSPIRE as of 08 Jul 2020

%\cite{Lobo:2009ip}
\bibitem{Lobo:2009ip}
F.~S.~N.~Lobo and M.~A.~Oliveira,
``Wormhole geometries in f(R) modified theories of gravity,''
Phys. Rev. D \textbf{80} (2009), 104012
%doi:10.1103/PhysRevD.80.104012
[arXiv:0909.5539 [gr-qc]].
%215 citations counted in INSPIRE as of 08 Jul 2020

%\cite{Harko:2013yb}
\bibitem{Harko:2013yb}
T.~Harko, F.~S.~N.~Lobo, M.~K.~Mak and S.~V.~Sushkov,
``Modified-gravity wormholes without exotic matter,''
Phys. Rev. D \textbf{87} (2013) no.6, 067504
%doi:10.1103/PhysRevD.87.067504
[arXiv:1301.6878 [gr-qc]].
%121 citations counted in INSPIRE as of 14 Jul 2020

%\cite{Bohmer:2011si}
\bibitem{Bohmer:2011si}
C.~G.~Boehmer, T.~Harko and F.~S.~N.~Lobo,
``Wormhole geometries in modified teleparralel gravity and the energy conditions,''
Phys. Rev. D \textbf{85} (2012), 044033
%doi:10.1103/PhysRevD.85.044033
[arXiv:1110.5756 [gr-qc]].
%129 citations counted in INSPIRE as of 08 Jul 2020

%\cite{Lobo:2007qi}
\bibitem{Lobo:2007qi}
F.~S.~N.~Lobo,
``A General class of braneworld wormholes,''
Phys. Rev. D \textbf{75} (2007), 064027
%doi:10.1103/PhysRevD.75.064027
[arXiv:gr-qc/0701133 [gr-qc]].
%98 citations counted in INSPIRE as of 08 Jul 2020

%\cite{MontelongoGarcia:2010xd}
\bibitem{MontelongoGarcia:2010xd}
N.~Montelongo Garcia and F.~S.~N.~Lobo,
``Nonminimal curvature-matter coupled wormholes with matter satisfying the null energy condition,''
Class. Quant. Grav. \textbf{28} (2011), 085018
%doi:10.1088/0264-9381/28/8/085018
[arXiv:1012.2443 [gr-qc]].
%95 citations counted in INSPIRE as of 08 Jul 2020

%\cite{Garcia:2010xb}
\bibitem{Garcia:2010xb}
N.~M.~Garcia and F.~S.~N.~Lobo,
``Wormhole geometries supported by a nonminimal curvature-matter coupling,''
Phys. Rev. D \textbf{82} (2010), 104018
%doi:10.1103/PhysRevD.82.104018
[arXiv:1007.3040 [gr-qc]].
%83 citations counted in INSPIRE as of 08 Jul 2020

%\cite{Lobo:2008zu}
\bibitem{Lobo:2008zu}
F.~S.~N.~Lobo,
``General class of wormhole geometries in conformal Weyl gravity,''
Class. Quant. Grav. \textbf{25} (2008), 175006
%doi:10.1088/0264-9381/25/17/175006
[arXiv:0801.4401 [gr-qc]].
%72 citations counted in INSPIRE as of 08 Jul 2020

%\cite{Mehdizadeh:2015jra}
\bibitem{Mehdizadeh:2015jra}
M.~R.~Mehdizadeh, M.~Kord Zangeneh and F.~S.~N.~Lobo,
``Einstein-Gauss-Bonnet traversable wormholes satisfying the weak energy condition,''
Phys. Rev. D \textbf{91} (2015) no.8, 084004
%doi:10.1103/PhysRevD.91.084004
[arXiv:1501.04773 [gr-qc]].
%66 citations counted in INSPIRE as of 14 Jul 2020

%\cite{Boehmer:2007rm}
\bibitem{Boehmer:2007rm}
C.~G.~Boehmer, T.~Harko and F.~S.~N.~Lobo,
``Conformally symmetric traversable wormholes,''
Phys. Rev. D \textbf{76} (2007), 084014
%doi:10.1103/PhysRevD.76.084014
[arXiv:0708.1537 [gr-qc]].
%56 citations counted in INSPIRE as of 08 Jul 2020

%\cite{Zangeneh:2015jda}
\bibitem{Zangeneh:2015jda}
M.~Kord Zangeneh, F.~S.~N.~Lobo and M.~H.~Dehghani,
``Traversable wormholes satisfying the weak energy condition in third-order Lovelock gravity,''
Phys. Rev. D \textbf{92} (2015) no.12, 124049
%doi:10.1103/PhysRevD.92.124049
[arXiv:1510.07089 [gr-qc]].
%46 citations counted in INSPIRE as of 08 Jul 2020

%\cite{Lobo:2010sb}
\bibitem{Lobo:2010sb}
F.~S.~N.~Lobo and M.~A.~Oliveira,
``General class of vacuum Brans-Dicke wormholes,''
Phys. Rev. D \textbf{81} (2010), 067501
%doi:10.1103/PhysRevD.81.067501
[arXiv:1001.0995 [gr-qc]].
%43 citations counted in INSPIRE as of 08 Jul 2020

%\cite{Harko:2013aya}
\bibitem{Harko:2013aya}
T.~Harko, F.~S.~N.~Lobo, M.~K.~Mak and S.~V.~Sushkov,
``Wormhole geometries in Eddington-Inspired Born–Infeld gravity,''
Mod. Phys. Lett. A \textbf{30} (2015) no.35, 1550190
%doi:10.1142/S0217732315501904
[arXiv:1307.1883 [gr-qc]].
%39 citations counted in INSPIRE as of 08 Jul 2020

%\cite{Korolev:2020ohi}
\bibitem{Korolev:2020ohi}
R.~Korolev, F.~S.~N.~Lobo and S.~V.~Sushkov,
``General constraints on Horndeski wormhole throats,''
Phys. Rev. D \textbf{101} (2020) no.12, 124057
%doi:10.1103/PhysRevD.101.124057
[arXiv:2004.12382 [gr-qc]].
%0 citations counted in INSPIRE as of 08 Jul 2020

%\cite{Capozziello:2012hr}
\bibitem{Capozziello:2012hr}
S.~Capozziello, T.~Harko, T.~S.~Koivisto, F.~S.~N.~Lobo and G.~J.~Olmo,
``Wormholes supported by hybrid metric-Palatini gravity,''
Phys. Rev. D \textbf{86} (2012), 127504
%doi:10.1103/PhysRevD.86.127504
[arXiv:1209.5862 [gr-qc]].
%67 citations counted in INSPIRE as of 08 Jul 2020

%\cite{Azizi:2015ina}
\bibitem{Azizi:2015ina}
T.~Azizi and N.~Borhani,
``Thermodynamics in hybrid metric-Palatini gravity,''
Astrophys. Space Sci. \textbf{357} (2015) no.2, 146.
%doi:10.1007/s10509-015-2383-7
%4 citations counted in INSPIRE as of 08 Jul 2020

%\cite{Fu:2016szo}
\bibitem{Fu:2016szo}
Q.~M.~Fu, L.~Zhao, B.~M.~Gu, K.~Yang and Y.~X.~Liu,
``Hybrid metric-Palatini brane system,''
Phys. Rev. D \textbf{94} (2016) no.2, 024020
%doi:10.1103/PhysRevD.94.024020
[arXiv:1601.06546 [gr-qc]].
%3 citations counted in INSPIRE as of 08 Jul 2020

%\cite{VargasdosSantos:2017ggl}
\bibitem{VargasdosSantos:2017ggl}
M.~Vargas dos Santos, J.~S.~Alcaniz, D.~F.~Mota and S.~Capozziello,
``Screening mechanisms in hybrid metric-Palatini gravity,''
Phys. Rev. D \textbf{97} (2018) no.10, 104010
%doi:10.1103/PhysRevD.97.104010
[arXiv:1712.03831 [gr-qc]].
%3 citations counted in INSPIRE as of 08 Jul 2020

%\cite{Kausar:2018ipo}
\bibitem{Kausar:2018ipo}
H.~R.~Kausar,
``Gravitational wave solutions in hybrid metric-Palatini theory,''
Astrophys. Space Sci. \textbf{363} (2018) no.11, 238.
%doi:10.1007/s10509-018-3458-z
%0 citations counted in INSPIRE as of 08 Jul 2020

%\cite{Harko:2020oxq}
\bibitem{Harko:2020oxq}
T.~Harko, F.~S.~N.~Lobo and H.~M.~R.~da Silva,
``Cosmic stringlike objects in hybrid metric-Palatini gravity,''
Phys. Rev. D \textbf{101} (2020) no.12, 124050
%doi:10.1103/PhysRevD.101.124050
[arXiv:2003.09751 [gr-qc]].
%2 citations counted in INSPIRE as of 23 Jul 2020

%\cite{Borowiec:2014wva}
\bibitem{Borowiec:2014wva}
A.~Borowiec, S.~Capozziello, M.~De Laurentis, F.~S.~N.~Lobo, A.~Paliathanasis, M.~Paolella and A.~Wojnar,
``Invariant solutions and Noether symmetries in Hybrid Gravity,''
Phys. Rev. D \textbf{91} (2015) no.2, 023517
%doi:10.1103/PhysRevD.91.023517
[arXiv:1407.4313 [gr-qc]].
%23 citations counted in INSPIRE as of 08 Jul 2020

%\cite{Tamanini:2013ltp}
\bibitem{Tamanini:2013ltp}
N.~Tamanini and C.~G.~Boehmer,
``Generalized hybrid metric-Palatini gravity,''
Phys. Rev. D \textbf{87} (2013) no.8, 084031
%doi:10.1103/PhysRevD.87.084031
[arXiv:1302.2355 [gr-qc]].
%39 citations counted in INSPIRE as of 15 Jul 2020

%\cite{Rosa:2017jld}
\bibitem{Rosa:2017jld}
J.~L.~Rosa, S.~Carloni, J.~P.~d.~Lemos and F.~S.~N.~Lobo,
``Cosmological solutions in generalized hybrid metric-Palatini gravity,''
Phys. Rev. D \textbf{95} (2017) no.12, 124035
%doi:10.1103/PhysRevD.95.124035
[arXiv:1703.03335 [gr-qc]].
%11 citations counted in INSPIRE as of 08 Jul 2020

%\cite{Rosa:2019ejh}
\bibitem{Rosa:2019ejh}
J.~L.~Rosa, S.~Carloni and J.~P.~S.~Lemos,
``Cosmological phase space of generalized hybrid metric-Palatini theories of gravity,''
Phys. Rev. D \textbf{01} (2020), 104056
%doi:10.1103/PhysRevD.101.104056
[arXiv:1908.07778 [gr-qc]].
%3 citations counted in INSPIRE as of 23 Jul 2020

%\cite{Rosa:2018jwp}
\bibitem{Rosa:2018jwp}
J.~L.~Rosa, J.~P.~S.~Lemos and F.~S.~N.~Lobo,
``Wormholes in generalized hybrid metric-Palatini gravity obeying the matter null energy condition everywhere,''
Phys. Rev. D \textbf{98} (2018) no.6, 064054
%doi:10.1103/PhysRevD.98.064054
[arXiv:1808.08975 [gr-qc]].
%12 citations counted in INSPIRE as of 08 Jul 2020

%\cite{Sa:2020qfd}
\bibitem{Sa:2020qfd}
P.~M.~Sá,
``Unified description of dark energy and dark matter within the generalized hybrid metric-Palatini theory of gravity,''
Universe \textbf{6} (2020) no.6, 78
%doi:10.3390/universe6060078
[arXiv:2002.09446 [gr-qc]].
%2 citations counted in INSPIRE as of 15 Jul 2020

%\cite{Sa:2020fvn}
\bibitem{Sa:2020fvn}
P.~M.~Sá,
``Triple unification of inflation, dark energy, and dark matter in two-scalar-field cosmology,''
[arXiv:2007.07109 [gr-qc]].
%0 citations counted in INSPIRE as of 23 Jul 2020

%\cite{Borowiec:2020lfx}
\bibitem{Borowiec:2020lfx}
A.~Borowiec and A.~Kozak,
``New class of hybrid metric-Palatini scalar-tensor theories of gravity,''
JCAP \textbf{07} (2020), 003
%doi:10.1088/1475-7516/2020/07/003
[arXiv:2003.02741 [gr-qc]].
%4 citations counted in INSPIRE as of 08 Jul 2020

%\cite{Koivisto:2013kwa}
\bibitem{Koivisto:2013kwa}
T.~S.~Koivisto and N.~Tamanini,
``Ghosts in pure and hybrid formalisms of gravity theories: A unified analysis,''
Phys. Rev. D \textbf{87} (2013) no.10, 104030
%doi:10.1103/PhysRevD.87.104030
[arXiv:1304.3607 [gr-qc]].
%21 citations counted in INSPIRE as of 08 Jul 2020

%\cite{Bombacigno:2019did}
\bibitem{Bombacigno:2019did}
F.~Bombacigno, F.~Moretti and G.~Montani,
``Scalar modes in extended hybrid metric-Palatini gravity: weak field phenomenology,''
Phys. Rev. D \textbf{100} (2019) no.12, 124036
%doi:10.1103/PhysRevD.100.124036
[arXiv:1907.11949 [gr-qc]].
%4 citations counted in INSPIRE as of 08 Jul 2020

%\cite{Rosa:2020uoi}
\bibitem{Rosa:2020uoi}
J.~L.~Rosa, J.~P.~S.~Lemos and F.~S.~N.~Lobo,
``Stability of Kerr black holes in generalized hybrid metric-Palatini gravity,''
Phys. Rev. D \textbf{101} (2020), 044055
%doi:10.1103/PhysRevD.101.044055
[arXiv:2003.00090 [gr-qc]].
%3 citations counted in INSPIRE as of 08 Jul 2020

%\cite{Mukohyama:2003nw}
\bibitem{Mukohyama:2003nw}
S.~Mukohyama and L.~Randall,
``A Dynamical approach to the cosmological constant,''
Phys. Rev. Lett. \textbf{92} (2004), 211302
%doi:10.1103/PhysRevLett.92.211302
[arXiv:hep-th/0306108 [hep-th]].
%85 citations counted in INSPIRE as of 08 Jul 2020

%\cite{Nojiri:2004bi}
\bibitem{Nojiri:2004bi}
S.~Nojiri and S.~D.~Odintsov,
``Gravity assisted dark energy dominance and cosmic acceleration,''
Phys. Lett. B \textbf{599} (2004), 137-142
%doi:10.1016/j.physletb.2004.08.045
[arXiv:astro-ph/0403622 [astro-ph]].
%237 citations counted in INSPIRE as of 08 Jul 2020

%\cite{Allemandi:2005qs}
\bibitem{Allemandi:2005qs}
G.~Allemandi, A.~Borowiec, M.~Francaviglia and S.~D.~Odintsov,
``Dark energy dominance and cosmic acceleration in first order formalism,''
Phys. Rev. D \textbf{72} (2005), 063505
%doi:10.1103/PhysRevD.72.063505
[arXiv:gr-qc/0504057 [gr-qc]].
%274 citations counted in INSPIRE as of 08 Jul 2020

%\cite{Sheikhahmadi:2019gzs}
\bibitem{Sheikhahmadi:2019gzs}
H.~Sheikhahmadi, A.~Mohammadi, A.~Aghamohammadi, T.~Harko, R.~Herrera, C.~Corda, A.~Abebe and K.~Saaidi,
``Constraining chameleon field driven warm inflation with Planck 2018 data,''
Eur. Phys. J. C \textbf{79} (2019) no.12, 1038
%doi:10.1140/epjc/s10052-019-7571-0
[arXiv:1907.10966 [gr-qc]].
%4 citations counted in INSPIRE as of 08 Jul 2020

%\cite{Harko:2010hw}
\bibitem{Harko:2010hw}
T.~Harko, T.~S.~Koivisto and F.~S.~N.~Lobo,
``Palatini formulation of modified gravity with a nonminimal curvature-matter coupling,''
Mod. Phys. Lett. A \textbf{26} (2011), 1467-1480
%doi:10.1142/S0217732311035869
[arXiv:1007.4415 [gr-qc]].
%65 citations counted in INSPIRE as of 08 Jul 2020

%\cite{Koivisto:2005yk}
\bibitem{Koivisto:2005yk}
T.~Koivisto,
``Covariant conservation of energy momentum in modified gravities,''
Class. Quant. Grav. \textbf{23} (2006), 4289-4296
%doi:10.1088/0264-9381/23/12/N01
[arXiv:gr-qc/0505128 [gr-qc]].
%243 citations counted in INSPIRE as of 08 Jul 2020

%\cite{Bertolami:2008ab}
\bibitem{Bertolami:2008ab}
O.~Bertolami, F.~S.~N.~Lobo and J.~Paramos,
``Non-minimum coupling of perfect fluids to curvature,''
Phys. Rev. D \textbf{78} (2008), 064036
%doi:10.1103/PhysRevD.78.064036
[arXiv:0806.4434 [gr-qc]].
%179 citations counted in INSPIRE as of 21 Jul 2020

%\cite{Faraoni:2009rk}
\bibitem{Faraoni:2009rk}
V.~Faraoni,
``The Lagrangian description of perfect fluids and modified gravity with an extra force,''
Phys. Rev. D \textbf{80} (2009), 124040
%doi:10.1103/PhysRevD.80.124040
[arXiv:0912.1249 [astro-ph.GA]].
%77 citations counted in INSPIRE as of 08 Jul 2020

%\cite{Bertolami:2013raa}
\bibitem{Bertolami:2013raa}
O.~Bertolami and J.~Páramos,
``Homogeneous spherically symmetric bodies with a non-minimal coupling between curvature and matter: the choice of the Lagrangian density for matter,''
Gen. Rel. Grav. \textbf{47} (2015) no.1, 1835
%doi:10.1007/s10714-014-1835-7
[arXiv:1306.1177 [gr-qc]].
%11 citations counted in INSPIRE as of 08 Jul 2020

%\cite{Minazzoli:2013bva}
\bibitem{Minazzoli:2013bva}
O.~Minazzoli,
``Conservation laws in theories with universal gravity/matter coupling,''
Phys. Rev. D \textbf{88} (2013), 027506
%doi:10.1103/PhysRevD.88.027506
[arXiv:1307.1590 [gr-qc]].
%26 citations counted in INSPIRE as of 08 Jul 2020

%\cite{Harko:2010zi}
\bibitem{Harko:2010zi}
T.~Harko,
``The matter Lagrangian and the energy-momentum tensor in modified gravity with non-minimal coupling between matter and geometry,''
Phys. Rev. D \textbf{81} (2010), 044021
%doi:10.1103/PhysRevD.81.044021
[arXiv:1001.5349 [gr-qc]].
%59 citations counted in INSPIRE as of 08 Jul 2020

%\cite{Faraoni:2007sn}
\bibitem{Faraoni:2007sn}
V.~Faraoni,
``A Viability criterion for modified gravity with an extra force,''
Phys. Rev. D \textbf{76} (2007), 127501
%doi:10.1103/PhysRevD.76.127501
[arXiv:0710.1291 [gr-qc]].
%99 citations counted in INSPIRE as of 08 Jul 2020

%\cite{Faraoni:2007yn}
\bibitem{Faraoni:2007yn}
V.~Faraoni,
``de Sitter space and the equivalence between f(R) and scalar-tensor gravity,''
Phys. Rev. D \textbf{75} (2007), 067302
%doi:10.1103/PhysRevD.75.067302
[arXiv:gr-qc/0703044 [gr-qc]].
%158 citations counted in INSPIRE as of 10 Jul 2020

%\cite{Teyssandier:1983zz}
\bibitem{Teyssandier:1983zz}
P.~Teyssandier and P.~Tourrenc,
``The Cauchy problem for the R+R**2 theories of gravity without torsion,''
J. Math. Phys. \textbf{24} (1983), 2793.
%doi:10.1063/1.525659
%234 citations counted in INSPIRE as of 23 Jul 2020

%\cite{Wands:1993uu}
\bibitem{Wands:1993uu}
D.~Wands,
``Extended gravity theories and the Einstein-Hilbert action,''
Class. Quant. Grav. \textbf{11} (1994), 269-280
%doi:10.1088/0264-9381/11/1/025
[arXiv:gr-qc/9307034 [gr-qc]].
%304 citations counted in INSPIRE as of 09 Jul 2020

%\cite{Whitt:1984pd}
\bibitem{Whitt:1984pd}
B.~Whitt,
``Fourth Order Gravity as General Relativity Plus Matter,''
Phys. Lett. B \textbf{145} (1984), 176-178.
%doi:10.1016/0370-2693(84)90332-0
%541 citations counted in INSPIRE as of 22 Jul 2020

%\cite{Harko:2008qz}
\bibitem{Harko:2008qz}
T.~Harko,
``Modified gravity with arbitrary coupling between matter and geometry,''
Phys. Lett. B \textbf{669} (2008), 376-379
%doi:10.1016/j.physletb.2008.10.007
[arXiv:0810.0742 [gr-qc]].
%112 citations counted in INSPIRE as of 08 Jul 2020

%\cite{Harko:2012ve}
\bibitem{Harko:2012ve}
T.~Harko and F.~S.~N.~Lobo,
``Geodesic deviation, Raychaudhuri equation, and tidal forces in modified gravity with an arbitrary curvature-matter coupling,''
%Phys. Rev. D \textbf{86} (2012), 124034
doi:10.1103/PhysRevD.86.124034
[arXiv:1210.8044 [gr-qc]].
%26 citations counted in INSPIRE as of 08 Jul 2020

%\cite{Wang:2012rw}
\bibitem{Wang:2012rw}
J.~Wang and K.~Liao,
``Energy conditions in f(R, L(m)) gravity,''
Class. Quant. Grav. \textbf{29} (2012), 215016
%doi:10.1088/0264-9381/29/21/215016
[arXiv:1212.4656 [physics.gen-ph]].
%37 citations counted in INSPIRE as of 08 Jul 2020

%\cite{Huang:2013dca}
\bibitem{Huang:2013dca}
R.~N.~Huang,
``The Wheeler-DeWitt equation of $f(R,L_m)$ gravity in minisuperspace,''
[arXiv:1304.5309 [gr-qc]].
%13 citations counted in INSPIRE as of 08 Jul 2020

%\cite{Tian:2014mta}
\bibitem{Tian:2014mta}
D.~W.~Tian and I.~Booth,
``Lessons from $f(R,R_c^2,R_m^2, L_m)$ gravity: Smooth Gauss-Bonnet limit, energy-momentum conservation and nonminimal coupling,''
Phys. Rev. D \textbf{90} (2014), 024059
%doi:10.1103/PhysRevD.90.024059
[arXiv:1404.7823 [gr-qc]].
%6 citations counted in INSPIRE as of 08 Jul 2020

%\cite{Harko:2012ar}
\bibitem{Harko:2012ar}
T.~Harko and F.~S.~N.~Lobo,
``Generalized dark gravity,''
Int. J. Mod. Phys. D \textbf{21} (2012), 1242019
%doi:10.1142/S0218271812420199
[arXiv:1205.3284 [gr-qc]].
%10 citations counted in INSPIRE as of 08 Jul 2020

%\cite{Haghani:2013oma}
\bibitem{Haghani:2013oma}
Z.~Haghani, T.~Harko, F.~S.~N.~Lobo, H.~R.~Sepangi and S.~Shahidi,
``Further matters in space-time geometry: $f(R, T, R_{\mu\nu} T^{\mu\nu})$ gravity,''
Phys. Rev. D \textbf{88} (2013) no.4, 044023
%doi:10.1103/PhysRevD.88.044023
[arXiv:1304.5957 [gr-qc]].
%144 citations counted in INSPIRE as of 08 Jul 2020

%\cite{Odintsov:2013iba}
\bibitem{Odintsov:2013iba}
S.~D.~Odintsov and D.~Sáez-Gómez,
``$f(R, T, R_{\mu\nu} T^{\mu\nu})$ gravity phenomenology and $\Lambda$CDM universe,''
Phys. Lett. B \textbf{725} (2013), 437-444
%doi:10.1016/j.physletb.2013.07.026
[arXiv:1304.5411 [gr-qc]].
%129 citations counted in INSPIRE as of 08 Jul 2020

%\cite{Nojiri:2009th}
\bibitem{Nojiri:2009th}
S.~Nojiri and S.~D.~Odintsov,
``Covariant Horava-like renormalizable gravity and its FRW cosmology,''
Phys. Rev. D \textbf{81} (2010), 043001
%doi:10.1103/PhysRevD.81.043001
[arXiv:0905.4213 [hep-th]].
%103 citations counted in INSPIRE as of 08 Jul 2020

%\cite{Horava:2009uw}
\bibitem{Horava:2009uw}
P.~Horava,
``Quantum Gravity at a Lifshitz Point,''
Phys. Rev. D \textbf{79} (2009), 084008
%doi:10.1103/PhysRevD.79.084008
[arXiv:0901.3775 [hep-th]].
%1924 citations counted in INSPIRE as of 22 Jul 2020

%\cite{Nojiri:2010kx}
\bibitem{Nojiri:2010kx}
S.~Nojiri and S.~D.~Odintsov,
``Covariant power-counting renormalizable gravity: Lorentz symmetry breaking and accelerating early-time FRW universe,''
Phys. Rev. D \textbf{83} (2011), 023001
%doi:10.1103/PhysRevD.83.023001
[arXiv:1007.4856 [hep-th]].
%31 citations counted in INSPIRE as of 08 Jul 2020

%\cite{Dolgov:2003px}
\bibitem{Dolgov:2003px}
A.~D.~Dolgov and M.~Kawasaki,
``Can modified gravity explain accelerated cosmic expansion?,''
Phys. Lett. B \textbf{573} (2003), 1-4
%doi:10.1016/j.physletb.2003.08.039
[arXiv:astro-ph/0307285 [astro-ph]].
%599 citations counted in INSPIRE as of 22 Jul 2020

%\cite{Faraoni:2006sy}
\bibitem{Faraoni:2006sy}
V.~Faraoni,
``Matter instability in modified gravity,''
Phys. Rev. D \textbf{74} (2006), 104017
%doi:10.1103/PhysRevD.74.104017
[arXiv:astro-ph/0610734 [astro-ph]].
%217 citations counted in INSPIRE as of 10 Jul 2020

%\cite{Sotiriou:2006sf}
\bibitem{Sotiriou:2006sf}
T.~P.~Sotiriou,
``Curvature scalar instability in f(R) gravity,''
Phys. Lett. B \textbf{645} (2007), 389-392
%doi:10.1016/j.physletb.2007.01.003
[arXiv:gr-qc/0611107 [gr-qc]].
%78 citations counted in INSPIRE as of 08 Jul 2020

%\cite{Bertolami:2009cd}
\bibitem{Bertolami:2009cd}
O.~Bertolami and M.~C.~Sequeira,
``Energy Conditions and Stability in $f(R)$ theories of gravity with non-minimal coupling to matter,''
Phys. Rev. D \textbf{79} (2009), 104010
%doi:10.1103/PhysRevD.79.104010
[arXiv:0903.4540 [gr-qc]].
%126 citations counted in INSPIRE as of 21 Jul 2020

%\cite{Wang:2012mws}
\bibitem{Wang:2012mws}
J.~Wang, Y.~B.~Wu, Y.~X.~Guo, W.~Q.~Yang and L.~Wang,
``Energy Conditions and Stability in generalized $f(R)$ gravity with arbitrary coupling between matter and geometry,''
Phys. Lett. B \textbf{689} (2010), 133-138
%doi:10.1016/j.physletb.2010.04.063
[arXiv:1212.4921 [gr-qc]].
%44 citations counted in INSPIRE as of 08 Jul 2020

%\cite{Harko:2015pma}
\bibitem{Harko:2015pma}
T.~Harko, F.~S.~N.~Lobo, J.~P.~Mimoso and D.~Pavón,
``Gravitational induced particle production through a nonminimal curvature–matter coupling,''
Eur. Phys. J. C \textbf{75} (2015), 386
%doi:10.1140/epjc/s10052-015-3620-5
[arXiv:1508.02511 [gr-qc]].
%38 citations counted in INSPIRE as of 24 Jul 2020

%\cite{Lobo:2015awa}
\bibitem{Lobo:2015awa}
T.~Harko, F.~S.~N.~Lobo, J.~P.~Mimoso and D.~Pavón,
``Irreversible matter creation processes through a nonminimal curvature-matter coupling,''
%doi:10.1142/9789813226609_0085
[arXiv:1508.03069 [gr-qc]].
%1 citations counted in INSPIRE as of 08 Jul 2020

\bibitem{Pri0}
I. Prigogine and J. G\'{e}h\'{e}niau, Proc. Natl. Acad. Sci. USA \textbf{83}, 6245 (1986).

\bibitem{Pri}
I. Prigogine {\it et al}, J. G\'{e}h\'{e}niau, E. Gunzig, and P. Nardone,
Proc. Natl. Acad. Sci. USA \textbf{85}, 7428 (1988).

%\cite{Calvao:1991wg}
\bibitem{Calvao:1991wg}
M.~O.~Calvao, J.~A.~S.~Lima and I.~Waga,
``On the thermodynamics of matter creation in cosmology,''
Phys. Lett. A \textbf{162} (1992), 223-226.
%doi:10.1016/0375-9601(92)90437-Q
%189 citations counted in INSPIRE as of 08 Jul 2020

%\cite{Mimoso:2013zhp}
\bibitem{Mimoso:2013zhp}
J.~P.~Mimoso and D.~Pavón,
``Entropy evolution of universes with initial and final de Sitter eras,''
Phys. Rev. D \textbf{87} (2013) no.4, 047302
%doi:10.1103/PhysRevD.87.047302
[arXiv:1302.1972 [gr-qc]].
%54 citations counted in INSPIRE as of 08 Jul 2020

%\cite{Bak:1999hd}
\bibitem{Bak:1999hd}
D.~Bak and S.~J.~Rey,
``Cosmic holography,''
Class. Quant. Grav. \textbf{17} (2000), L83
%doi:10.1088/0264-9381/17/15/101
[arXiv:hep-th/9902173 [hep-th]].
%327 citations counted in INSPIRE as of 08 Jul 2020

%\cite{Brown:1992kc}
\bibitem{Brown:1992kc}
J.~D.~Brown,
``Action functionals for relativistic perfect fluids,''
Class. Quant. Grav. \textbf{10} (1993), 1579-1606
%doi:10.1088/0264-9381/10/8/017
[arXiv:gr-qc/9304026 [gr-qc]].
%246 citations counted in INSPIRE as of 21 Jul 2020

%\cite{Steigl:2005fk}
\bibitem{Steigl:2005fk}
R.~Steigl and F.~Hinterleitner,
``Factor ordering in standard quantum cosmology,''
Class. Quant. Grav. \textbf{23} (2006), 3879-3894
%doi:10.1088/0264-9381/23/11/013
[arXiv:gr-qc/0511149 [gr-qc]].
%17 citations counted in INSPIRE as of 08 Jul 2020

%\cite{Baez:1995sj}
\bibitem{Baez:1995sj}
J.~Baez and J.~P.~Muniain,
``Gauge fields, knots and gravity,''
Singapore: World Scientific (1994) 465.
%8 citations counted in INSPIRE as of 08 Jul 2020

%\cite{Abbott:2020khf}
\bibitem{Abbott:2020khf}
R.~Abbott \textit{et al.} [LIGO Scientific and Virgo],
``GW190814: Gravitational Waves from the Coalescence of a 23 Solar Mass Black Hole with a 2.6 Solar Mass Compact Object,''
Astrophys. J. \textbf{896} (2020) no.2, L44
%doi:10.3847/2041-8213/ab960f
[arXiv:2006.12611 [astro-ph.HE]].
%51 citations counted in INSPIRE as of 24 Jul 2020

%\cite{Tsokaros:2020hli}
\bibitem{Tsokaros:2020hli}
A.~Tsokaros, M.~Ruiz and S.~L.~Shapiro,
``GW190814: Spin and equation of state of a neutron star companion,''
[arXiv:2007.05526 [astro-ph.HE]].
%5 citations counted in INSPIRE as of 23 Jul 2020

%\cite{Parker:2009uva}
\bibitem{Parker:2009uva}
L.~E.~Parker and D.~Toms,
``Quantum Field Theory in Curved Spacetime: Quantized Fields and Gravity,''
(Cambridge Monographs on Mathematical Physics). Cambridge: Cambridge University  (2009)
%doi:10.1017/CBO9780511813924
%96 citations counted in INSPIRE as of 17 Jul 2020

%\cite{Haghani:2017vqx}
\bibitem{Haghani:2017vqx}
Z.~Haghani, T.~Harko and S.~Shahidi,
``The Einstein dark energy model,''
Phys. Dark Univ. \textbf{21} (2018), 27-39
%doi:10.1016/j.dark.2018.05.006
[arXiv:1707.00939 [gr-qc]].
%6 citations counted in INSPIRE as of 08 Jul 2020

%\cite{Lobato:2018vpq}
\bibitem{Lobato:2018vpq}
R.~V.~Lobato, G.~A.~Carvalho, A.~G.~Martins and P.~H.~R.~S.~Moraes,
``Energy nonconservation as a link between $f(R,T)$ gravity and noncommutative quantum theory,''
Eur. Phys. J. Plus \textbf{134} (2019) no.4, 132
%doi:10.1140/epjp/i2019-12638-6
[arXiv:1803.08630 [gr-qc]].
%7 citations counted in INSPIRE as of 08 Jul 2020

%\cite{Dzhunushaliev:2013nea}
\bibitem{Dzhunushaliev:2013nea}
V.~Dzhunushaliev, V.~Folomeev, B.~Kleihaus and J.~Kunz,
``Modified gravity from the quantum part of the metric,''
Eur. Phys. J. C \textbf{74} (2014), 2743
%doi:10.1140/epjc/s10052-014-2743-4
[arXiv:1312.0225 [gr-qc]].
%17 citations counted in INSPIRE as of 08 Jul 2020

%\cite{Yang:2015jla}
\bibitem{Yang:2015jla}
R.~Yang,
``Effects of quantum fluctuations of metric on the universe,''
Phys. Dark Univ. \textbf{13} (2016), 87-91
%doi:10.1016/j.dark.2016.04.007
[arXiv:1506.02889 [gr-qc]].
%4 citations counted in INSPIRE as of 08 Jul 2020

%\cite{Liu:2016qfx}
\bibitem{Liu:2016qfx}
X.~Liu, T.~Harko and S.~D.~Liang,
``Cosmological implications of modified gravity induced by quantum metric fluctuations,''
Eur. Phys. J. C \textbf{76} (2016) no.8, 420
%doi:10.1140/epjc/s10052-016-4275-6
[arXiv:1607.04874 [gr-qc]].
%12 citations counted in INSPIRE as of 08 Jul 2020

\end{thebibliography}
\end{document}